\documentclass[fleqn,usenatbib]{mnras}

\usepackage{newtxtext,newtxmath}
\usepackage{comment}
\usepackage{longtable}
\usepackage{tablefootnote}
\usepackage{threeparttablex}
\usepackage{graphicx}
\usepackage{subcaption}
\usepackage{txfonts}
\usepackage{verbatim}
\usepackage{hyperref}
\usepackage{tabularx}
\usepackage{array}
\usepackage{adjustbox}
\usepackage{pdflscape}
\usepackage[dvipsnames]{xcolor}
\newcommand{\lsun}{$\log{(L/L_\odot)}$}
\newcommand{\msun}{$M$/$M_{\odot}\,$}

\usepackage{multicol}
\usepackage{graphicx}
\usepackage{natbib}
\defcitealias{Desomma2020a}{DS20a}
\defcitealias{Desomma2020b}{DS20b}
\defcitealias{Desomma2021}{DS21}
\defcitealias{Desomma2022}{DS22}

\usepackage[bottom]{footmisc}
\graphicspath{{./}{figures/}}
\usepackage[T1]{fontenc}

\DeclareRobustCommand{\VAN}[3]{#2}
\let\VANthebibliography\thebibliography
\def\thebibliography{\DeclareRobustCommand{\VAN}[3]{##3}\VANthebibliography}

\usepackage{graphicx}	
\usepackage{amsmath}	
\usepackage{amssymb}	

\title[Classical Cepheid Pulsation properties in the Rubin-LSST filters]{Classical Cepheid Pulsation properties in the Rubin-LSST filters}

\author[G. De Somma et al.]{
Giulia De Somma,$^{1}$$^{2}$$^{3}$\thanks{E-mail: giulia.desomma@inaf.it}
Marcella Marconi,$^{1}$
Santi Cassisi,$^{2}$$^{4}$ 
Roberto Molinaro,$^{1}$
\newauthor{Anupam Bhardwaj},$^{1}$
Vincenzo Ripepi,$^{1}$
Ilaria Musella,$^{1}$
Adriano Pietrinferni,$^{2}$
\newauthor{Teresa Sicignano},$^{5}$$^{1}$
Erasmo Trentin,$^{6}$$^{7}$$^{1}$
Silvio Leccia,$^{1}$
\\
$^{1}$ INAF-Osservatorio Astronomico di Capodimonte, Via Moiariello 16, 80131 Napoli, Italy\\
$^{2}$ INAF-Osservatorio Astronomico d'Abruzzo, Via Maggini sn, 64100 Teramo, Italy\\
$^{3}$ Istituto Nazionale di Fisica Nucleare (INFN)-Sez. di Napoli, Compl. Univ.di Monte S. Angelo, Edificio G, Via Cinthia, 80126 Napoli, Italy\\
$^{4}$ Istituto Nazionale di Fisica Nucleare (INFN) - Sezione di Pisa, Universit\'a di Pisa, Largo Pontecorvo 3, 56127 Pisa, Italy\\
$^{5}$ Scuola Superiore Meridionale, Largo S. Marcellino 10, 80138 Napoli, Italy \\ 
$^{6}$ Leibniz-Institut f\"ur Astrophysik Potsdam (AIP), An der Sternwarte 16, D-14482 Potsdam, Germany \\
$^{7}$ Institut für Physik und Astronomie, Universität Potsdam, Haus 28, Karl-Liebknecht-Str. 24/25, D-14476 Golm, Potsdam, Germany}

\date{Accepted 202 Month day. Received 202 Month day; in original form 202 Month day}

\pubyear{2021}

\begin{document}
\label{firstpage}
\pagerange{\pageref{firstpage}--\pageref{lastpage}}
\maketitle


\begin{abstract}
Homogeneous multi-wavelength observations of classical Cepheids from the forthcoming Rubin-LSST have the potential to significantly contribute to our understanding of the evolutionary and pulsation properties of these pulsating stars. Updated pulsation models for Classical Cepheid stars have been computed under various assumptions about chemical compositions, including relatively low metallicity ($Z$ = $0.004$ with $Y$ =$0.25$ and $Z$=$0.008$ with $Y$ =$0.25$), solar metallicity ($Z$=$0.02$ with $Y$=$0.28$), and supersolar metallicity environments ($Z$ = $0.03$ with $Y$ = $0.28$).

From the predicted periods, intensity-weighted mean magnitudes, and colors, we have derived the first theoretical pulsation relations in the Rubin-LSST filters (ugrizy),  including period-luminosity-color, period-Wesenheit, and period-age-color relations.
We find that the coefficients of these relations are almost insensitive to the efficiency of superadiabatic convection but are significantly affected by the assumption of the mass-luminosity relation and the adopted chemical composition. Metal-dependent versions of these relations are also derived, representing valuable tools for individual distance determinations and correction for metallicity effects on the cosmic distance scale.

\end{abstract}




\section{Introduction}
Classical Cepheids (CCs) are massive and intermediate-mass stars (ranging from $\sim3$ to $\sim13$ $M_{\odot}$) that cross the pulsation instability strip (IS) in the Hertzsprung–Russel (H-R) diagram during the so-called blue loop phase corresponding to the core helium burning phase.
These radially pulsating stars play a fundamental role as primary distance indicators within the Local Group and beyond, owing to their characteristic Period-Luminosity (PL) and Period-Luminosity-Color (PLC) relations. 
These well-established relations are crucial for calibrating secondary distance indicators and, in turn, accurately determining the Hubble constant through the cosmic distance ladder \citep[see][and reference therein]{Brout2022, BroutDS2022, ScolnicRiess2023}

A sub-percent precise determination of the Hubble constant has become an ultimate goal in modern astrophysics, due to the significant discrepancy between the values derived from early and late Universe observations known as the 'Hubble constant tension debate' \citep[see][and reference therein]{Freedman2019, Freedman2023, Riess2021, Riess2022, RiessBreuval2023}.
As a consequence, a sound investigation of any systematic uncertainties, that may impact the calibration of PL and Period-Wesenheit (PW) relations becomes crucial to understanding possible biases in Cepheid-based distance determinations. These unknown systematic uncertainties may propagate to the local measurements of the Hubble constant and help understand the origin of the aforementioned discrepancy.

Stellar evolution predicts a Mass-Luminosity (ML) relation for the central helium-burning phase of massive and intermediate-mass stars. This relation is influenced by the chemical composition and several non-canonical physical processes, such as rotation, core convective overshooting during the core hydrogen burning phase, and mass loss efficiency.
It is well known that the correlation between period and luminosity combined with the ML relation and the anti-correlation between mass and age implies the existence of a Period-Age (PA) relation \citep[][]{Bono2005, Marconi2006}. On this basis, Cepheids are currently used not only as distance but also as age indicators so becoming important stellar population tracers \citep[see e.g.][and reference therein]{Anderson2016, Efremov2003} and \citep[][hereinafter DS20a, DS20b, DS21, and DS22, respectively]{Desomma2020a, Desomma2020b, Desomma2021, Desomma2022}. 

The Rubin LSST survey will perform repetitive imaging of the southern night sky in six filters (u, g, r, i, z, and y) with excellent image quality and, under ideal conditions, it should be able to detect point sources with a signal-to-noise ratio of 10 down to a magnitude of about 24.5 mag in the r-band. This capability enables the acquisition of highly detailed and well-sampled multiband light curves for various classes of pulsating stars, including CCs. Consequently, establishing a theoretical framework in the context of the Rubin-LSST photometric system, mandatory to extract any possible information from the expected observations, is essential. To this purpose, for the first time in this work, we provide accurate CC pulsation relations in the Rubin-LSST filters.

The structure of the paper is as follows:
in Section 2 we present the adopted theoretical scenario, that is transferred into the Rubin-LSST filters as described in Section 3. The derived PLC, Period-Weseneheit (PW), PA, and Period-Age-Color (PAC) relationships are provided in Section 4. The conclusions and some final remarks in Section 5 close the paper.

\section{The theoretical evolution and pulsation scenario}

To derive the first theoretical pulsation relations in the Rubin-LSST filters we rely on the extensive set of nonlinear and convective models presented by \citetalias[][]{Desomma2020a, Desomma2021, Desomma2022}. This set of models was built by analyzing the full amplitude behavior of selected pulsating stellar envelopes in both the fundamental (F) and the first overtone (FO) modes.
Four chemical compositions were taken into account:
\begin{enumerate}
 \item $Z$ = $0.004$ $Y$ = $0.25$
 \item $Z$ = $0.008$ $Y$ = $0.25$
 \item $Z$ = $0.02$ $Y$ = $0.28$
 \item $Z$ = $0.03$ $Y$ = $0.28$
\end{enumerate}
For each chemical composition, a wide mass range was considered (from $3$ to $11$ $M_{\odot}$ with a step of $1$ $M_{\odot}$) for both F and FO pulsation modes. Additionally, various luminosity levels were investigated for each mass to take into account the uncertainty in the CC ML relation due to the issues related to the non-canonical physical processes mentioned in the previous section.
Specifically, three luminosity levels were adopted: a \lq{canonical level}\rq, labeled "case A", which neglects core convective overshooting, rotation, and mass loss, and two noncanonical levels, labeled "case B" and "case C." The noncanonical levels were obtained by increasing the canonical luminosity level by $\Delta\log(L$/$L_\odot)$=$~0.2 \;$dex and $\Delta\log(L$/$L_\odot)$=$~0.4 \;$dex, respectively \citep[][]{BonoTorn2000}.

For each combination of chemical composition, mass, and luminosity, the modal stability was explored across a wide range of effective temperatures ($T_{eff}$) (approximately from 3600 to 7200 K with a step of 100 K) and three values of the mixing length parameter ($\alpha_{ml}$ = $1.5$, $1.7$, and $1.9$). $\alpha_{ml}$ is a free parameter in the Mixing Length Theory (MLT) adopted in the code to close the nonlinear system of dynamical and convective equations \citep[see e.g.][]{Stell1982, BonoStell1994},  determining the super-adiabatic convection efficiency.

The impact of the adopted mixing length parameter is significant for both the pulsation amplitude and the location of the IS boundaries, particularly the red boundary. Increasing the $\alpha_{ml}$ value from 1.5 to 1.9 results in a narrower IS for all the investigated chemical compositions. This change is characterized by the blue boundary shifting towards redder temperature by about 100-200 K and the red boundary shifting toward bluer temperatures by about 200-300 K. The higher quenching effect of convection on pulsation occurring when increasing the mixing length parameter causes the temperature shift. The more evident effect on the IS red boundary occurs because, in the Color-Magnitude Diagram (CMD), the cooler a star's position is, the greater the extension of the super-adiabatic region (\citetalias{Desomma2020a}, \citealt[][]{Dicriscienzo2004, Fiorentino2007}).

The computed pulsation models are combined with the stellar evolutionary framework provided by the latest version of the BaSTI (a Bag of Stellar Tracks and Isochrones) stellar evolutionary library \footnote{The whole BaSTI database can be accessed at the URL: http://basti-iac.oa-abruzzo.inaf.it.}, in its release accounting for a solar-scaled heavy element distribution
\citep[see][for more details]{Hidalgo18}.
The BaSTI library provides evolutionary predictions for various heavy element mixtures for both canonical stellar models (no overshooting during the core H-burning phase) and noncanonical ones accounting for moderate core convective overshooting. Once again we refer to \cite{Hidalgo18} for a detailed discussion about this topic. 

In this analysis, we specifically selected stellar models with masses in the mass range from 3 to 13 $M_\odot$ and considered three chemical compositions: $Z$=$0.004$, $Y$=$0.252$; $Z$=$0.008$, $Y$=$0.257$; and $Z$=$0.03$, $Y$=$0.284$. These chemical compositions closely match those used for the pulsational computations, with negligible differences in the initial He abundances, which have no impact on the present investigation.

\section{From Bolometric to Multi-photometric band Light Curves}

\begin{figure*}[ht]
  \subcaptionbox*{}[.45\linewidth]{%
    \includegraphics[width=\linewidth]{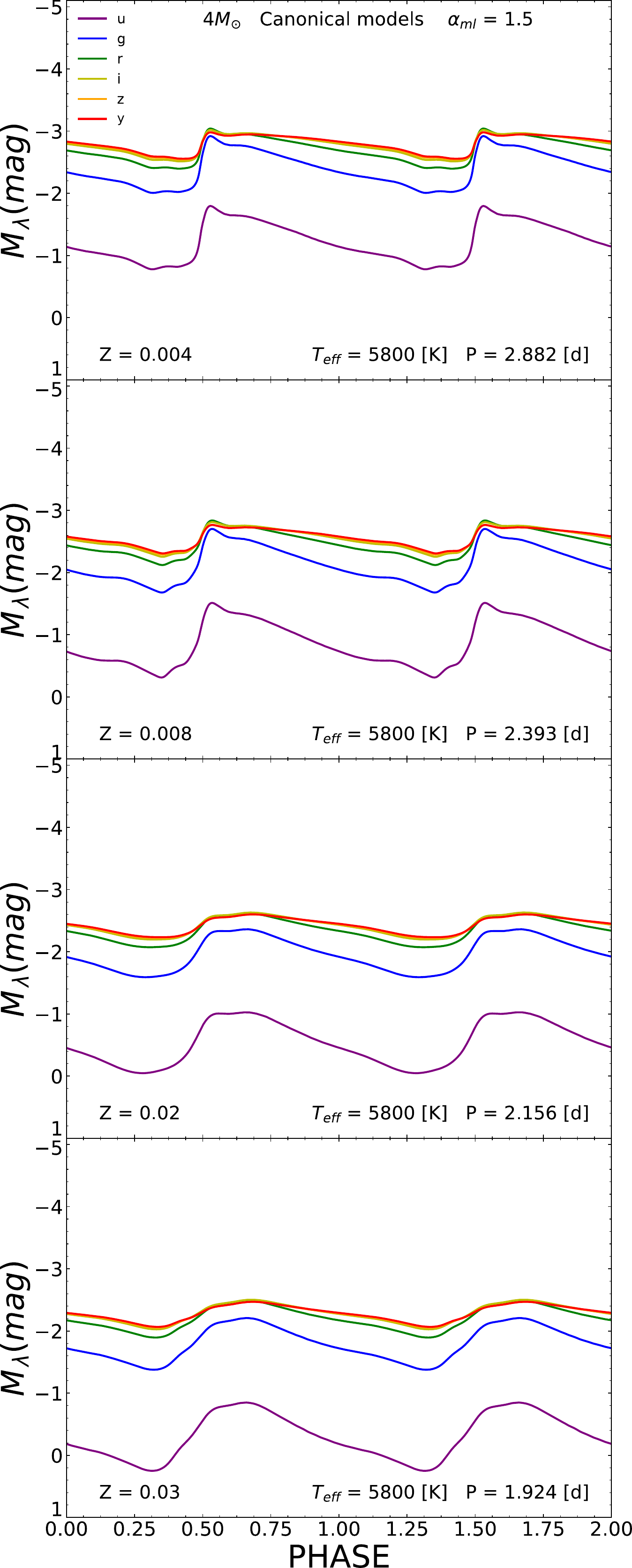}%
  }%
  \hfill
  \subcaptionbox*{}[.45\linewidth]{%
    \includegraphics[width=\linewidth]{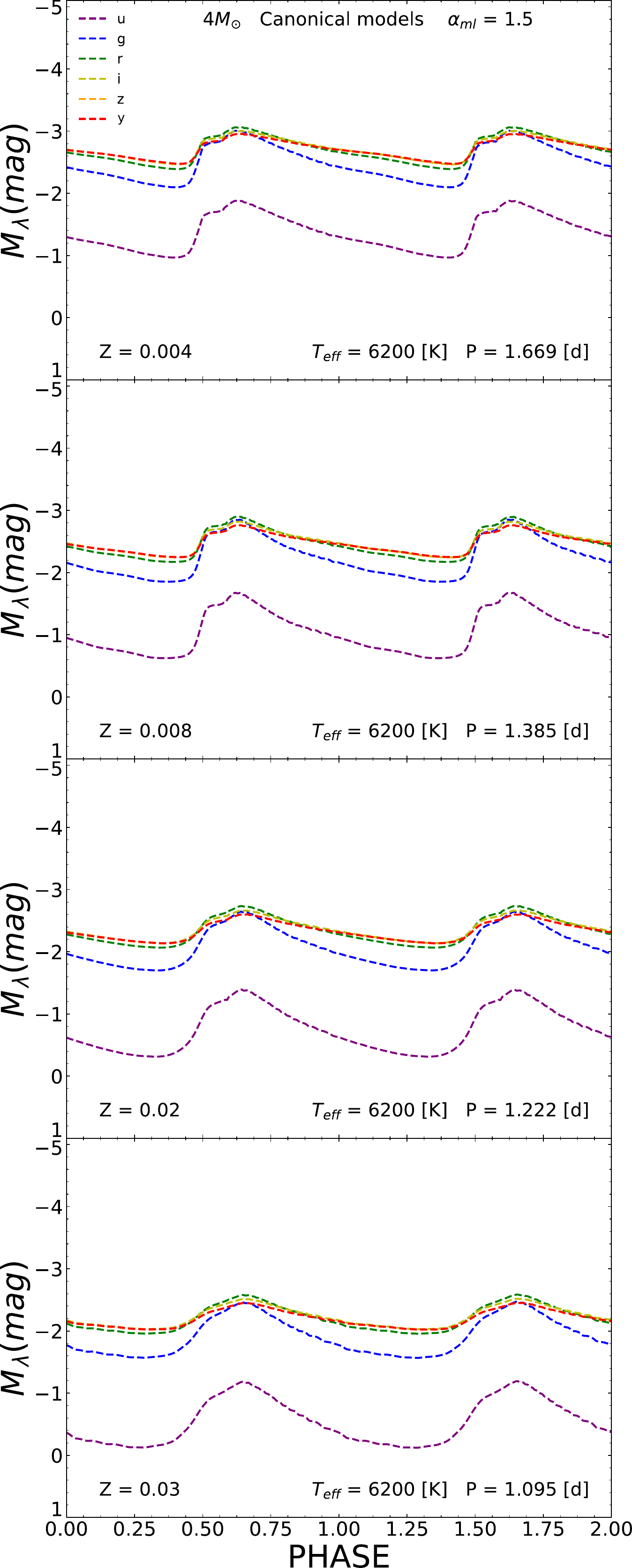}%
  }
  \caption{Left panels: Theoretical light curves in the various Rubin-LSST passbands for a canonical 4.0 $M_\odot$ F-mode model with $\alpha_{ml}$ = $1.5$, located at the center of the IS. From top to bottom, each panel corresponds to a different metallicity (see labels). The pulsational period and the effective temperature are also labeled. Right panels: As in the left panels but for FO-mode models.\\ The complete atlas of light curves in the Rubin-LSST photometric system, for the various assumptions concerning metallicity, mass-luminosity relation, and super-adiabatic convection efficiency, is available upon request.}
  \label{fig:lc_lsst}
\end{figure*}

By computing nonlinear pulsation models, we can predict variations of all pulsation observables, including luminosity, radius, radial velocity, surface temperature, and gravity, along the entire pulsation cycle. 
In particular, in this work, a theoretical atlas of bolometric light- and radial velocity curves for each investigated chemical composition, is derived.
To provide these prescriptions in the photometric planes of the Rubin-LSST survey, the bolometric light curves were converted in the Rubin-LSST photometric bands by using the bolometric correction (BC) tabulations as provided by \cite{Chen2019}. \footnote{The bolometric correction tables provided by \cite{Chen2019} are based on the most updated LSST passbands, available at the following URL:\url{http://stev.oapd.inaf.it/cmd_3.7/photsys.html’}}
To employ these tables, we utilized a proprietary C code that interpolates BC tables along the $T_{eff}$-$\log$(g) direction, with input from the effective temperature and gravity model curves. 
When our models' chemical composition does not align with a specific value in \citet{Chen2019} grid, we also interpolate along the Z direction, by considering two BC tables embracing the considered chemical composition.
Figure \ref{fig:lc_lsst} displays a subset of light curves converted in the Rubin LSST photometric plane corresponding to F-mode (left panels) and FO-mode (right panels) models, respectively, assuming a canonical ML level for a 4 $M_{\odot}$ and  $\alpha_{ml}$ = $1.5$. All the selected models correspond to variables located close to the center of the IS. The various panels refer to different metallicity values (see labels), from $Z$ = $0.004$ (top panel) to $Z$ = $0.03$ (bottom panel), while the blue, green, yellow, orange, and red lines represent light curves in the u, g, r, i, z, and y bands, respectively.  The effective temperature (in Kelvin) and the pulsation period (in days) are also labeled in each panel. Following the behavior shown by the bolometric light curves in \citetalias[][]{Desomma2020a}, the amplitude and the morphology of predicted Rubin-LSST light curves depend on the assumed metal content: the shape becomes smoother and the amplitude reduces as the metallicity increases. Moreover, as expected, the Rubin-LSST light curves exhibit a reduction in pulsation amplitudes and a narrower instability strip, as the filter wavelength increases. 

The transformed light curves are then used to derive a compilation of intensity-weighted mean magnitudes and pulsation amplitudes in the Rubin-LSST filters. These results are reported in Tables \ref{meanmag_amp_F} and \ref{meanmag_amp_FO} for the F and FO modes.

\begin{figure*}
\centering
\includegraphics[width=1.0\textwidth]{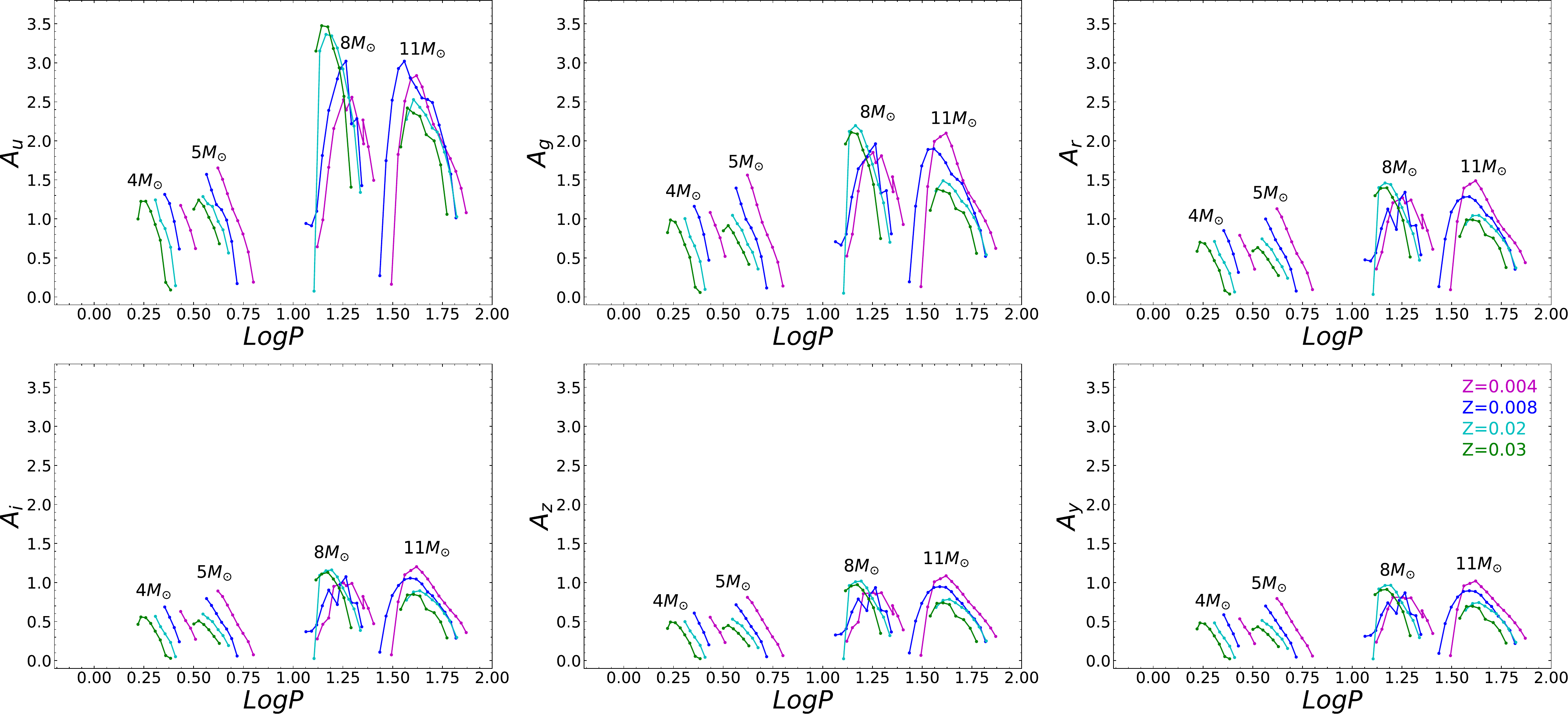}
\caption{From top left to bottom right the F-mode theoretical amplitudes for 4.0, 5.0, 8.0, and 11.0 $M_\odot$ models, computed by adopting a mixing length parameter $\alpha_{ml}$ = $1.5$ and the canonical ML relation, for the u, g, r, i, z, and y Rubin-LSST bands are displayed. These amplitudes correspond to different metallicities: Z=0.004 (dashed magenta line), Z=0.008 (dashed blue line), Z=0.02 (dashed cyan line), and Z=0.03 (dashed green line).}
\label{fig:amp_Z_F}
\end{figure*}

\begin{figure*}
\centering
\includegraphics[width=1.0\textwidth]{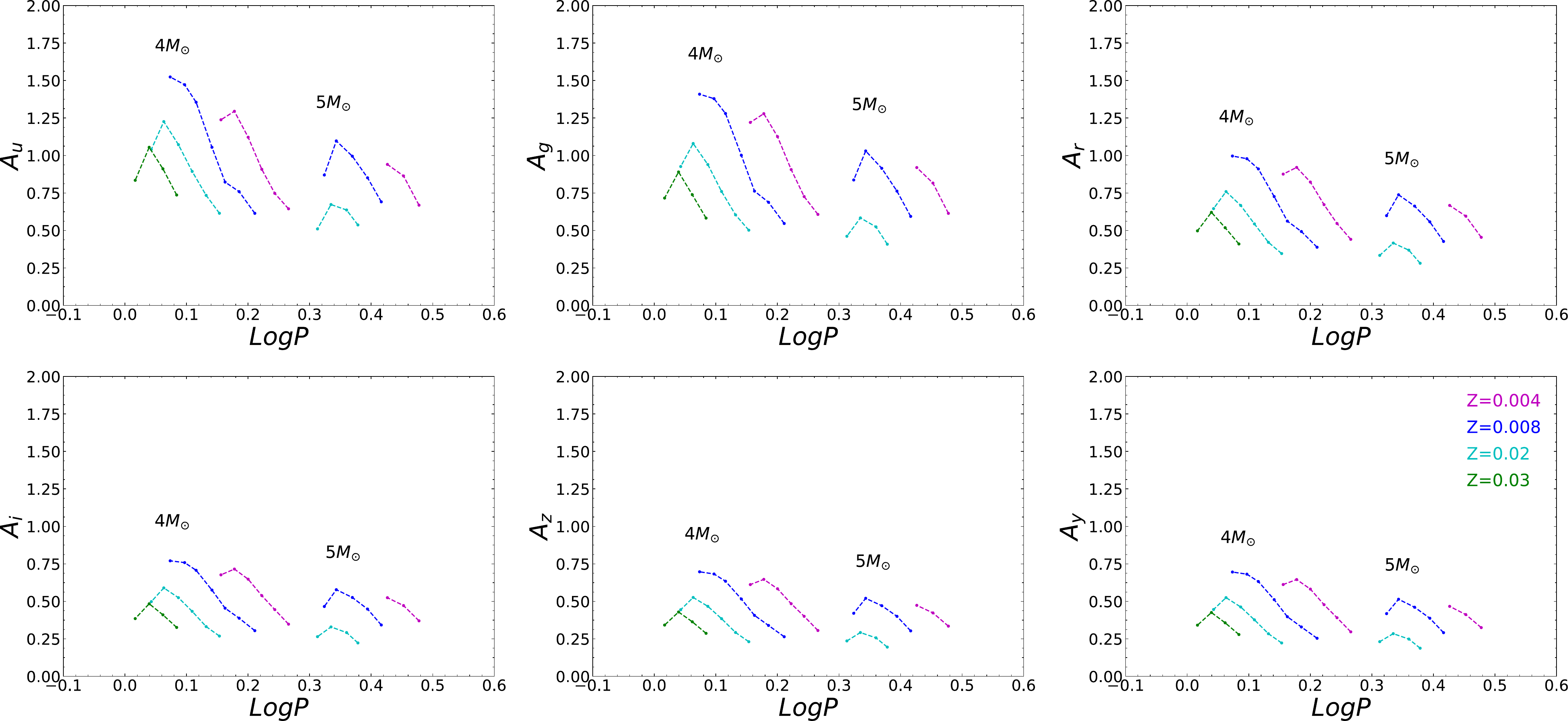}
\caption{The same as Fig \ref{fig:amp_Z_F} but for FO-mode theoretical amplitudes for 4.0 and 5.0 $M_\odot$. However, please note that the adopted scale between the two figures differs.}
\label{fig:amp_Z_FO}
\end{figure*}

In Figures \ref{fig:amp_Z_F} and \ref{fig:amp_Z_FO}, multiband pulsation amplitudes are plotted against the pulsation period for F- and FO-mode models, respectively, for the labeled values of metal abundances and stellar masses. In these plots, a canonical ML relation and $\alpha_{ml}$ = $1.5$ are assumed. We notice that in agreement with earlier theoretical findings \citep[see e.g.][]{BonCastellani2000}, pulsation amplitudes tend to decrease as the metallicity increases, as an effect of the higher opacity. Furthermore, at lower luminosity levels, F-mode amplitudes exhibit a more linear relationship with pulsation periods, whereas at higher luminosities, within a range where the FO-mode pulsation becomes less and less efficient, the F-mode amplitudes tend to mirror the bell-shaped pattern typical of the FO mode. 

\begin{table*}
\caption{\label{meanmag_amp_F} Intensity-weighted mean magnitudes and amplitudes in the Rubin LSST filters for the computed F-mode models with $Z$=$0.004$, $Z$=$0.008$, $Z$=$0.02$, and $Z$=$0.03$. The complete table is accessible in a machine-readable format.}
\centering
\scalebox{0.90}{
\begin{tabular}{ccccccccccccccccccc}
\hline\hline
Z & Y & \msun & \lsun & $T_{eff}$[K] & $\alpha_{ml}$ & ML & <u> & $A_{u}$ & <g> & $A_{g}$ & <r> & $A_{r}$ & <i> & $A_{i}$ & <z> & $A_{z}$ & <y> & $A_{y}$ \\
\hline\hline
 0.004 &  0.25 &   3.0 &  2.49 &  5900 &  1.5 &  A & $-$0.295 &  0.761 & $-$1.403 &  0.715 & $-$1.666 &  0.493 & $-$1.728 &  0.369 & $-$1.725 &  0.314 & $-$1.728 &  0.306 \\
 0.004 &  0.25 &   3.0 &  2.49 &  6000 &  1.5 &  A & $-$0.322 &  1.041 & $-$1.426 &  1.011 & $-$1.666 &  0.710 & $-$1.715 &  0.539 & $-$1.706 &  0.461 & $-$1.709 &  0.457 \\
... & ... & ... & ... & ... & ... & ... & ... & ... & ... & ... & ... & ... & ... & ... & ... & ... & ... & ... \\
 0.008 &  0.25 &   3.0 &  2.39 &  6000 &  1.5 &  A &  0.007 &  1.075 & $-$1.177 &  0.978 & $-$1.441 &  0.675 & $-$1.488 &  0.511 & $-$1.475 &  0.431 & $-$1.477 &  0.420 \\
 0.008 &  0.25 &   3.0 &  2.59 &  5700 &  1.5 &  B & $-$0.342 &  0.609 & $-$1.601 &  0.492 & $-$1.941 &  0.331 & $-$2.027 &  0.252 & $-$2.035 &  0.211 & $-$2.044 &  0.198 \\
... & ... & ... & ... & ... & ... & ... & ... & ... & ... & ... & ... & ... & ... & ... & ... & ... & ... & ... \\
 0.02 &  0.28 &   3.0 &  2.32 &  5900 &  1.5 &  A &  0.354 &  0.200 & $-$0.981 &  0.155 & $-$1.316 &  0.102 & $-$1.375 &  0.078 & $-$1.361 &  0.065 & $-$1.362 &  0.062 \\
 0.02 &  0.28 &   3.0 &  2.32 &  6000 &  1.5 &  A &  0.305 &  0.555 & $-$1.008 &  0.440 & $-$1.316 &  0.302 & $-$1.362 &  0.235 & $-$1.342 &  0.202 & $-$1.341 &  0.195 \\

... & ... & ... & ... & ... & ... & ... & ... & ... & ... & ... & ... & ... & ... & ... & ... & ... & ... & ...\\0.03 &  0.28 &   4.0 &  2.68 &  5400 &  1.5 &  A & $-$0.034 &  0.090 & $-$1.683 &  0.058 & $-$2.205 &  0.038 & -2.338 &  0.029 & $-$2.364 &  0.025 & $-$2.379 &  0.023 \\
 0.03 &  0.28 &   4.0 &  2.68 &  5500 &  1.5 &  A & $-$0.125 &  0.187 & $-$1.723 &  0.125 & $-$2.212 &  0.082 & $-$2.331 &  0.065 & $-$2.350 &  0.055 & $-$2.362 &  0.052 \\
... & ... & ... & ... & ... & ... & ... & ... & ... & ... & ... & ... & ... & ... & ... & ... & ... & ... & ... \\
\hline\hline
\end{tabular}}
\end{table*}

\begin{table*}
\caption{\label{meanmag_amp_FO} The same as Table \ref{meanmag_amp_F} but for the computed FO-mode models. The complete table is accessible in a machine-readable format.}
\centering
\scalebox{0.90}{
\begin{tabular}{ccccccccccccccccccc}
\hline\hline
Z & Y & \msun & \lsun & $T_{eff}$[K] & $\alpha_{ml}$ & ML & <u> & $A_{u}$ & <g> & $A_{g}$ & <r> & $A_{r}$ & <i> & $A_{i}$ & <z> & $A_{z}$ & <y> & $A_{y}$\\
\hline\hline
 0.004 &  0.25 &  3.0 &  2.49 &  6100 &  1.5 &  A & $-$0.361 &  0.642 & $-$1.451 &  0.647 & $-$1.669 &  0.471 & $-$1.706 &  0.368 & $-$1.688 &  0.319 & $-$1.687 &  0.313 \\
 0.004 &  0.25 &  3.0 &  2.49 &  6200 &  1.5 &  A & $-$0.384 &  0.796 & $-$1.473 &  0.813 & $-$1.668 &  0.594 & $-$1.692 &  0.465 & $-$1.668 &  0.406 & $-$1.667 &  0.402 \\
... & ... & ... & ... & ... & ... & ... & ... & ... & ... & ... & ... & ... & ... & ... & ... & ... & ... & ... \\
0.008 &  0.25 &  3.0 &  2.39 &  6100 &  1.5 &  A & $-$0.040 &  0.692 & $-$1.205 &  0.642 & $-$1.444 &  0.449 & $-$1.478 &  0.351 & $-$1.456 &  0.304 & $-$1.455 &  0.298 \\
 0.008 &  0.25 &  3.0 &  2.39 &  6200 &  1.5 &  A & $-$0.069 &  0.881 & $-$1.227 &  0.835 & $-$1.442 &  0.584 & $-$1.463 &  0.443 & $-$1.435 &  0.378 & $-$1.433 &  0.377 \\
... & ... & ... & ... & ... & ... & ... & ... & ... & ... & ... & ... & ... & ... & ... & ... & ... & ... & ...  \\
0.02 &  0.28 &  3.0 &  2.32 &  6200 &  1.5 &  A &  0.212 &  0.727 & $-$1.061 &  0.621 & $-$1.316 &  0.424 & $-$1.335 &  0.321 & $-$1.301 &  0.271 & $-$1.298 &  0.266 \\
 0.02 &  0.28 &  3.0 &  2.32 &  6300 &  1.5 &  A &  0.175 &  0.892 & $-$1.087 &  0.771 & $-$1.314 &  0.526 & $-$1.318 &  0.410 & $-$1.278 &  0.356 & $-$1.275 &  0.352 \\
... & ... & ... & ... & ... & ... & ... & ... & ... & ... & ... & ... & ... & ... & ... & ... & ... & ... & ... \\
0.03 &  0.28 &  4.0 &  2.68 &  6000 &  1.5 &  A & $-$0.493 &  0.737 & $-$1.895 &  0.583 & $-$2.226 &  0.410 & $-$2.272 &  0.326 & $-$2.252 &  0.287 & $-$2.253 &  0.279 \\
 0.03 &  0.28 &  4.0 &  2.68 &  6100 &  1.5 &  A & $-$0.552 &  0.911 & $-$1.927 &  0.739 & $-$2.227 &  0.519 & $-$2.258 &  0.410 & $-$2.231 &  0.364 & $-$2.232 &  0.357 \\
... & ... & ... & ... & ... & ... & ... & ... & ... & ... & ... & ... & ... & ... & ... & ... & ... & ... & ...  \\
\hline\hline
\end{tabular}}
\end{table*}

\clearpage

\clearpage

\section{Pulsation Relationships}
Relying on the derived intensity-weighted mean magnitudes and colors, we can now obtain pulsation relations in the Rubin-LSST filters for CCs.

\subsection{Period-luminosity-Color and Period Wesenheit relations}

The predicted intensity-weighted mean magnitudes and colors, allow us to obtain the first theoretical Period-Luminosity-Color and Period-Wesenheit relations in the Rubin-LSST filters. Period-luminosity relations, which are statistical relations affected by the finite width of the instability strip \citep[see][for a detailed discussion]{MadoreFreedman1991, BCM1999}, will be presented in a separate paper devoted to multi-band synthetic relations (Musella et al. in prep.).
To reduce the effects related to the finite width of the strip, we can introduce a color term and derive the period-luminosity-color and the period-Wesenheit relations. The significant advantage of using the PW relation is that it is reddening-free by definition because the magnitude is replaced by the Wesenheit function, which includes the color term with a coefficient that is the exact ratio between the total and selective absorption. Its formulation, for example in the B-V color, is W(B-V) = V $-$ $\gamma$ (B$-$V) where $\gamma$ = $A_{V}$/ E(B$-$V).
In this study, the PLC and PW relations in the Rubin-LSST filters have been derived considering various assumptions on the chemical composition, the ML relation, and super-adiabatic convection, for both the F and FO modes.
Tables \ref{plc_smc}, \ref{plc_lmc}, \ref{plc_mw}, and \ref{plc_supersolar} in Appendix A, report the coefficients of the fitted PLC relations in the form of $M_{\lambda}$ = $a$ + $b \log P$ + $c$ CI for  $Z$=$0.004$, $Z$=$0.008$, $Z$=$0.02$, and $Z$=$0.03$, respectively. Similarly, Tables \ref{pw_smc}, \ref{pw_lmc}, \ref{pw_mw}, and \ref{pw_supersolar} in Appendix B show the coefficients of the PW relations for the same chemical compositions.

The Wesenheit functions for the selected filter combinations are defined as follows:

\begin{enumerate}
    \item $W$(g, u$-$g) = $g_{LSST}$ $-$ 3.100 $\cdot$ ($u_{LSST}$-$g_{LSST}$)
    \item $W$(r, g$-$r) = $r_{LSST}$ $-$ 2.796 $\cdot$ ($g_{LSST}$-$r_{LSST}$)
    \item $W$(i, g$-$i) = $i_{LSST}$ $-$ 1.287 $\cdot$ ($g_{LSST}$-$i_{LSST}$)
    \item $W$(z, i$-$z) = $z_{LSST}$ $-$ 3.204 $\cdot$ ($i_{LSST}$-$z_{LSST}$)
    \item $W$(y, g$-$y) = $y_{LSST}$ $-$ 0.560 $\cdot$ ($g_{LSST}$-$y_{LSST}$)
\end{enumerate}

The color term coefficients are derived from the extinction law of \citet[][]{Cardelli1989}, under the assumption of $R_{V}$=$3.1$. The effective band wavelengths for all the LSST filters, necessary to derive the reddening vector, were sourced from the SVO database, available at the URL:\url{http://svo2.cab.inta-csic.es/theory/fps/index.php}, and presented in Table \ref{effective_lambda_lsst}.

\begin{table}
\caption{\label{effective_lambda_lsst} The effective band wavelengths for u, g, r, i, z, y LSST filters are taken from the SVO database.}
\centering
\begin{tabular}{cc}
\hline\hline
band & effective wavelength value (\AA) \\
\hline
u & 3751.20\\
g & 4740.66\\ 
r & 6172.34\\ 
i & 7500.97\\ 
z & 8678.90\\
y & 9711.82\\
\end{tabular}
\end{table}

\begin{figure*}
 \centering
 \vspace*{5pt}%
 \hspace*{\fill}%
  \begin{subfigure}{0.56\textwidth}     
    \centering
    \includegraphics[width=\textwidth]{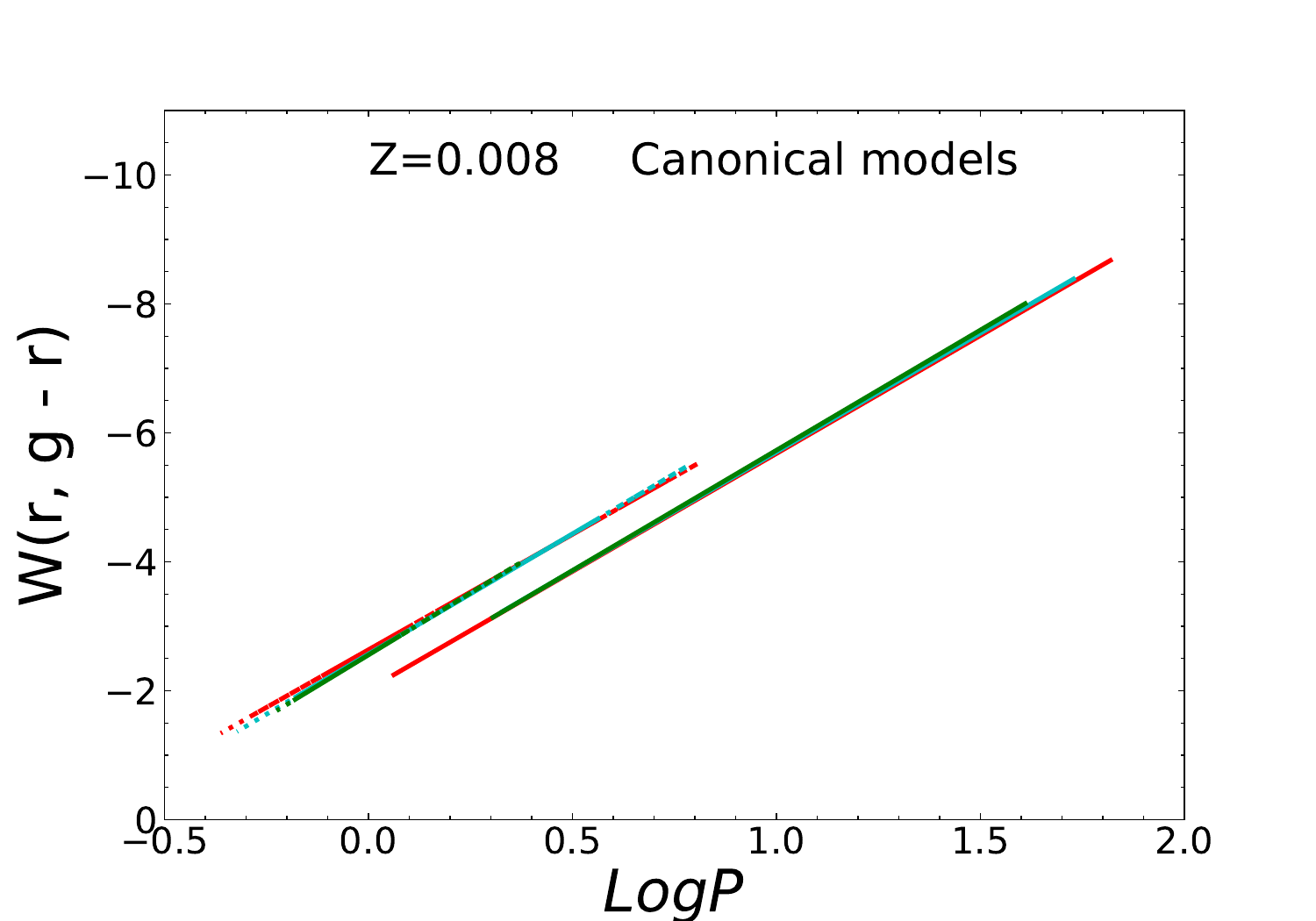}%
    \captionsetup{skip=12pt}%
    \label{fig:4.1}
  \end{subfigure}
  \hspace*{\fill}

  \vspace*{8pt}%

  \hspace*{\fill}%
   \begin{subfigure}{0.56\textwidth}        
    \centering
    \includegraphics[width=\textwidth]{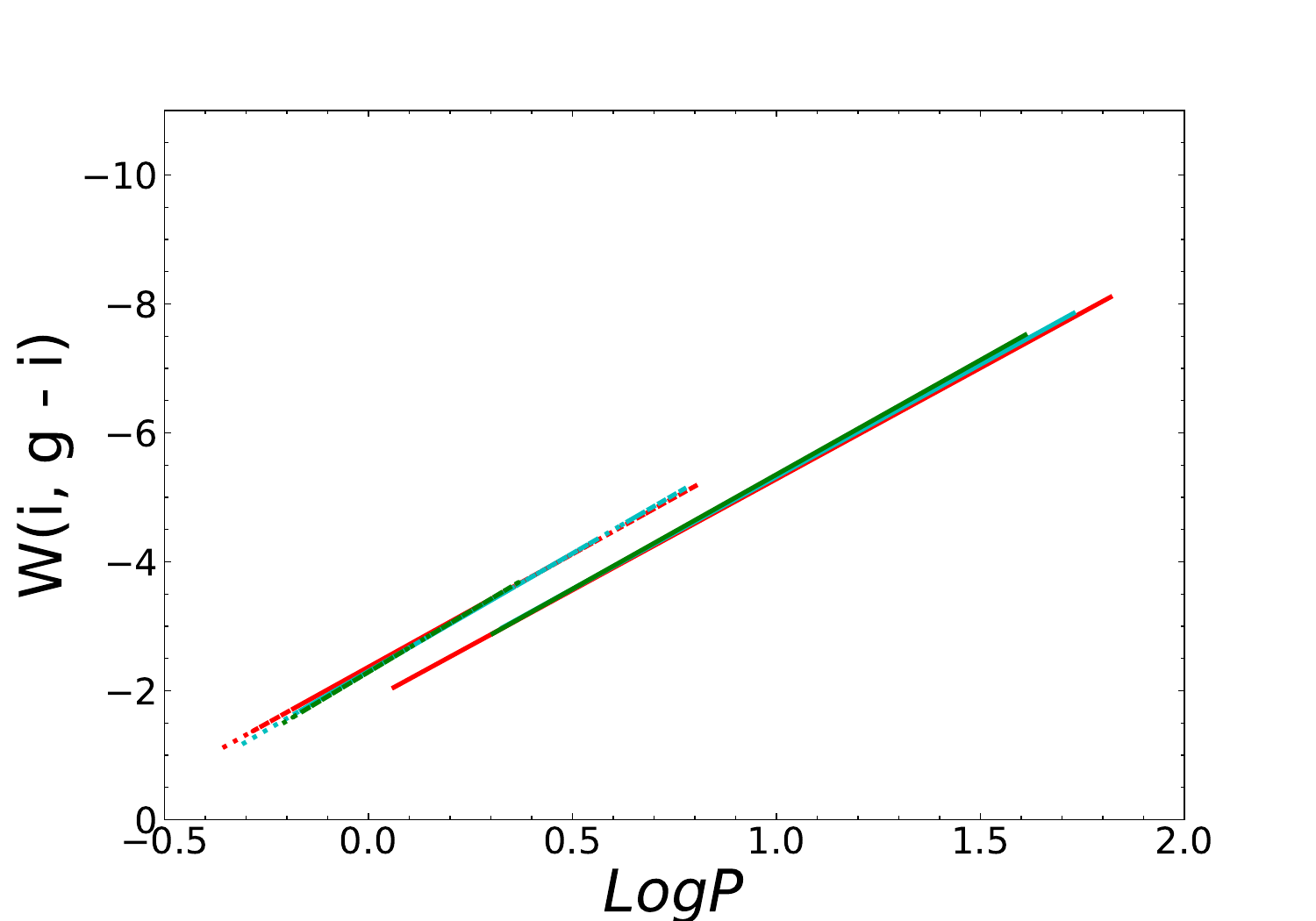}%
    \captionsetup{skip=12pt}%
    \label{fig:4.2}
  \end{subfigure}
  \hspace*{\fill}

  \vspace*{8pt}%

  \hspace*{\fill}%
  \begin{subfigure}{0.56\textwidth}     
    \centering
    \includegraphics[width=\textwidth]{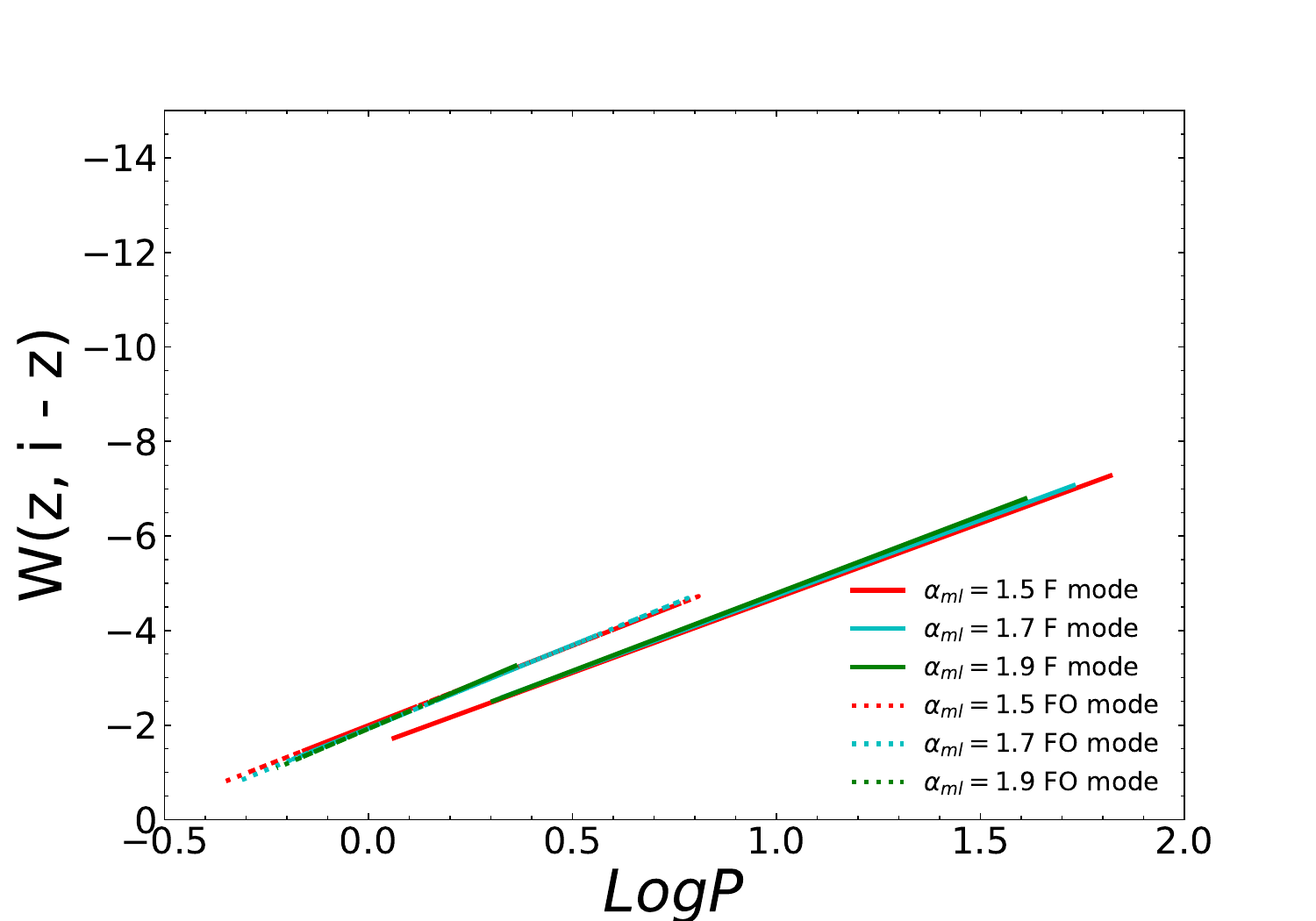}%
    \captionsetup{skip=12pt}%
    \label{fig:4.3}
  \end{subfigure}
  \hspace*{\fill}%
  \captionsetup{skip=8pt}
  \caption{The Wesenheit functions, $W$(r, g$-$r) = $r$ $-$ $2.796 \cdot $(r, g$-$r) (top panel), $W$(i, g$-$i) = $i$ $-$ $1.287 \cdot $(i, g$-$i) (middle panel) and $W$(z, i$-$z) = $z$ $-$ $3.204 \cdot $(z, i$-$z) (bottom panel), are plotted as a function of the logarithmic period for the F- and FO-mode CC models with $Z$=$0.008$.}
  \label{fig:was_F_FO_alfa}
\end{figure*}

Figure \ref{fig:was_F_FO_alfa} shows the PW relations $W$(r,g$-$r) (top panel), W(i, g$-$i) (middle panel), W(z,i$-$z) (bottom panel) for both F and FO-mode models with $Z$=$0.008$. As obtained for other photometric systems \citepalias[see e.g.][]{Desomma2020a, Desomma2022}, these PW relations appear almost insensitive to variations in the mixing length parameter. This occurrence is due to the negligible dependence of PW relation on the IS width and the pulsation amplitudes.

\begin{figure*}
\centering
    \vbox{
    \hbox{
    \includegraphics[width=0.55\textwidth]{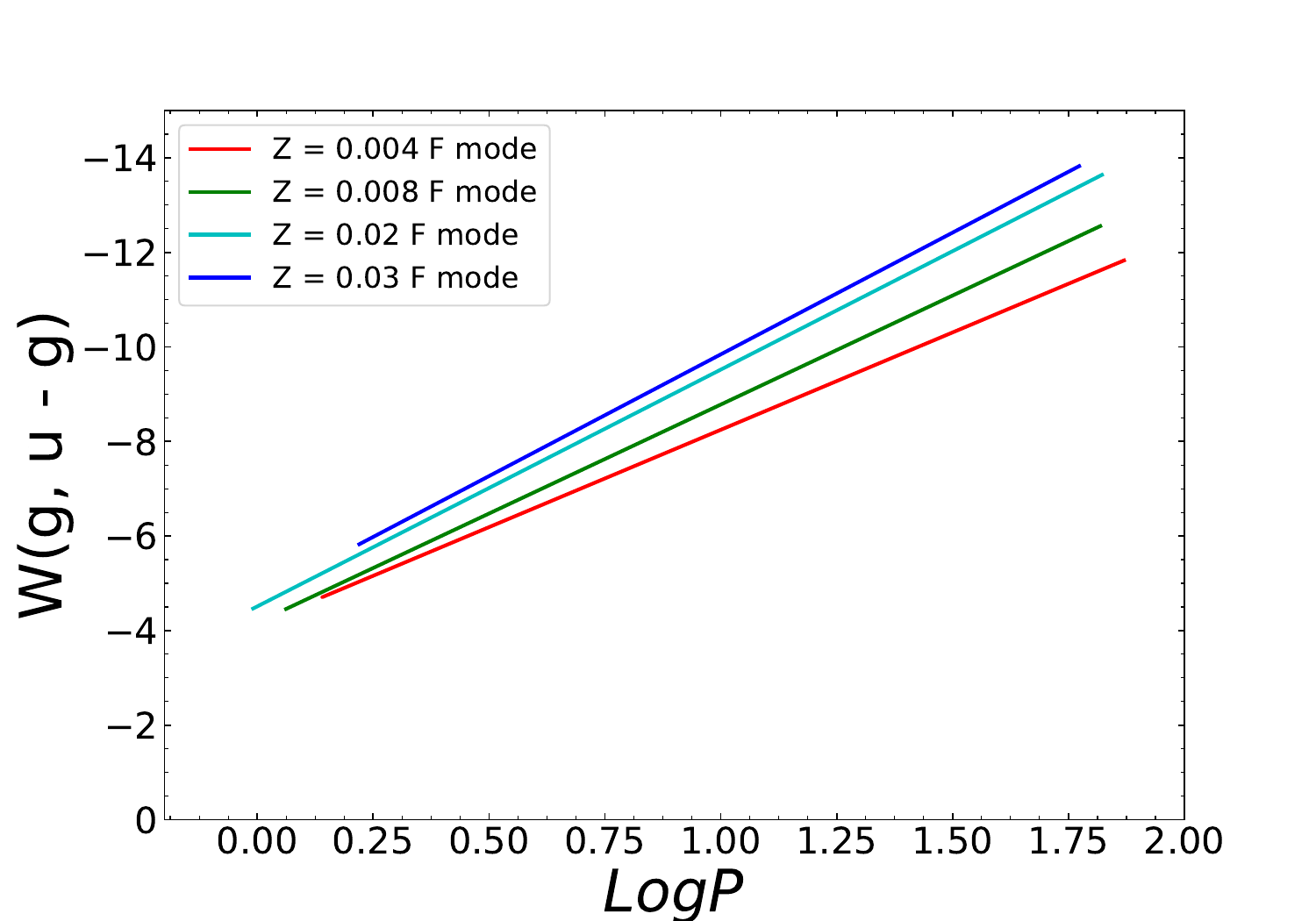}
    \includegraphics[width=0.55\textwidth]{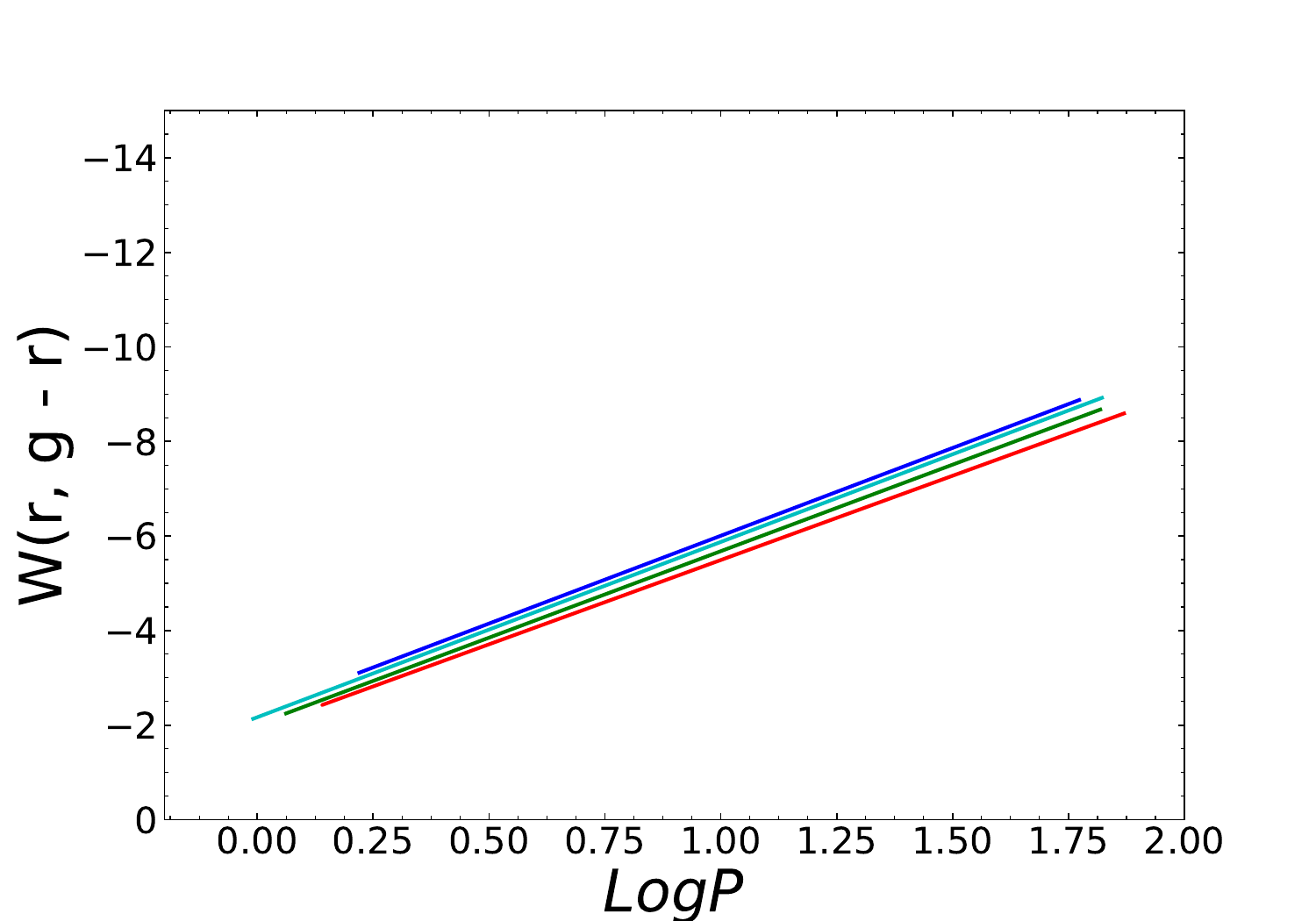}
    }
    \hbox{
    \includegraphics[width=0.55\textwidth]{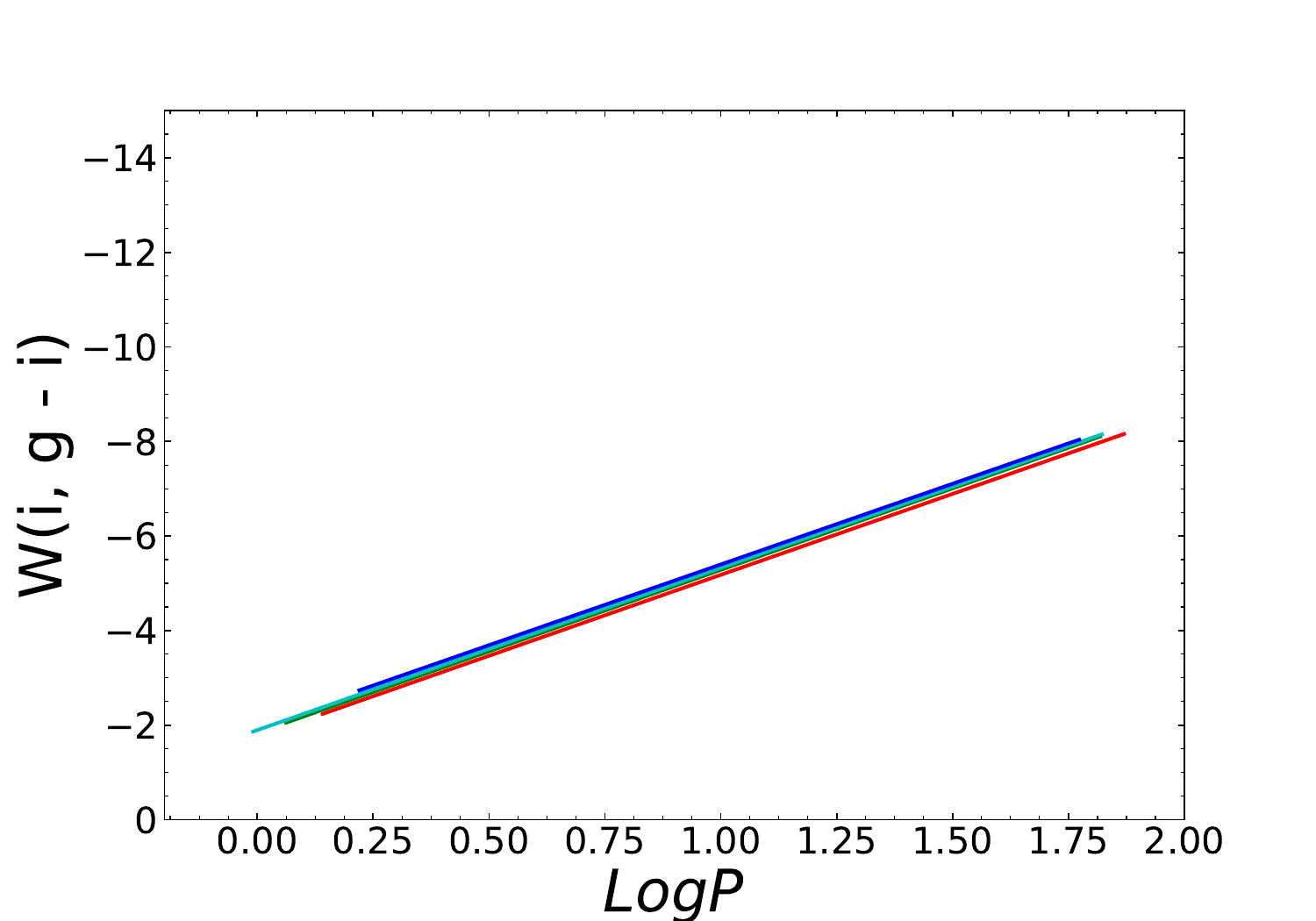}
    \includegraphics[width=0.55\textwidth]{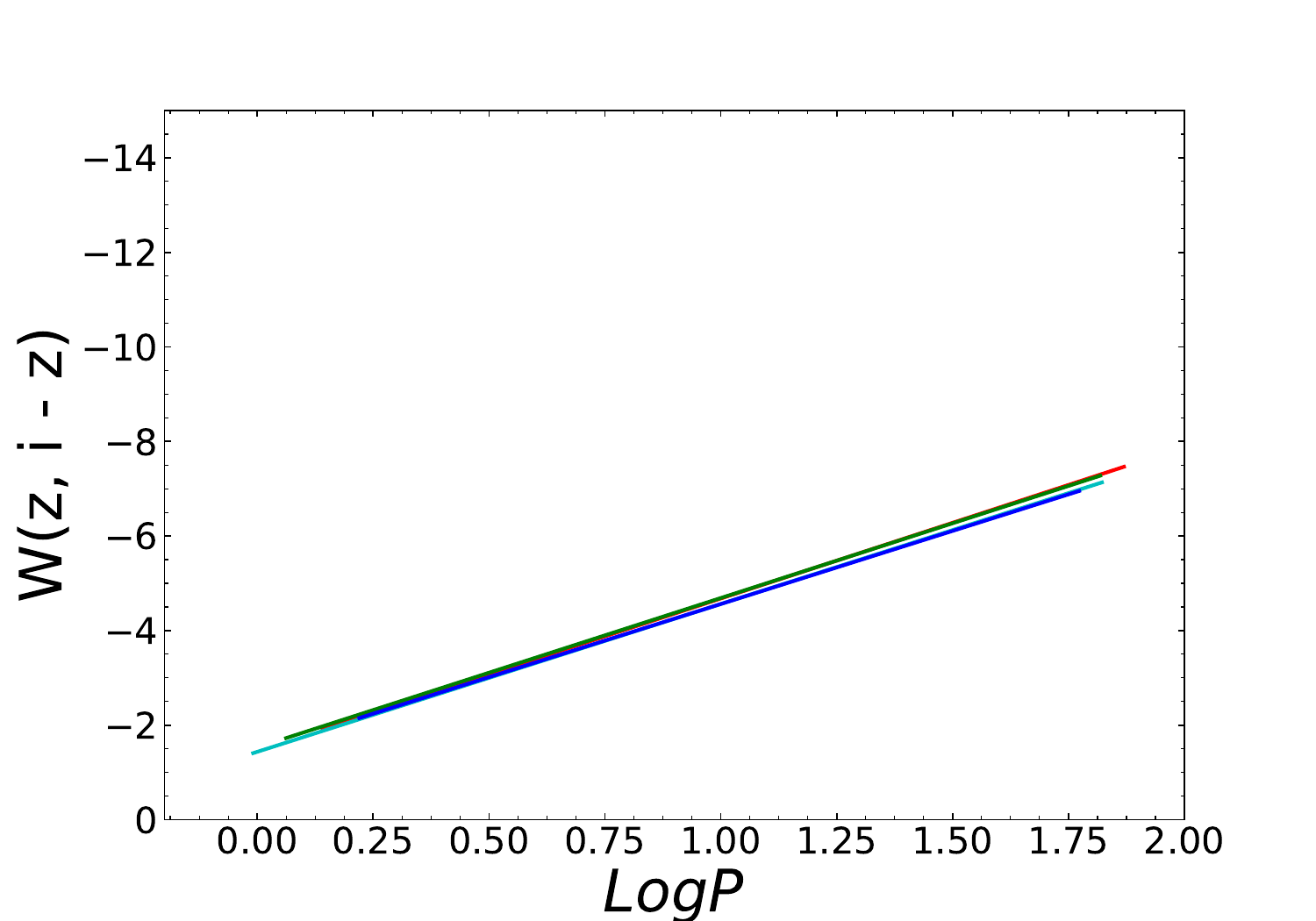}
    }
    \hbox{
    \includegraphics[width=0.55\textwidth]{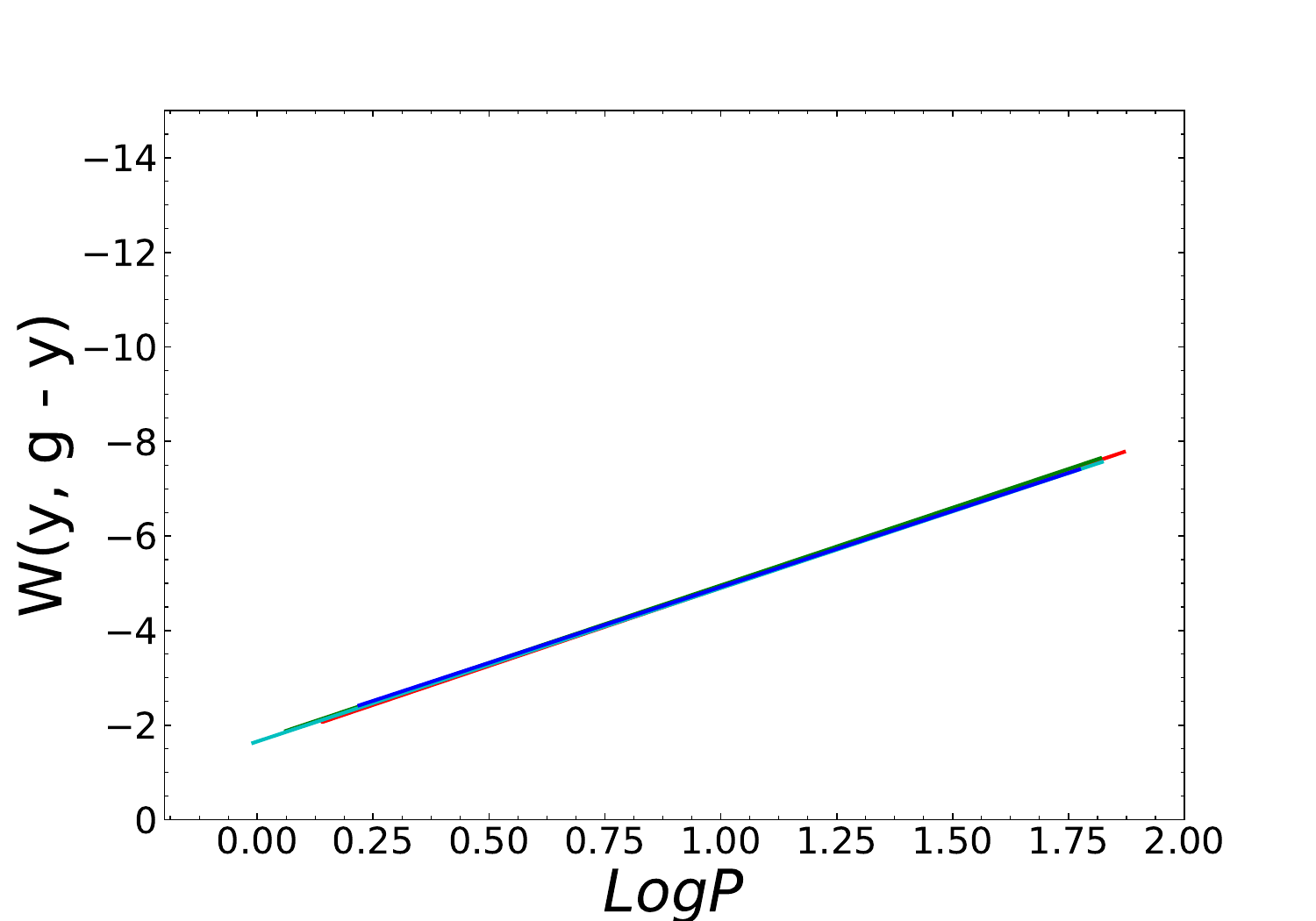}
    }}
    \caption{\label{fig:was_F} The Wesenheit function, $W$(g, u$-$g) = $g$ $-$ $3.100 \cdot $(u$-$g) (top left panel), $W$(r, g$-$r) = $r$ $-$ $2.796 \cdot $(g$-$r) (top right panel),  $W$(i, g$-$i) = $i$ $-$ $1.287 \cdot $(g$-$i) (middle left panel), $W$(z, i$-$z) = $z$ $-$ $3.204\cdot $(i$-$z) (middle right panel) and $W$(y, g$-$y) = $y$ $-$ $0.560 \cdot $(g$-$y) (bottom left panel) are plotted as a function of the logarithmic period for F-mode CCs derived for $Z$=$0.004$, $Z$=$0.008$, $Z$=$0.02$, and $Z$=$0.03$ (see labels on the top left panel).}
\end{figure*}

Figure \ref{fig:was_F} displays the projected F-mode Wesenheit relations in selected Rubin-LSST filters for the labeled metal abundances. Similarly to what occurs in other filter systems, we notice that the effect of metallicity depends on the assumed band combination, primarily contributing to variations in the zero-point of the relations. However, for the g, u$-$g combination (top panel) the slope also appears to depend on the assumed metal content, with the relations getting steeper as the metallicity increases.
We also notice that the effect of metallicity is negligible when near-infrared filters are combined with $g$ or $i$.

\cleardoublepage

\subsection{Metal-dependent Period Wesenheit relations}
In table \ref{pwz_lsst} in Appendix B, we present the Period-Wesenheit-Metallicity (PWZ) relation in selected Rubin-LSST filters for all the adopted assumptions about the ML relation and super-adiabatic convective efficiency, for both F and FO-mode.
We notice that the coefficient of the metallicity term is notably non-negligible. In particular, for the PWZ relations in the optical Rubin-LSST filters, it becomes even more negative, reaching $-0.6$ dex for the r, g$-$r combination. This value is lower than other theoretical and observational estimates found by different authors who used the Wesenheit function in various filter combinations, where the metallicity coefficient typically falls between $-0.3$ and $-0.1$ dex \citep[see e.g.][and reference therein]{Anderson2016, Bhardwaj2024, Desomma2022, Narloch2023, RiessBreuval2022, Ripepi2022, Trentin2023, Trentin2023ar}. 
On the other hand, when near-infrared filters are involved the coefficient of the metallicity term for F-mode models becomes very small.
We excluded the $g$, $u$-$g$ combination filter due to the aforementioned dependence of both the zero point and slope of the relations on the metallicity for both F and FO-mode models.

\subsection{The Color-Color Relations}

\begin{figure*}
\centering
\includegraphics[width=1.0\textwidth]{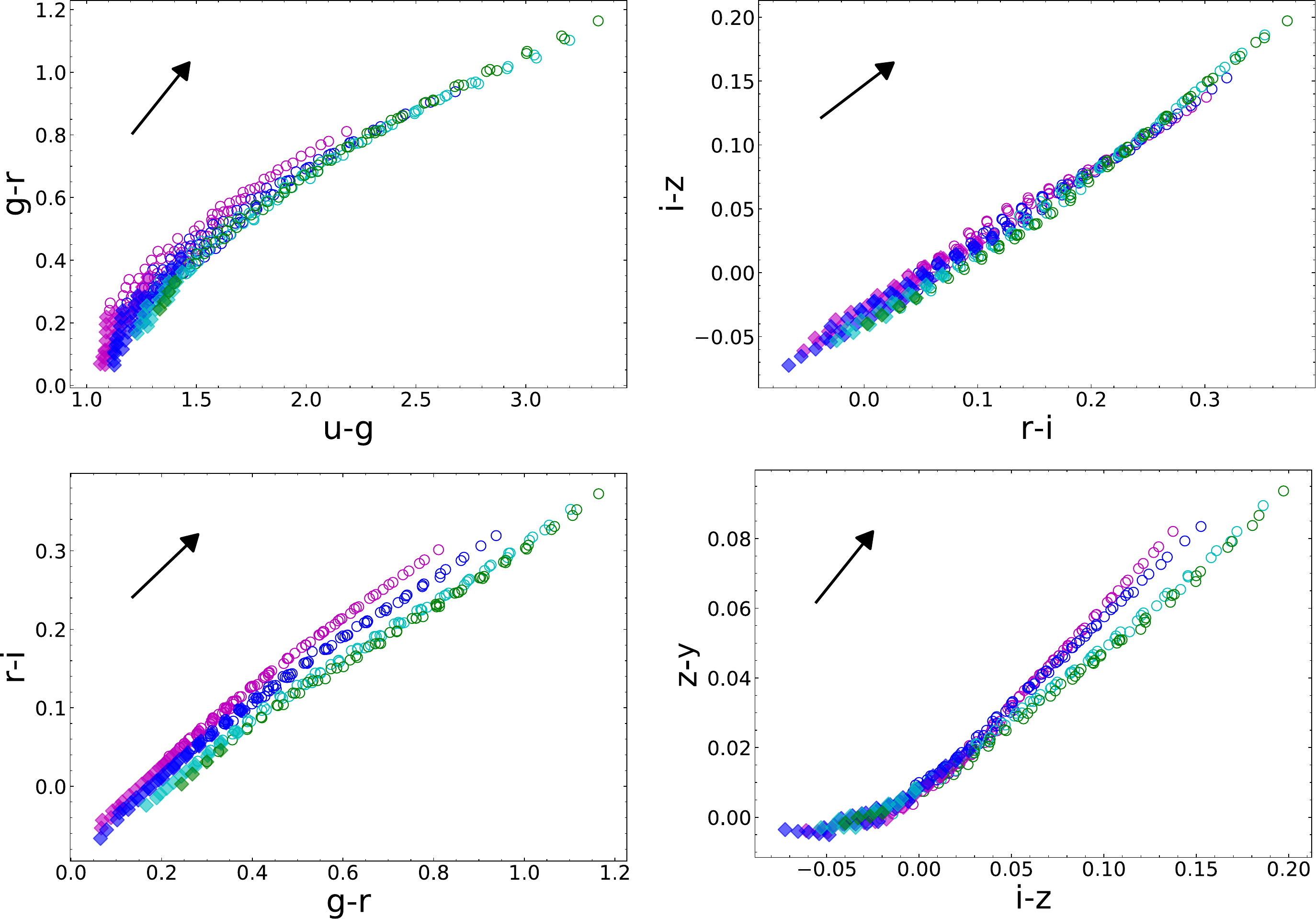}
\caption{The distribution of CCs in the color-color plane for the analyzed chemical compositions, as indicated by the labels on the plot. FO and F-mode pulsators are represented by filled diamonds and open circles, respectively. In every panel, the black arrow indicates the reddening vector computed based on the extinction law by \citet{Cardelli1989}}
\label{fig:color_color}
\end{figure*}

By adopting the multi-filter intensity-weighted mean colors, we also investigated the behavior of pulsation models in color–color diagrams. Figure \ref{fig:color_color} displays the projected distribution of CCs, with F-mode indicated by open circles and FO-mode by filled diamonds, across multiple color–color diagrams corresponding to the labeled Z values. In each panel, the black arrow represents the reddening vector calculated using the \citet{Cardelli1989} extinction law.

We notice that no color-color combination seems to be suitable for individual reddening determinations as their slope is not significantly different from the reddening vector. As for the metallicity effect, we observe that the least-affected relation emerges in the i$-$z vs r$-$i plane. Moreover, the metallicity dependence is best disentangled in the r$-$i vs g$-$r plane, with the effect increasing towards redder colors.

\subsection{Period-Age-Color and Period-Age-Color-Metallicity relations}

As stated in the introduction, Cepheid variables are reliable age indicators for young stellar populations, due to the existence of a PA relation. Despite their numerous advantages \citep[for a detailed discussion, see e.g.][]{Bono2005}, Cepheid ages derived from the PA relation are based on the statistical assumption that the Cepheid IS has a negligible temperature width and do not account for the pulsator's position within the strip. The occurrence of a very narrow IS holds for short-period pulsators, while for periods longer than 10 days the adoption of Period-Age-Color (PAC) relations is recommended. PAC relations take into consideration the individual positions of Cepheids in color within the IS and are also less affected by luminosity variations caused by evolutionary effects.

To derive PAC relations in the Rubin-LSST filters, we adopted the same evolutionary scenario as presented in \citetalias{Desomma2020b, Desomma2021}. In the latter papers, the authors have shown that both the PAC and Period-Age-Color-Metallicity (PACZ) relations are minimally affected by the choice of the mixing length parameter. Therefore, in deriving these relations, we utilized only the standard convective efficiency value, $\alpha_{ml}$ = $1.5$. However, concerning the ML relation and to maintain consistency with the adopted evolutionary scenario, which does not predict overluminosities exceeding +0.4 dex compared to the canonical ML case, the relations were derived by assuming cases A and B for the ML relation. The C case, as discussed in Section 2, represents our canonical ML relation increased by $\Delta\log(L$/$L_\odot)$=$~0.4 \;$ dex and was neglected.

For each adopted chemical composition, we have combined our predictions about the boundaries of the IS with the evolutionary stellar models, to identify the portion of these tracks within the IS during the blue loop phase (corresponding to the central He-burning stage). Hence we applied our Period-Mass-Luminosity-Temperature (PLMT) relation to estimate the pulsational period along the evolutionary path inside the IS. By performing a statistical linear regression among the predicted periods, stellar age, and colors - repeated for various colors in the LSST photometric system - we derived the first theoretical PAC relation in the Rubin-LSST bands for all the adopted chemical compositions. The results for F and FO-mode models can be found in Table \ref{pac_f_fo_allz} in Appendix C.

Following the evidence of a non-negligible metallicity effect on both stellar evolution and pulsation models, we also derived PACZ relations. The coefficients of these relations are reported in Table \ref{pacz_f_fo_allz} for both F and FO-modes in Appendix C. 

\section{Conclusions and Future Developments}
The Vera C. Rubin Observatory holds the promise of revolutionizing our understanding of the Universe, undoubtedly leading to numerous breakthroughs and discoveries in different fields of astrophysics research. Its continuous and wide-area survey of the night sky will enable the unprecedented monitoring of variable stars with enhanced frequency and accuracy, providing a wealth of data for in-depth study of known variable stars and the discovery of new ones, ultimately advancing our comprehension of stellar variability. 
In this study, we built the first theoretical scenario for Classical Cepheids in the Rubin-LSST filters and established precise theoretical pulsation relations, useful to trace the intrinsic stellar properties and infer the distances of observed samples.
In particular, the main results of our investigations are the following:
\begin{enumerate}
\item A theoretical atlas of Cepheid light curves in the Rubin-LSST filters has been produced for the various assumptions on the ML relation and the efficiency of super-adiabatic convection.  
\item Intensity-weighted mean magnitudes and colors, as well as pulsation amplitudes, have been derived. The latter have been investigated as a function of the pulsation period varying the model parameters, including the metal abundance.
\item Theoretical PLC and PW relations in the Rubin-LSST filters have been provided. The metal dependence of the PW relation has also been investigated. Metallicity was found to affect both the slope and the zero-point in the case of the $g$,$u$-$g$ PW relation, which is the most sensitive to variations in the metal abundance. For the filter combinations not involving the $u$ band, the metallicity effect essentially contributes to the PW zero-point so that PWZ relations were derived. 
\item The behavior of the pulsation models in the Rubin-LSST color-color diagram was investigated, but no combination appeared to be useful for individual reddening estimations.
\item The first theoretical PAC and PACZ relations in the Rubin-LSST filters were finally derived.
\end{enumerate}

\clearpage

\appendix

\section{The PLC relations in the Rubin-LSST filters} 
The PLC coefficients in the Rubin-LSST filters for the various investigated chemical compositions are listed in Tables \ref{plc_smc}-\ref{plc_supersolar}.

\begin{table*}
\caption{\label{plc_smc}The coefficients of the PLC relations $M_{\lambda}$ = $a$ + $b \log P$ + $c$ CI, for $Z$=$0.004$, $Y$= $0.25$. The last two columns represent the root-mean-square deviation ($\sigma$) and the R-squared ($R^2$) coefficients.}
\centering
\scalebox{0.75}{
\begin{tabular}{cccccccccccccc}
\hline\hline
Z&Y&color&mode&$\alpha_{ml}$&ML&a&b&c&$\sigma_{a}$&$\sigma_{b}$&$\sigma_{c}$&$\sigma$&$R^2$\\
\hline\hline
 0.004 &  0.25 &  u$-$r &   F &  1.5 &  A &  $-$2.922 &  $-$3.881 &    2.368 &  0.034 &  0.029 &  0.029 &  0.057 &  0.996 \\
 0.004 &  0.25 &  u$-$g  &   F &  1.5 &  A &  $-$4.005 &  $-$4.039 &    2.951 &  0.094 &  0.063 &  0.102 &  0.115 &  0.988 \\
 0.004 &  0.25 &  g$-$r  &   F &  1.5 &  A &  $-$1.974 &  $-$3.703 &    3.273 &  0.012 &  0.014 &  0.038 &  0.031 &  0.999 \\
 0.004 &  0.25 &  g$-$i &   F &  1.5 &  A &  $-$1.814 &  $-$3.706 &    1.959 &  0.011 &  0.013 &  0.025 &  0.029 &  0.999 \\
 0.004 &  0.25 &  i$-$z &   F &  1.5 &  A &  $-$1.102 &  $-$3.914 &   12.574 &  0.011 &  0.014 &  0.151 &  0.026 &  1.000 \\
 0.004 &  0.25 &  u$-$r &   F &  1.7 &  A &  $-$3.058 &  $-$3.891 &    2.442 &  0.037 &  0.026 &  0.032 &  0.045 &  0.998 \\
 0.004 &  0.25 &  u$-$g  &   F &  1.7 &  A &  $-$4.353 &  $-$4.058 &    3.200 &  0.113 &  0.061 &  0.119 &  0.094 &  0.993 \\
 0.004 &  0.25 &  g$-$r  &   F &  1.7 &  A &  $-$1.981 &  $-$3.718 &    3.304 &  0.010 &  0.013 &  0.040 &  0.025 &  1.000 \\
 0.004 &  0.25 &  g$-$i &   F &  1.7 &  A &  $-$1.811 &  $-$3.722 &    1.973 &  0.009 &  0.012 &  0.026 &  0.023 &  1.000 \\
 0.004 &  0.25 &  i$-$z &   F &  1.7 &  A &  $-$1.082 &  $-$3.931 &   12.946 &  0.011 &  0.014 &  0.186 &  0.023 &  1.000 \\
 0.004 &  0.25 &  u$-$r &   F &  1.9 &  A &  $-$3.232 &  $-$3.921 &    2.552 &  0.049 &  0.029 &  0.044 &  0.038 &  0.999 \\
 0.004 &  0.25 &  u$-$g  &   F &  1.9 &  A &  $-$4.780 &  $-$4.079 &    3.514 &  0.174 &  0.074 &  0.180 &  0.085 &  0.995 \\
 0.004 &  0.25 &  g$-$r  &   F &  1.9 &  A &  $-$1.982 &  $-$3.745 &    3.352 &  0.010 &  0.013 &  0.049 &  0.020 &  1.000 \\
 0.004 &  0.25 &  g$-$i &   F &  1.9 &  A &  $-$1.799 &  $-$3.747 &    1.986 &  0.009 &  0.013 &  0.033 &  0.019 &  1.000 \\
 0.004 &  0.25 &  i$-$z &   F &  1.9 &  A &  $-$1.053 &  $-$3.956 &   13.362 &  0.016 &  0.021 &  0.321 &  0.026 &  1.000 \\
 0.004 &  0.25 &  u$-$r &   F &  1.5 &  B &  $-$2.730 &  $-$3.809 &    2.298 &  0.036 &  0.030 &  0.029 &  0.070 &  0.995 \\
 0.004 &  0.25 &  u$-$g  &   F &  1.5 &  B &  $-$3.708 &  $-$3.918 &    2.741 &  0.081 &  0.054 &  0.084 &  0.122 &  0.988 \\
 0.004 &  0.25 &  g$-$r  &   F &  1.5 &  B &  $-$1.841 &  $-$3.679 &    3.261 &  0.017 &  0.018 &  0.047 &  0.046 &  0.999 \\
 0.004 &  0.25 &  g$-$i &   F &  1.5 &  B &  $-$1.689 &  $-$3.679 &    1.961 &  0.015 &  0.017 &  0.031 &  0.043 &  0.999 \\
 0.004 &  0.25 &  i$-$z &   F &  1.5 &  B &  $-$1.048 &  $-$3.804 &   12.121 &  0.019 &  0.020 &  0.230 &  0.048 &  0.999 \\
 0.004 &  0.25 &  u$-$r &   F &  1.7 &  B &  $-$2.829 &  $-$3.829 &    2.360 &  0.031 &  0.025 &  0.028 &  0.051 &  0.997 \\
 0.004 &  0.25 &  u$-$g  &   F &  1.7 &  B &  $-$3.920 &  $-$3.925 &    2.882 &  0.091 &  0.056 &  0.099 &  0.107 &  0.992 \\
 0.004 &  0.25 &  g$-$r  &   F &  1.7 &  B &  $-$1.842 &  $-$3.700 &    3.315 &  0.009 &  0.011 &  0.032 &  0.024 &  1.000 \\
 0.004 &  0.25 &  g$-$i &   F &  1.7 &  B &  $-$1.680 &  $-$3.701 &    1.992 &  0.008 &  0.011 &  0.022 &  0.024 &  1.000 \\
 0.004 &  0.25 &  i$-$z &   F &  1.7 &  B &  $-$1.008 &  $-$3.841 &   12.697 &  0.016 &  0.018 &  0.230 &  0.036 &  0.999 \\
 0.004 &  0.25 &  u$-$r &   F &  1.9 &  B &  $-$2.897 &  $-$3.814 &    2.380 &  0.034 &  0.026 &  0.032 &  0.044 &  0.998 \\
 0.004 &  0.25 &  u$-$g  &   F &  1.9 &  B &  $-$4.001 &  $-$3.853 &    2.863 &  0.109 &  0.060 &  0.119 &  0.097 &  0.994 \\
 0.004 &  0.25 &  g$-$r  &   F &  1.9 &  B &  $-$1.852 &  $-$3.721 &    3.381 &  0.007 &  0.009 &  0.031 &  0.017 &  1.000 \\
 0.004 &  0.25 &  g$-$i &   F &  1.9 &  B &  $-$1.679 &  $-$3.725 &    2.036 &  0.007 &  0.011 &  0.025 &  0.021 &  1.000 \\
 0.004 &  0.25 & i$-$z &   F &  1.9 &  B &  $-$0.979 &  $-$3.865 &   13.117 &  0.019 &  0.023 &  0.319 &  0.037 &  0.999 \\
 0.004 &  0.25 &  u$-$r &   F &  1.5 &  C &  $-$2.726 &  $-$3.821 &    2.386 &  0.045 &  0.029 &  0.034 &  0.070 &  0.996 \\
 0.004 &  0.25 &  u$-$g  &   F &  1.5 &  C &  $-$3.917 &  $-$3.975 &    3.036 &  0.102 &  0.053 &  0.099 &  0.120 &  0.990 \\
 0.004 &  0.25 &  g$-$r  &   F &  1.5 &  C &  $-$1.712 &  $-$3.668 &    3.293 &  0.020 &  0.018 &  0.054 &  0.049 &  0.999 \\
 0.004 &  0.25 & g$-$i &   F &  1.5 &  C &  $-$1.557 &  $-$3.661 &    1.965 &  0.018 &  0.017 &  0.036 &  0.046 &  0.999 \\
 0.004 &  0.25 &  i$-$z &   F &  1.5 &  C &  $-$0.973 &  $-$3.729 &   11.878 &  0.023 &  0.022 &  0.290 &  0.057 &  0.998 \\
 0.004 &  0.25 &  u$-$r &   F &  1.7 &  C &  $-$2.709 &  $-$3.778 &    2.345 &  0.038 &  0.027 &  0.032 &  0.058 &  0.997 \\
 0.004 &  0.25 &  u$-$g  &   F &  1.7 &  C &  $-$3.823 &  $-$3.859 &    2.859 &  0.102 &  0.056 &  0.105 &  0.115 &  0.991 \\
 0.004 &  0.25 &  g$-$r  &   F &  1.7 &  C &  $-$1.705 &  $-$3.669 &    3.282 &  0.012 &  0.013 &  0.039 &  0.030 &  0.999 \\
 0.004 &  0.25 &  g$-$i &   F &  1.7 &  C &  $-$1.543 &  $-$3.666 &    1.960 &  0.011 &  0.013 &  0.027 &  0.030 &  1.000 \\
 0.004 &  0.25 &  i$-$z &   F &  1.7 &  C &  $-$0.947 &  $-$3.731 &   11.871 &  0.020 &  0.021 &  0.285 &  0.047 &  0.999 \\
 0.004 &  0.25 &  u$-$r &   F &  1.9 &  C &  $-$2.733 &  $-$3.772 &    2.350 &  0.035 &  0.026 &  0.032 &  0.045 &  0.998 \\
 0.004 &  0.25 &  u$-$g  &   F &  1.9 &  C &  $-$3.837 &  $-$3.812 &    2.820 &  0.109 &  0.059 &  0.115 &  0.099 &  0.993 \\
 0.004 &  0.25 &  g$-$r  &   F &  1.9 &  C &  $-$1.704 &  $-$3.685 &    3.316 &  0.008 &  0.010 &  0.033 &  0.019 &  1.000 \\
 0.004 &  0.25 &  g$-$i &   F &  1.9 &  C &  $-$1.536 &  $-$3.686 &    1.987 &  0.009 &  0.012 &  0.027 &  0.023 &  1.000 \\
 0.004 &  0.25 &  i$-$z &   F &  1.9 &  C &  $-$0.912 &  $-$3.757 &   12.169 &  0.021 &  0.024 &  0.340 &  0.043 &  0.999 \\
 0.004 &  0.25 &  u$-$r &  FO &  1.5 &  A &  $-$4.632 &  $-$3.692 &    3.056 &  0.807 &  0.256 &  0.649 &  0.195 &  0.957 \\
 0.004 &  0.25 &  u$-$g  &  FO &  1.5 &  A &  $-$3.409 &  $-$2.992 &    1.336 &  2.596 &  0.426 &  2.361 &  0.254 &  0.935 \\
 0.004 &  0.25 &  g$-$r  &  FO &  1.5 &  A &  $-$2.583 &  $-$3.620 &    3.541 &  0.114 &  0.167 &  0.746 &  0.170 &  0.974 \\
  0.004 &  0.25 &  g$-$i &  FO &  1.5 &  A &  $-$2.347 &  $-$3.641 &    2.018 &  0.076 &  0.163 &  0.472 &  0.165 &  0.977 \\
 0.004 &  0.25 &  i$-$z &  FO &  1.5 &  A &  $-$1.545 &  $-$3.997 &   14.269 &  0.115 &  0.209 &  3.105 &  0.155 &  0.980 \\
 0.004 &  0.25 &  u$-$r &  FO &  1.7 &  A &  $-$5.369 &  $-$4.166 &    3.706 &  0.400 &  0.123 &  0.325 &  0.050 &  0.997 \\
 0.004 &  0.25 &  u$-$g  &  FO &  1.7 &  A &  $-$5.635 &  $-$3.590 &    3.404 &  3.039 &  0.481 &  2.774 &  0.145 &  0.977 \\
 0.004 &  0.25 &  g$-$r  &  FO &  1.7 &  A &  $-$2.509 &  $-$3.856 &    3.482 &  0.031 &  0.047 &  0.217 &  0.028 &  0.999 \\
 0.004 &  0.25 &  g$-$i &  FO &  1.7 &  A &  $-$2.264 &  $-$3.856 &    1.903 &  0.016 &  0.039 &  0.118 &  0.024 &  0.999 \\
 0.004 &  0.25 &  i$-$z &  FO &  1.7 &  A &  $-$1.528 &  $-$4.146 &   12.430 &  0.017 &  0.029 &  0.430 &  0.012 &  1.000 \\
 0.004 &  0.25 &  u$-$r &  FO &  1.9 &  A &  $-$5.433 &  $-$4.182 &    3.762 &  0.282 &  0.079 &  0.229 &  0.021 &  0.999 \\
 0.004 &  0.25 &  u$-$g  &  FO &  1.9 &  A &  $-$6.528 &  $-$3.750 &    4.232 &  3.581 &  0.521 &  3.271 &  0.101 &  0.988 \\
 0.004 &  0.25 &  g$-$r  &  FO &  1.9 &  A &  $-$2.494 &  $-$3.861 &    3.381 &  0.021 &  0.029 &  0.150 &  0.011 &  1.000 \\
 0.004 &  0.25 &  g$-$i &  FO &  1.9 &  A &  $-$2.255 &  $-$3.861 &    1.839 &  0.014 &  0.030 &  0.101 &  0.012 &  1.000 \\
 0.004 &  0.25 & i$-$z &  FO &  1.9 &  A &  $-$1.507 &  $-$4.170 &   12.674 &  0.033 &  0.053 &  0.868 &  0.013 &  1.000 \\
 0.004 &  0.25 &  u$-$r &  FO &  1.5 &  B &  $-$4.334 &  $-$3.810 &    2.989 &  0.711 &  0.286 &  0.582 &  0.138 &  0.953 \\
 0.004 &  0.25 &  u$-$g  &  FO &  1.5 &  B &  $-$7.801 &  $-$3.826 &    5.529 &  2.919 &  0.563 &  2.682 &  0.192 &  0.920 \\
 0.004 &  0.25 &  g$-$r  &  FO &  1.5 &  B &  $-$2.280 &  $-$3.614 &    2.841 &  0.111 &  0.211 &  0.711 &  0.128 &  0.968 \\
 0.004 &  0.25 &  g$-$i &  FO &  1.5 &  B &  $-$2.079 &  $-$3.624 &    1.527 &  0.082 &  0.209 &  0.457 &  0.127 &  0.971 \\
 0.004 &  0.25 &  i$-$z &  FO &  1.5 &  B &  $-$1.497 &  $-$3.861 &   10.029 &  0.132 &  0.257 &  3.159 &  0.124 &  0.973 \\
 0.004 &  0.25 &  u$-$r &  FO &  1.7 &  B &  $-$4.106 &  $-$3.611 &    2.767 &  1.179 &  0.433 &  0.969 &  0.148 &  0.940 \\
 0.004 &  0.25 &  u$-$g  &  FO &  1.7 &  B &  $-$4.230 &  $-$3.054 &    2.191 &  4.644 &  0.781 &  4.255 &  0.187 &  0.915 \\
 0.004 &  0.25 &  g$-$r  &  FO &  1.7 &  B &  $-$2.282 &  $-$3.497 &    2.620 &  0.157 &  0.310 &  1.122 &  0.139 &  0.959 \\
 0.004 &  0.25 &  g$-$i &  FO &  1.7 &  B &  $-$2.088 &  $-$3.508 &    1.374 &  0.106 &  0.307 &  0.718 &  0.137 &  0.963 \\
 0.004 &  0.25 &  i$-$z &  FO &  1.7 &  B &  $-$1.560 &  $-$3.712 &    8.733 &  0.209 &  0.390 &  4.942 &  0.136 &  0.965 \\
 0.004 &  0.25 &  u$-$r &  FO &  1.9 &  B &  $-$5.886 &  $-$4.328 &    4.240 &  0.650 &  0.185 &  0.529 &  0.020 &  0.999 \\
 0.004 &  0.25 &  u$-$g  &  FO &  1.9 &  B &  $-$1.198 &  $-$2.981 &   $-$0.522 &  8.710 &  1.330 &  7.978 &  0.094 &  0.982 \\
 0.004 &  0.25 &  g$-$r  &  FO &  1.9 &  B &  $-$2.378 &  $-$3.835 &    3.459 &  0.013 &  0.019 &  0.095 &  0.003 &  1.000 \\
 0.004 &  0.25 & g$-$i &  FO &  1.9 &  B &  $-$2.130 &  $-$3.828 &    1.859 &  0.009 &  0.021 &  0.070 &  0.004 &  1.000 \\
 0.004 &  0.25 &  i$-$z &  FO &  1.9 &  B &  $-$1.390 &  $-$4.116 &   12.510 &  0.006 &  0.010 &  0.164 &  0.001 &  1.000 \\
 0.004 &  0.25 &  u$-$r &  FO &  1.5 &  C &  $-$4.754 &  $-$4.040 &    3.429 &  0.446 &  0.161 &  0.377 &  0.015 &  0.999 \\
 0.004 &  0.25 &  u$-$g  &  FO &  1.5 &  C &  $-$9.304 &  $-$4.214 &    6.975 &  4.896 &  0.909 &  4.536 &  0.056 &  0.987 \\
 0.004 &  0.25 &  g$-$r  &  FO &  1.5 &  C &  $-$2.222 &  $-$3.744 &    3.289 &  0.017 &  0.034 &  0.147 &  0.004 &  1.000 \\
 0.004 &  0.25 &  g$-$i &  FO &  1.5 &  C &  $-$1.986 &  $-$3.749 &    1.769 &  0.008 &  0.026 &  0.072 &  0.003 &  1.000 \\
 0.004 &  0.25 &  i$-$z &  FO &  1.5 &  C &  $-$1.288 &  $-$4.080 &   11.967 &  0.014 &  0.022 &  0.305 &  0.002 &  1.000 \\
\hline\hline
\end{tabular}}
\end{table*}

\begin{table*}
\caption{\normalsize{\label{plc_lmc}}The coefficients of the PLC relations $M_{\lambda}$ = $a$ + $b \log P$ + $c$ CI, for $Z$=$0.008$, $Y$= $0.25$. The last two columns represent the root-mean-square deviation ($\sigma$) and the R-squared ($R^2$) coefficients.}
\centering
\scalebox{0.75}{
\begin{tabular}{cccccccccccccc}
\hline\hline
Z&Y&color&mode&$alpha_{ml}$&ML&a&b&c&$\sigma_{a}$&$\sigma_{b}$&$\sigma_{c}$&$\sigma$&$R^2$\\
\hline\hline
 0.008 &  0.25 &  u$-$r &   F &  1.5 &  A &   $-$2.726 &   $-$3.848 &   2.078 &  0.031 &  0.032 &  0.024 &  0.059 &  0.995 \\
 0.008 &  0.25 &  u$-$g  &   F &  1.5 &  A &   $-$3.370 &   $-$3.885 &   2.125 &  0.071 &  0.062 &  0.071 &  0.113 &  0.987 \\
 0.008 &  0.25 &  g$-$r  &   F &  1.5 &  A &   $-$2.050 &   $-$3.764 &   3.087 &  0.012 &  0.016 &  0.036 &  0.031 &  0.999 \\
 0.008 &  0.25 &  g$-$i &   F &  1.5 &  A &   $-$1.895 &   $-$3.755 &   1.904 &  0.011 &  0.015 &  0.025 &  0.030 &  0.999 \\
 0.008 &  0.25 &  i$-$z &   F &  1.5 &  A &   $-$1.146 &   $-$3.884 &  11.686 &  0.008 &  0.010 &  0.100 &  0.018 &  1.000 \\
 0.008 &  0.25 &  u$-$r &   F &  1.7 &  A &   $-$2.798 &   $-$3.853 &   2.106 &  0.038 &  0.038 &  0.034 &  0.046 &  0.996 \\
 0.008 &  0.25 &  u$-$g  &   F &  1.7 &  A &   $-$3.543 &   $-$3.887 &   2.221 &  0.097 &  0.076 &  0.103 &  0.089 &  0.991 \\
 0.008 &  0.25 &  g$-$r  &   F &  1.7 &  A &   $-$2.043 &   $-$3.763 &   3.028 &  0.011 &  0.017 &  0.044 &  0.022 &  1.000 \\
 0.008 &  0.25 &  g$-$i &   F &  1.7 &  A &   $-$1.886 &   $-$3.754 &   1.860 &  0.010 &  0.017 &  0.031 &  0.022 &  1.000 \\
 0.008 &  0.25 &  i$-$z &   F &  1.7 &  A &   $-$1.141 &   $-$3.898 &  12.159 &  0.011 &  0.015 &  0.172 &  0.017 &  1.000 \\
 0.008 &  0.25 &  u$-$r &   F &  1.9 &  A &   $-$2.929 &   $-$3.891 &   2.177 &  0.067 &  0.056 &  0.061 &  0.041 &  0.997 \\
 0.008 &  0.25 &  u$-$g  &   F &  1.9 &  A &   $-$3.802 &   $-$3.885 &   2.360 &  0.179 &  0.111 &  0.186 &  0.078 &  0.993 \\
 0.008 &  0.25 &  g$-$r  &   F &  1.9 &  A &   $-$2.030 &   $-$3.812 &   3.066 &  0.012 &  0.022 &  0.067 &  0.018 &  1.000 \\
 0.008 &  0.25 &  g$-$i &   F &  1.9 &  A &   $-$1.857 &   $-$3.803 &   1.868 &  0.009 &  0.021 &  0.044 &  0.017 &  1.000 \\
 0.008 &  0.25 &  i$-$z &   F &  1.9 &  A &   $-$1.111 &   $-$3.925 &  12.537 &  0.022 &  0.033 &  0.447 &  0.023 &  1.000 \\
 0.008 &  0.25 &  u$-$r &   F &  1.5 &  B &   $-$2.518 &   $-$3.766 &   2.006 &  0.030 &  0.032 &  0.023 &  0.074 &  0.993 \\
 0.008 &  0.25 &  u$-$g  &   F &  1.5 &  B &   $-$3.037 &   $-$3.744 &   1.908 &  0.060 &  0.056 &  0.058 &  0.129 &  0.986 \\
 0.008 &  0.25 &  g$-$r  &   F &  1.5 &  B &   $-$1.930 &   $-$3.743 &   3.120 &  0.014 &  0.018 &  0.039 &  0.042 &  0.999 \\
 0.008 &  0.25 &  g$-$i &   F &  1.5 &  B &   $-$1.773 &   $-$3.736 &   1.933 &  0.013 &  0.017 &  0.027 &  0.041 &  0.999 \\
 0.008 &  0.25 &  i$-$z &   F &  1.5 &  B &   $-$1.054 &   $-$3.804 &  11.253 &  0.017 &  0.019 &  0.181 &  0.043 &  0.999 \\
 0.008 &  0.25 &  u$-$r &   F &  1.7 &  B &   $-$2.593 &   $-$3.767 &   2.033 &  0.026 &  0.029 &  0.023 &  0.052 &  0.996 \\
 0.008 &  0.25 &  u$-$g  &   F &  1.7 &  B &   $-$3.134 &   $-$3.693 &   1.913 &  0.062 &  0.057 &  0.066 &  0.104 &  0.991 \\
 0.008 &  0.25 &  g$-$r  &   F &  1.7 &  B &   $-$1.946 &   $-$3.779 &   3.219 &  0.011 &  0.018 &  0.043 &  0.032 &  0.999 \\
 0.008 &  0.25 &  g$-$i &   F &  1.7 &  B &   $-$1.778 &   $-$3.779 &   2.012 &  0.011 &  0.018 &  0.031 &  0.033 &  0.999 \\
  0.008 &  0.25 &  i$-$z &   F &  1.7 &  B &   $-$1.010 &   $-$3.863 &  12.121 &  0.014 &  0.018 &  0.185 &  0.030 &  1.000 \\
 0.008 &  0.25 &  u$-$r &   F &  1.9 &  B &   $-$2.690 &   $-$3.784 &   2.075 &  0.036 &  0.036 &  0.034 &  0.040 &  0.998 \\
 0.008 &  0.25 &  u$-$g  &   F &  1.9 &  B &   $-$3.302 &   $-$3.674 &   1.980 &  0.094 &  0.071 &  0.102 &  0.081 &  0.995 \\
 0.008 &  0.25 &  g$-$r  &   F &  1.9 &  B &   $-$1.938 &   $-$3.813 &   3.245 &  0.011 &  0.023 &  0.065 &  0.025 &  1.000 \\
 0.008 &  0.25 &  g$-$i &   F &  1.9 &  B &   $-$1.760 &   $-$3.817 &   2.031 &  0.011 &  0.026 &  0.051 &  0.028 &  1.000 \\
 0.008 &  0.25 &  i$-$z &   F &  1.9 &  B &   $-$0.973 &   $-$3.906 &  12.875 &  0.024 &  0.034 &  0.428 &  0.034 &  0.999 \\
 0.008 &  0.25 &  u$-$r &   F &  1.5 &  C &   $-$2.395 &   $-$3.751 &   2.016 &  0.033 &  0.028 &  0.023 &  0.067 &  0.996 \\
 0.008 &  0.25 &  u$-$g  &   F &  1.5 &  C &   $-$2.981 &   $-$3.759 &   1.982 &  0.067 &  0.049 &  0.060 &  0.116 &  0.990 \\
 0.008 &  0.25 &  g$-$r  &   F &  1.5 &  C &   $-$1.765 &   $-$3.710 &   3.039 &  0.014 &  0.015 &  0.039 &  0.038 &  0.999 \\
 0.008 &  0.25 &  g$-$i &   F &  1.5 &  C &   $-$1.613 &   $-$3.700 &   1.872 &  0.014 &  0.015 &  0.027 &  0.038 &  0.999 \\
 0.008 &  0.25 &  i$-$z &   F &  1.5 &  C &   $-$0.980 &   $-$3.698 &  10.608 &  0.019 &  0.020 &  0.228 &  0.050 &  0.999 \\
 0.008 &  0.25 &  u$-$r &   F &  1.7 &  C &   $-$2.500 &   $-$3.750 &   2.053 &  0.028 &  0.025 &  0.022 &  0.050 &  0.997 \\
 0.008 &  0.25 &  u$-$g  &   F &  1.7 &  C &   $-$3.139 &   $-$3.730 &   2.033 &  0.067 &  0.050 &  0.065 &  0.098 &  0.993 \\
 0.008 &  0.25 &  g$-$r  &   F &  1.7 &  C &   $-$1.799 &   $-$3.726 &   3.120 &  0.011 &  0.014 &  0.036 &  0.027 &  1.000 \\
 0.008 &  0.25 &  g$-$i &   F &  1.7 &  C &   $-$1.634 &   $-$3.717 &   1.926 &  0.011 &  0.014 &  0.026 &  0.028 &  1.000 \\
 0.008 &  0.25 &  i$-$z &   F &  1.7 &  C &   $-$0.984 &   $-$3.701 &  10.931 &  0.018 &  0.020 &  0.233 &  0.040 &  0.999 \\
 0.008 &  0.25 &  u$-$r &   F &  1.9 &  C &   $-$2.522 &   $-$3.712 &   2.032 &  0.031 &  0.033 &  0.030 &  0.042 &  0.998 \\
 0.008 &  0.25 &  u$-$g  &   F &  1.9 &  C &   $-$3.100 &   $-$3.600 &   1.897 &  0.078 &  0.064 &  0.085 &  0.083 &  0.995 \\
 0.008 &  0.25 &  g$-$r  &   F &  1.9 &  C &   $-$1.814 &   $-$3.750 &   3.188 &  0.013 &  0.025 &  0.067 &  0.031 &  0.999 \\
 0.008 &  0.25 &  g$-$i &   F &  1.9 &  C &   $-$1.639 &   $-$3.748 &   1.982 &  0.013 &  0.027 &  0.052 &  0.033 &  0.999 \\
 0.008 &  0.25 &  i$-$z &   F &  1.9 &  C &   $-$0.948 &   $-$3.736 &  11.390 &  0.024 &  0.031 &  0.371 &  0.038 &  0.999 \\
 0.008 &  0.25 &  u$-$r &  FO &  1.5 &  A &   $-$4.604 &   $-$3.797 &   2.813 &  0.534 &  0.185 &  0.393 &  0.152 &  0.977 \\
 0.008 &  0.25 &  u$-$g  &  FO &  1.5 &  A &   $-$5.468 &   $-$3.494 &   2.964 &  1.588 &  0.327 &  1.344 &  0.217 &  0.959 \\
 0.008 &  0.25 &  g$-$r  &  FO &  1.5 &  A &   $-$2.743 &   $-$3.700 &   3.403 &  0.088 &  0.112 &  0.476 &  0.122 &  0.989 \\
 0.008 &  0.25 &  g$-$i &  FO &  1.5 &  A &   $-$2.487 &   $-$3.712 &   1.967 &  0.057 &  0.105 &  0.296 &  0.115 &  0.991 \\
 0.008 &  0.25 &  i$-$z &  FO &  1.5 &  A &   $-$1.607 &   $-$4.059 &  14.245 &  0.062 &  0.113 &  1.677 &  0.093 &  0.994 \\
 0.008 &  0.25 &  u$-$r &  FO &  1.7 &  A &   $-$4.658 &   $-$3.991 &   2.912 &  0.493 &  0.170 &  0.366 &  0.097 &  0.989 \\
 0.008 &  0.25 &  u$-$g  &  FO &  1.7 &  A &   $-$6.961 &   $-$3.948 &   4.303 &  1.721 &  0.351 &  1.469 &  0.152 &  0.978 \\
 0.008 &  0.25 &  g$-$r  &  FO &  1.7 &  A &   $-$2.632 &   $-$3.806 &   3.165 &  0.079 &  0.104 &  0.445 &  0.080 &  0.994 \\
 0.008 &  0.25 &  g$-$i &  FO &  1.7 &  A &   $-$2.387 &   $-$3.808 &   1.788 &  0.051 &  0.098 &  0.281 &  0.076 &  0.995 \\
 0.008 &  0.25 &  i$-$z &  FO &  1.7 &  A &   $-$1.578 &   $-$4.107 &  12.924 &  0.068 &  0.117 &  1.798 &  0.067 &  0.997 \\
 0.008 &  0.25 &  u$-$r &  FO &  1.9 &  A &   $-$5.960 &   $-$4.372 &   3.901 &  0.180 &  0.051 &  0.135 &  0.011 &  1.000 \\
 0.008 &  0.25 &  u$-$g  &  FO &  1.9 &  A &  $-$26.520 &   $-$6.910 &  21.057 &  5.143 &  0.785 &  4.405 &  0.056 &  0.993 \\
 0.008 &  0.25 &  g$-$r  &  FO &  1.9 &  A &   $-$2.641 &   $-$3.932 &   3.324 &  0.032 &  0.041 &  0.195 &  0.014 &  1.000 \\
 0.008 &  0.25 &  g$-$i &  FO &  1.9 &  A &   $-$2.369 &   $-$3.918 &   1.799 &  0.017 &  0.035 &  0.108 &  0.012 &  1.000 \\
 0.008 &  0.25 &  i$-$z &  FO &  1.9 &  A &   $-$1.596 &   $-$4.165 &  11.477 &  0.022 &  0.034 &  0.538 &  0.009 &  1.000 \\
 0.008 &  0.25 &  u$-$r &  FO &  1.5 &  B &   $-$4.998 &   $-$4.250 &   3.295 &  0.218 &  0.092 &  0.165 &  0.040 &  0.997 \\
 0.008 &  0.25 &  u$-$g  &  FO &  1.5 &  B &   $-$9.896 &   $-$4.705 &   6.946 &  1.654 &  0.396 &  1.422 &  0.119 &  0.975 \\
 0.008 &  0.25 &  g$-$r  &  FO &  1.5 &  B &   $-$2.489 &   $-$3.883 &   3.324 &  0.024 &  0.042 &  0.145 &  0.025 &  0.999 \\
 0.008 &  0.25 &  g$-$i &  FO &  1.5 &  B &   $-$2.227 &   $-$3.865 &   1.836 &  0.013 &  0.033 &  0.075 &  0.020 &  0.999 \\
 0.008 &  0.25 &  i$-$z &  FO &  1.5 &  B &   $-$1.473 &   $-$4.082 &  11.610 &  0.009 &  0.017 &  0.220 &  0.009 &  1.000 \\
 0.008 &  0.25 &  u$-$r &  FO &  1.7 &  B &   $-$5.124 &   $-$4.281 &   3.392 &  0.171 &  0.065 &  0.129 &  0.018 &  0.999 \\
 0.008 &  0.25 &  u$-$g  &  FO &  1.7 &  B &  $-$11.049 &   $-$4.911 &   7.949 &  2.460 &  0.541 &  2.113 &  0.092 &  0.986 \\
 0.008 &  0.25 &  g$-$r  &  FO &  1.7 &  B &   $-$2.470 &   $-$3.865 &   3.204 &  0.018 &  0.028 &  0.107 &  0.011 &  1.000 \\
 0.008 &  0.25 &  g$-$i &  FO &  1.7 &  B &   $-$2.215 &   $-$3.848 &   1.760 &  0.011 &  0.026 &  0.066 &  0.011 &  1.000 \\
 0.008 &  0.25 &  i$-$z &  FO &  1.7 &  B &   $-$1.449 &   $-$4.110 &  12.102 &  0.009 &  0.016 &  0.226 &  0.005 &  1.000 \\
 0.008 &  0.25 &  u$-$r &  FO &  1.9 &  B &   $-$6.002 &   $-$4.391 &   4.037 &  0.137 &  0.046 &  0.103 &  0.006 &  1.000 \\
 0.008 &  0.25 &  u$-$g  &  FO &  1.9 &  B &  $-$57.390 &  $-$12.283 &  47.635 &  9.010 &  1.515 &  7.723 &  0.035 &  0.994 \\
 0.008 &  0.25 &  g$-$r  &  FO &  1.9 &  B &   $-$2.497 &   $-$3.845 &   3.275 &  0.024 &  0.040 &  0.147 &  0.008 &  1.000 \\
 0.008 &  0.25 &  g$-$i &  FO &  1.9 &  B &   $-$2.227 &   $-$3.829 &   1.768 &  0.014 &  0.038 &  0.091 &  0.007 &  1.000 \\
 0.008 &  0.25 &  i$-$z &  FO &  1.9 &  B &   $-$1.442 &   $-$4.096 &  12.068 &  0.027 &  0.049 &  0.671 &  0.007 &  1.000 \\
 0.008 &  0.25 &  u$-$r &  FO &  1.5 &  C &   $-$4.803 &   $-$4.143 &   3.230 &  0.229 &  0.115 &  0.180 &  0.025 &  0.997 \\
 0.008 &  0.25 &  u$-$g  &  FO &  1.5 &  C &  $-$10.836 &   $-$4.872 &   7.833 &  1.569 &  0.421 &  1.365 &  0.064 &  0.982 \\
 0.008 &  0.25 &  g$-$r  &  FO &  1.5 &  C &   $-$2.379 &   $-$3.799 &   3.331 &  0.024 &  0.057 &  0.166 &  0.016 &  0.999 \\
 0.008 &  0.25 &  g$-$i &  FO &  1.5 &  C &   $-$2.110 &   $-$3.793 &   1.836 &  0.012 &  0.043 &  0.082 &  0.012 &  1.000 \\
 0.008 &  0.25 &  i$-$z &  FO &  1.5 &  C &   $-$1.350 &   $-$4.041 &  11.481 &  0.014 &  0.026 &  0.282 &  0.006 &  1.000 \\
\hline\hline
\end{tabular}}
\end{table*}

\begin{table*}
\caption{\label{plc_mw}The coefficients of the PLC relations $M_{\lambda}$ = $a$ + $b \log P$ + $c$ CI, for $Z$=$0.02$, $Y$= $0.28$. The last two columns represent the root-mean-square deviation ($\sigma$) and the R-squared ($R^2$) coefficients.}
\centering
\scalebox{0.75}{
\begin{tabular}{cccccccccccccc}
\hline\hline
Z&Y&color&mode&$alpha_{ml}$&ML&a&b&c&$\sigma_{a}$&$\sigma_{b}$&$\sigma_{c}$&$\sigma$&$R^2$\\
\hline\hline
0.02 &  0.28 &  u$-$r &   F &  1.5 &  A &   $-$2.663 &  $-$3.722 &   1.876 &  0.028 &  0.030 &  0.020 &  0.046 &  0.996 \\
 0.02 &  0.28 &  u$-$g &   F &  1.5 &  A &   $-$2.991 &  $-$3.588 &   1.575 &  0.058 &  0.053 &  0.051 &  0.084 &  0.994 \\
 0.02 &  0.28 &  g$-$r &   F &  1.5 &  A &   $-$2.221 &  $-$3.824 &   3.074 &  0.016 &  0.025 &  0.054 &  0.036 &  0.999 \\
 0.02 &  0.28 &  g$-$i &   F &  1.5 &  A &   $-$2.034 &  $-$3.822 &   1.963 &  0.014 &  0.024 &  0.038 &  0.034 &  0.999 \\
 0.02 &  0.28 &  i$-$z &   F &  1.5 &  A &   $-$1.128 &  $-$3.819 &   9.926 &  0.008 &  0.013 &  0.116 &  0.019 &  1.000 \\
 0.02 &  0.28 &  u$-$r &   F &  1.7 &  A &   $-$2.635 &  $-$3.671 &   1.823 &  0.050 &  0.050 &  0.038 &  0.035 &  0.998 \\
 0.02 &  0.28 &  u$-$g &   F &  1.7 &  A &   $-$2.912 &  $-$3.439 &   1.417 &  0.093 &  0.076 &  0.085 &  0.058 &  0.997 \\
 0.02 &  0.28 &  g$-$r &   F &  1.7 &  A &   $-$2.208 &  $-$3.900 &   3.124 &  0.024 &  0.047 &  0.114 &  0.028 &  1.000 \\
 0.02 &  0.28 &  g$-$i &   F &  1.7 &  A &   $-$2.031 &  $-$3.951 &   2.100 &  0.021 &  0.049 &  0.087 &  0.028 &  1.000 \\
 0.02 &  0.28 &  i$-$z &   F &  1.7 &  A &   $-$1.086 &  $-$3.896 &  10.999 &  0.018 &  0.035 &  0.366 &  0.021 &  1.000 \\
 0.02 &  0.28 &  u$-$r &   F &  1.9 &  A &   $-$3.642 &  $-$3.930 &   2.450 &  0.359 &  0.129 &  0.235 &  0.008 &  1.000 \\
 0.02 &  0.28 &  u$-$g &   F &  1.9 &  A &   $-$5.624 &  $-$4.048 &   3.504 &  1.151 &  0.293 &  0.905 &  0.017 &  0.999 \\
 0.02 &  0.28 &  g$-$r &   F &  1.9 &  A &   $-$2.215 &  $-$3.823 &   3.071 &  0.040 &  0.035 &  0.157 &  0.003 &  1.000 \\
 0.02 &  0.28 &  g$-$i &   F &  1.9 &  A &   $-$1.941 &  $-$3.822 &   1.783 &  0.020 &  0.023 &  0.070 &  0.002 &  1.000 \\
 0.02 &  0.28 &  i$-$z &   F &  1.9 &  A &   $-$1.060 &  $-$3.960 &  10.356 &  0.052 &  0.095 &  1.499 &  0.006 &  1.000 \\
 0.02 &  0.28 &  u$-$r &   F &  1.5 &  B &   $-$2.511 &  $-$3.713 &   1.874 &  0.020 &  0.024 &  0.015 &  0.044 &  0.997 \\
 0.02 &  0.28 &  u$-$g &   F &  1.5 &  B &   $-$2.776 &  $-$3.573 &   1.532 &  0.043 &  0.045 &  0.039 &  0.086 &  0.993 \\
 0.02 &  0.28 &  g$-$r &   F &  1.5 &  B &   $-$2.131 &  $-$3.822 &   3.194 &  0.017 &  0.027 &  0.057 &  0.047 &  0.999 \\
 0.02 &  0.28 &  g$-$i &   F &  1.5 &  B &   $-$1.931 &  $-$3.797 &   2.011 &  0.014 &  0.025 &  0.037 &  0.043 &  0.999 \\
 0.02 &  0.28 &  i$-$z &   F &  1.5 &  B &   $-$1.034 &  $-$3.734 &   9.511 &  0.010 &  0.014 &  0.116 &  0.026 &  1.000 \\
 0.02 &  0.28 &  u$-$r &   F &  1.7 &  B &   $-$2.567 &  $-$3.737 &   1.892 &  0.024 &  0.031 &  0.020 &  0.033 &  0.998 \\
 0.02 &  0.28 &  u$-$g &   F &  1.7 &  B &   $-$2.768 &  $-$3.494 &   1.459 &  0.046 &  0.050 &  0.046 &  0.057 &  0.997 \\
 0.02 &  0.28 &  g$-$r &   F &  1.7 &  B &   $-$2.193 &  $-$3.962 &   3.503 &  0.031 &  0.063 &  0.140 &  0.057 &  0.998 \\
 0.02 &  0.28 &  g$-$i &   F &  1.7 &  B &   $-$1.978 &  $-$3.944 &   2.255 &  0.024 &  0.059 &  0.095 &  0.053 &  0.999 \\
 0.02 &  0.28 &  i$-$z &   F &  1.7 &  B &   $-$1.019 &  $-$3.750 &   9.773 &  0.015 &  0.024 &  0.210 &  0.025 &  1.000 \\
 0.02 &  0.28 &  u$-$r &   F &  1.9 &  B &   $-$3.670 &  $-$4.001 &   2.569 &  0.210 &  0.084 &  0.138 &  0.007 &  1.000 \\
 0.02 &  0.28 &  u$-$g &   F &  1.9 &  B &   $-$6.376 &  $-$4.306 &   4.219 &  0.732 &  0.205 &  0.579 &  0.014 &  0.999 \\
 0.02 &  0.28 &  g$-$r &   F &  1.9 &  B &   $-$2.062 &  $-$3.814 &   3.064 &  0.028 &  0.028 &  0.109 &  0.003 &  1.000 \\
 0.02 &  0.28 &  g$-$i &   F &  1.9 &  B &   $-$1.785 &  $-$3.804 &   1.761 &  0.015 &  0.021 &  0.055 &  0.002 &  1.000 \\
 0.02 &  0.28 &  i$-$z &   F &  1.9 &  B &   $-$0.906 &  $-$3.965 &  10.597 &  0.008 &  0.016 &  0.226 &  0.001 &  1.000 \\
 0.02 &  0.28 &  u$-$r &   F &  1.5 &  C &   $-$2.384 &  $-$3.707 &   1.881 &  0.023 &  0.023 &  0.014 &  0.047 &  0.997 \\
 0.02 &  0.28 &  u$-$g &   F &  1.5 &  C &   $-$2.688 &  $-$3.612 &   1.583 &  0.047 &  0.043 &  0.037 &  0.090 &  0.993 \\
 0.02 &  0.28 &  g$-$r &   F &  1.5 &  C &   $-$1.980 &  $-$3.773 &   3.115 &  0.018 &  0.023 &  0.049 &  0.046 &  0.999 \\
 0.02 &  0.28 &  g$-$i &   F &  1.5 &  C &   $-$1.786 &  $-$3.740 &   1.947 &  0.016 &  0.021 &  0.033 &  0.043 &  0.999 \\
 0.02 &  0.28 &  i$-$z &   F &  1.5 &  C &   $-$0.962 &  $-$3.624 &   8.907 &  0.014 &  0.016 &  0.139 &  0.036 &  0.999 \\
 0.02 &  0.28 &  u$-$r &   F &  1.7 &  C &   $-$2.429 &  $-$3.703 &   1.883 &  0.031 &  0.034 &  0.024 &  0.038 &  0.997 \\
 0.02 &  0.28 &  u$-$g &   F &  1.7 &  C &   $-$2.758 &  $-$3.567 &   1.570 &  0.066 &  0.062 &  0.063 &  0.072 &  0.996 \\
 0.02 &  0.28 & g$-$r &   F &  1.7 &  C &   $-$1.954 &  $-$3.753 &   3.008 &  0.027 &  0.047 &  0.108 &  0.050 &  0.998 \\
 0.02 &  0.28 &  g$-$i &   F &  1.7 &  C &   $-$1.766 &  $-$3.726 &   1.885 &  0.024 &  0.047 &  0.078 &  0.050 &  0.999 \\
 0.02 &  0.28 &  i$-$z &   F &  1.7 &  C &   $-$0.955 &  $-$3.621 &   8.873 &  0.017 &  0.025 &  0.236 &  0.030 &  1.000 \\
 0.02 &  0.28 &  u$-$r &   F &  1.9 &  C &   $-$3.003 &  $-$4.021 &   2.267 &  0.439 &  0.326 &  0.329 &  0.035 &  0.994 \\
 0.02 &  0.28 &  u$-$g &   F &  1.9 &  C &   $-$3.626 &  $-$3.727 &   2.164 &  1.160 &  0.620 &  1.006 &  0.065 &  0.988 \\
 0.02 &  0.28 &  g$-$r &   F &  1.9 &  C &   $-$1.896 &  $-$3.972 &   3.203 &  0.032 &  0.064 &  0.168 &  0.008 &  1.000 \\
 0.02 &  0.28 &  g$-$i &   F &  1.9 &  C &   $-$1.635 &  $-$3.913 &   1.872 &  0.009 &  0.026 &  0.048 &  0.003 &  1.000 \\
 0.02 &  0.28 &  i$-$z &   F &  1.9 &  C &   $-$0.782 &  $-$3.908 &  10.932 &  0.045 &  0.081 &  0.972 &  0.011 &  1.000 \\
 0.02 &  0.28 &  u$-$r &  FO &  1.5 &  A &   $-$4.977 &  $-$4.292 &   2.909 &  0.088 &  0.031 &  0.057 &  0.010 &  1.000 \\
 0.02 &  0.28 &  u$-$g &  FO &  1.5 &  A &  $-$10.889 &  $-$5.123 &   6.990 &  0.642 &  0.147 &  0.492 &  0.032 &  0.998 \\
 0.02 &  0.28 &  g$-$r &  FO &  1.5 &  A &   $-$2.742 &  $-$3.968 &   2.945 &  0.016 &  0.017 &  0.064 &  0.007 &  1.000 \\
 0.02 &  0.28 &  g$-$i &  FO &  1.5 &  A &   $-$2.456 &  $-$3.947 &   1.633 &  0.010 &  0.015 &  0.038 &  0.007 &  1.000 \\
 0.02 &  0.28 &  i$-$z &  FO &  1.5 &  A &   $-$1.647 &  $-$4.127 &  10.763 &  0.017 &  0.033 &  0.488 &  0.012 &  1.000 \\
 0.02 &  0.28 &  u$-$r &  FO &  1.7 &  A &   $-$5.394 &  $-$4.314 &   3.178 &  0.181 &  0.052 &  0.119 &  0.006 &  1.000 \\
 0.02 &  0.28 & u$-$g &  FO &  1.7 &  A &  $-$13.239 &  $-$5.315 &   8.833 &  2.661 &  0.490 &  2.060 &  0.031 &  0.996 \\
 0.02 &  0.28 &  g$-$r &  FO &  1.7 &  A &   $-$2.791 &  $-$3.960 &   3.061 &  0.024 &  0.021 &  0.099 &  0.003 &  1.000 \\
 0.02 &  0.28 & g$-$i &  FO &  1.7 &  A &   $-$2.486 &  $-$3.948 &   1.678 &  0.016 &  0.020 &  0.063 &  0.003 &  1.000 \\
 0.02 &  0.28 & i$-$z &  FO &  1.7 &  A &   $-$1.635 &  $-$4.202 &  11.056 &  0.012 &  0.020 &  0.318 &  0.002 &  1.000 \\
 0.02 &  0.28 &  u$-$r &  FO &  1.5 &  B &   $-$5.370 &  $-$4.496 &   3.259 &  0.140 &  0.058 &  0.092 &  0.012 &  0.999 \\
 0.02 &  0.28 &  u$-$g &  FO &  1.5 &  B &  $-$14.281 &  $-$5.972 &   9.719 &  0.412 &  0.101 &  0.318 &  0.013 &  0.999 \\
 0.02 &  0.28 &  g$-$r &  FO &  1.5 &  B &   $-$2.668 &  $-$4.046 &   3.234 &  0.036 &  0.051 &  0.153 &  0.015 &  0.999 \\
 0.02 &  0.28 &  g$-$i &  FO &  1.5 &  B &   $-$2.358 &  $-$4.010 &   1.807 &  0.024 &  0.047 &  0.095 &  0.014 &  0.999 \\
 0.02 &  0.28 &  i$-$z &  FO &  1.5 &  B &   $-$1.464 &  $-$4.175 &  12.222 &  0.018 &  0.038 &  0.495 &  0.010 &  1.000 \\
 0.02 &  0.28 &  u$-$r &  FO &  1.7 &  B &    0.090 &   2.148 &  $-$0.215 &  2.616 &  3.012 &  1.663 &  0.001 &  1.000 \\
 0.02 &  0.28 &  u$-$g &  FO &  1.7 &  B &   $-$0.564 &   1.595 &  $-$0.758 &  0.319 &  0.166 &  0.244 &  0.000 &  1.000 \\
 0.02 &  0.28 & g$-$r &  FO &  1.7 &  B &   $-$1.731 &   0.326 &  $-$0.334 &  0.037 &  0.158 &  0.139 &  0.000 &  0.963 \\
 0.02 &  0.28 &  g$-$i &  FO &  1.7 &  B &   $-$1.712 &   0.136 &  $-$0.458 &  0.027 &  0.159 &  0.091 &  0.000 &  1.000 \\
 0.02 &  0.28 &  i$-$z &  FO &  1.7 &  B &   $-$1.768 &  $-$1.475 &   1.795 &  0.021 &  0.214 &  0.736 &  0.000 &  1.000 \\
 0.02 &  0.28 &  u$-$r &  FO &  1.5 &  C &   $-$5.233 &  $-$4.136 &   3.221 &  0.133 &  0.052 &  0.088 &  0.007 &  1.000 \\
 0.02 &  0.28 & u$-$g &  FO &  1.5 &  C &  $-$11.430 &  $-$4.722 &   7.539 &  0.220 &  0.047 &  0.169 &  0.005 &  1.000 \\
 0.02 &  0.28 &  g$-$r &  FO &  1.5 &  C &   $-$2.644 &  $-$3.891 &   3.561 &  0.038 &  0.055 &  0.174 &  0.008 &  1.000 \\
 0.02 &  0.28 &  g$-$i &  FO &  1.5 &  C &   $-$2.292 &  $-$3.874 &   1.981 &  0.021 &  0.045 &  0.093 &  0.007 &  1.000 \\
 0.02 &  0.28 & i$-$z &  FO &  1.5 &  C &   $-$1.348 &  $-$4.043 &  12.082 &  0.008 &  0.016 &  0.191 &  0.002 &  1.000 \\
\hline\hline
\end{tabular}}
\end{table*}

\begin{table*}
\caption{\label{plc_supersolar}The coefficients of the PLC relations $M_{\lambda}$ = $a$ + $b \log P$ + $c$ CI, for $Z$=$0.03$, $Y$= $0.28$. The last two columns represent the root-mean-square deviation ($\sigma$) and the R-squared ($R^2$) coefficients.}
\centering
\scalebox{0.75}{
\begin{tabular}{cccccccccccccc}
\hline\hline
Z&Y&color&mode&$\alpha_{ml}$&ML&a&b&c&$\sigma_{a}$&$\sigma_{b}$&$\sigma_{c}$&$\sigma$&$R^2$\\
\hline\hline
0.03 &  0.28 &  u$-$r &   F &  1.5 &  A &  $-$2.757 &  $-$3.707 &   1.859 &  0.035 &  0.039 &  0.025 &  0.045 &  0.995 \\
 0.03 &  0.28 &  u$-$g  &   F &  1.5 &  A &  $-$3.017 &  $-$3.520 &   1.506 &  0.068 &  0.065 &  0.059 &  0.080 &  0.993 \\
 0.03 &  0.28 &  g$-$r  &   F &  1.5 &  A &  $-$2.366 &  $-$3.864 &   3.124 &  0.020 &  0.034 &  0.072 &  0.037 &  0.999 \\
 0.03 &  0.28 &  g$-$i &   F &  1.5 &  A &  $-$2.145 &  $-$3.835 &   1.958 &  0.017 &  0.033 &  0.050 &  0.035 &  0.999 \\
 0.03 &  0.28 &  i$-$z &   F &  1.5 &  A &  $-$1.185 &  $-$3.765 &   9.029 &  0.007 &  0.012 &  0.097 &  0.014 &  1.000 \\
 0.03 &  0.28 &  u$-$r &   F &  1.7 &  A &  $-$2.884 &  $-$3.848 &   1.932 &  0.101 &  0.107 &  0.075 &  0.035 &  0.996 \\
 0.03 &  0.28 &  u$-$g  &   F &  1.7 &  A &  $-$2.968 &  $-$3.444 &   1.404 &  0.184 &  0.163 &  0.165 &  0.061 &  0.996 \\
 0.03 &  0.28 &  g$-$r  &   F &  1.7 &  A &  $-$2.446 &  $-$4.107 &   3.537 &  0.071 &  0.138 &  0.308 &  0.038 &  0.999 \\
 0.03 &  0.28 &  g$-$i &   F &  1.7 &  A &  $-$2.199 &  $-$4.059 &   2.245 &  0.059 &  0.144 &  0.235 &  0.040 &  0.999 \\
 0.03 &  0.28 &  i$-$z &   F &  1.7 &  A &  $-$1.184 &  $-$3.775 &   9.353 &  0.014 &  0.031 &  0.289 &  0.012 &  1.000 \\
 0.03 &  0.28 &  u$-$r &   F &  1.5 &  B &  $-$2.619 &  $-$3.691 &   1.858 &  0.018 &  0.025 &  0.013 &  0.034 &  0.997 \\
 0.03 &  0.28 &  u$-$g  &   F &  1.5 &  B &  $-$2.812 &  $-$3.531 &   1.486 &  0.036 &  0.045 &  0.034 &  0.065 &  0.994 \\
 0.03 &  0.28 &  g$-$r  &   F &  1.5 &  B &  $-$2.330 &  $-$3.828 &   3.241 &  0.021 &  0.037 &  0.069 &  0.047 &  0.998 \\
 0.03 &  0.28 &  g$-$i &   F &  1.5 &  B &  $-$2.061 &  $-$3.755 &   1.942 &  0.013 &  0.024 &  0.032 &  0.031 &  0.999 \\
 0.03 &  0.28 &  i$-$z &   F &  1.5 &  B &  $-$1.092 &  $-$3.684 &   8.661 &  0.008 &  0.012 &  0.080 &  0.016 &  1.000 \\
 0.03 &  0.28 &  u$-$r &   F &  1.7 &  B &  $-$2.704 &  $-$3.811 &   1.919 &  0.031 &  0.045 &  0.026 &  0.026 &  0.997 \\
 0.03 &  0.28 &  u$-$g  &   F &  1.7 &  B &  $-$2.843 &  $-$3.571 &   1.494 &  0.041 &  0.054 &  0.044 &  0.033 &  0.999 \\
 0.03 &  0.28 &  g$-$r  &   F &  1.7 &  B &  $-$2.415 &  $-$4.023 &   3.628 &  0.050 &  0.108 &  0.221 &  0.054 &  0.998 \\
 0.03 &  0.28 &  g$-$i &   F &  1.7 &  B &  $-$2.144 &  $-$3.914 &   2.208 &  0.034 &  0.086 &  0.128 &  0.045 &  0.999 \\
 0.03 &  0.28 &  i$-$z &   F &  1.7 &  B &  $-$1.103 &  $-$3.661 &   8.573 &  0.011 &  0.017 &  0.133 &  0.011 &  1.000 \\
 0.03 &  0.28 &  u$-$r &   F &  1.5 &  C &  $-$2.459 &  $-$3.686 &   1.857 &  0.017 &  0.018 &  0.010 &  0.030 &  0.998 \\
 0.03 &  0.28 &  u$-$g  &   F &  1.5 &  C &  $-$2.633 &  $-$3.626 &   1.533 &  0.034 &  0.036 &  0.026 &  0.059 &  0.995 \\
 0.03 &  0.28 &  g$-$r  &   F &  1.5 &  C &  $-$2.221 &  $-$3.723 &   3.128 &  0.021 &  0.025 &  0.048 &  0.040 &  0.998 \\
 0.03 &  0.28 &  g$-$i &   F &  1.5 &  C &  $-$1.957 &  $-$3.650 &   1.859 &  0.013 &  0.016 &  0.022 &  0.026 &  0.999 \\
 0.03 &  0.28 &  i$-$z &   F &  1.5 &  C &  $-$1.003 &  $-$3.605 &   8.334 &  0.012 &  0.014 &  0.097 &  0.023 &  1.000 \\
 0.03 &  0.28 &  u$-$r &   F &  1.7 &  C &  $-$2.517 &  $-$3.693 &   1.865 &  0.021 &  0.026 &  0.015 &  0.022 &  0.998 \\
 0.03 &  0.28 &  u$-$g  &   F &  1.7 &  C &  $-$2.664 &  $-$3.619 &   1.519 &  0.031 &  0.035 &  0.029 &  0.030 &  0.999 \\
 0.03 &  0.28 &  g$-$r  &   F &  1.7 &  C &  $-$2.288 &  $-$3.723 &   3.174 &  0.038 &  0.055 &  0.118 &  0.045 &  0.998 \\
 0.03 &  0.28 &  g$-$i &   F &  1.7 &  C &  $-$2.044 &  $-$3.655 &   1.932 &  0.027 &  0.043 &  0.067 &  0.037 &  0.999 \\
 0.03 &  0.28 &  i$-$z &   F &  1.7 &  C &  $-$1.009 &  $-$3.583 &   8.181 &  0.012 &  0.017 &  0.129 &  0.015 &  1.000 \\
\hline\hline
\end{tabular}}
\end{table*}

\section{The PW and PWZ relations in the Rubin-LSST filters} 
The PW coefficients in the Rubin-LSST filters for the various investigated chemical compositions are listed in Tables \ref{pw_smc}–\ref{pw_supersolar}. The PWZ relation coefficients are reported in table \ref{pwz_lsst}.

\begin{table*}
\caption{\label{pw_smc}The coefficients of the PW ($W$ = $a$ + $b$ ($\log P$)) relations for $Z$=$0.004$, $Y$= $0.25$. The last two columns represent the root-mean-square deviation ($\sigma$) and the R-squared ($R^2$) coefficients.}
\centering
\scalebox{0.75}{
\begin{tabular}{cccccccccccc}
\hline\hline
Z&Y&band&mode&$\alpha_{ml}$&ML&a&b&$\sigma_{a}$&$\sigma_{b}$&$\sigma$&$R^2$\\
\hline\hline
0.004 &  0.25 &  g, u$-$g &   F &  1.5 &  A &  $-$4.128 &  $-$4.119 &  0.042 &  0.033 &  0.117 &  0.995 \\
 0.004 &  0.25 &  r, g$-$r  &   F &  1.5 &  A &  $-$1.927 &  $-$3.565 &  0.019 &  0.015 &  0.054 &  0.999 \\
 0.004 &  0.25 &  i, g$-$i &   F &  1.5 &  A &  $-$1.754 &  $-$3.425 &  0.034 &  0.026 &  0.094 &  0.995 \\
 0.004 &  0.25 &  z, i$-$z &   F &  1.5 &  A &  $-$1.474 &  $-$3.204 &  0.066 &  0.051 &  0.182 &  0.980 \\
 0.004 &  0.25 &  y, g$-$y &   F &  1.5 &  A &  $-$1.601 &  $-$3.304 &  0.052 &  0.040 &  0.142 &  0.989 \\
 0.004 &  0.25 &  g, u$-$g &   F &  1.7 &  A &  $-$4.264 &  $-$4.013 &  0.036 &  0.028 &  0.094 &  0.997 \\
 0.004 &  0.25 &  r, g$-$r  &   F &  1.7 &  A &  $-$1.923 &  $-$3.589 &  0.017 &  0.014 &  0.045 &  0.999 \\
 0.004 &  0.25 &  i, g$-$i &   F &  1.7 &  A &  $-$1.738 &  $-$3.468 &  0.030 &  0.024 &  0.078 &  0.997 \\
 0.004 &  0.25 &  z, i$-$z &   F &  1.7 &  A &  $-$1.444 &  $-$3.275 &  0.058 &  0.046 &  0.151 &  0.987 \\
 0.004 &  0.25 &  y, g$-$y &   F &  1.7 &  A &  $-$1.579 &  $-$3.360 &  0.045 &  0.036 &  0.118 &  0.993 \\
 0.004 &  0.25 &  g, u$-$g &   F &  1.9 &  A &  $-$4.389 &  $-$3.922 &  0.035 &  0.029 &  0.089 &  0.997 \\
 0.004 &  0.25 &  r, g$-$r  &   F &  1.9 &  A &  $-$1.915 &  $-$3.617 &  0.014 &  0.012 &  0.037 &  0.999 \\
 0.004 &  0.25 &  i, g$-$i &   F &  1.9 &  A &  $-$1.719 &  $-$3.512 &  0.023 &  0.020 &  0.060 &  0.998 \\
 0.004 &  0.25 &  z, i$-$z &   F &  1.9 &  A &  $-$1.421 &  $-$3.336 &  0.045 &  0.038 &  0.116 &  0.993 \\
 0.004 &  0.25 &  y, g$-$y &   F &  1.9 &  A &  $-$1.562 &  $-$3.411 &  0.035 &  0.030 &  0.091 &  0.996 \\
 0.004 &  0.25 &  g, u$-$g &   F &  1.5 &  B &  $-$4.002 &  $-$4.112 &  0.047 &  0.032 &  0.133 &  0.994 \\
 0.004 &  0.25 &  r, g$-$r  &   F &  1.5 &  B &  $-$1.800 &  $-$3.544 &  0.023 &  0.016 &  0.065 &  0.998 \\
 0.004 &  0.25 &  i, g$-$i &   F &  1.5 &  B &  $-$1.638 &  $-$3.397 &  0.037 &  0.026 &  0.107 &  0.995 \\
 0.004 &  0.25 &  z, i$-$z &   F &  1.5 &  B &  $-$1.383 &  $-$3.160 &  0.069 &  0.048 &  0.199 &  0.979 \\
 0.004 &  0.25 &  y, g$-$y &   F &  1.5 &  B &  $-$1.500 &  $-$3.267 &  0.054 &  0.037 &  0.156 &  0.988 \\
 0.004 &  0.25 &  g, u$-$g &   F &  1.7 &  B &  $-$4.103 &  $-$4.034 &  0.038 &  0.027 &  0.110 &  0.996 \\
 0.004 &  0.25 &  r, g$-$r  &   F &  1.7 &  B &  $-$1.801 &  $-$3.553 &  0.017 &  0.012 &  0.049 &  0.999 \\
 0.004 &  0.25 &  i, g$-$i &   F &  1.7 &  B &  $-$1.637 &  $-$3.414 &  0.030 &  0.021 &  0.088 &  0.997 \\
 0.004 &  0.25 &  z, i$-$z &   F &  1.7 &  B &  $-$1.391 &  $-$3.182 &  0.058 &  0.040 &  0.168 &  0.987 \\
 0.004 &  0.25 &  y, g$-$y &   F &  1.7 &  B &  $-$1.505 &  $-$3.284 &  0.045 &  0.032 &  0.131 &  0.992 \\
 0.004 &  0.25 &  g, u$-$g &   F &  1.9 &  B &  $-$4.208 &  $-$3.961 &  0.035 &  0.025 &  0.100 &  0.997 \\
 0.004 &  0.25 &  r, g$-$r  &   F &  1.9 &  B &  $-$1.806 &  $-$3.564 &  0.015 &  0.011 &  0.043 &  0.999 \\
 0.004 &  0.25 &  i, g$-$i &   F &  1.9 &  B &  $-$1.636 &  $-$3.434 &  0.027 &  0.019 &  0.076 &  0.998 \\
 0.004 &  0.25 &  z, i$-$z &   F &  1.9 &  B &  $-$1.388 &  $-$3.211 &  0.050 &  0.036 &  0.142 &  0.991 \\
 0.004 &  0.25 &  y, g$-$y &   F &  1.9 &  B &  $-$1.505 &  $-$3.308 &  0.039 &  0.028 &  0.112 &  0.995 \\
 0.004 &  0.25 &  g, u$-$g &   F &  1.5 &  C &  $-$3.977 &  $-$4.003 &  0.045 &  0.031 &  0.120 &  0.995 \\
 0.004 &  0.25 &  r, g$-$r  &   F &  1.5 &  C &  $-$1.642 &  $-$3.550 &  0.026 &  0.018 &  0.069 &  0.998 \\
 0.004 &  0.25 &  i, g$-$i &   F &  1.5 &  C &  $-$1.453 &  $-$3.429 &  0.040 &  0.028 &  0.107 &  0.995 \\
 0.004 &  0.25 &  z, i$-$z &   F &  1.5 &  C &  $-$1.163 &  $-$3.230 &  0.074 &  0.051 &  0.196 &  0.980 \\
 0.004 &  0.25 &  y, g$-$y &   F &  1.5 &  C &  $-$1.298 &  $-$3.318 &  0.059 &  0.040 &  0.155 &  0.988 \\
 0.004 &  0.25 &  g, u$-$g &   F &  1.7 &  C &  $-$4.036 &  $-$3.969 &  0.044 &  0.030 &  0.118 &  0.996 \\
 0.004 &  0.25 &  r, g$-$r  &   F &  1.7 &  C &  $-$1.658 &  $-$3.542 &  0.019 &  0.013 &  0.052 &  0.999 \\
 0.004 &  0.25 &  i, g$-$i &   F &  1.7 &  C &  $-$1.486 &  $-$3.412 &  0.033 &  0.023 &  0.090 &  0.997 \\
 0.004 &  0.25 &  z, i$-$z &   F &  1.7 &  C &  $-$1.235 &  $-$3.190 &  0.063 &  0.043 &  0.172 &  0.986 \\
 0.004 &  0.25 &  y, g$-$y &   F &  1.7 &  C &  $-$1.353 &  $-$3.287 &  0.050 &  0.034 &  0.134 &  0.992 \\
 0.004 &  0.25 &  g, u$-$g &   F &  1.9 &  C &  $-$4.083 &  $-$3.938 &  0.040 &  0.027 &  0.104 &  0.997 \\
 0.004 &  0.25 &  r, g$-$r  &   F &  1.9 &  C &  $-$1.663 &  $-$3.547 &  0.016 &  0.011 &  0.043 &  0.999 \\
 0.004 &  0.25 &  i, g$-$i &   F &  1.9 &  C &  $-$1.494 &  $-$3.415 &  0.030 &  0.020 &  0.077 &  0.998 \\
 0.004 &  0.25 &  z, i$-$z &   F &  1.9 &  C &  $-$1.256 &  $-$3.186 &  0.057 &  0.038 &  0.147 &  0.990 \\
 0.004 &  0.25 &  y, g$-$y &   F &  1.9 &  C &  $-$1.370 &  $-$3.286 &  0.044 &  0.030 &  0.115 &  0.994 \\
  0.004 &  0.25 &  g, u$-$g &  FO &  1.5 &  A &  $-$5.347 &  $-$3.294 &  0.063 &  0.134 &  0.257 &  0.953 \\
 0.004 &  0.25 &  r, g$-$r  &  FO &  1.5 &  A &  $-$2.478 &  $-$3.480 &  0.042 &  0.090 &  0.172 &  0.980 \\
 0.004 &  0.25 &  i, g$-$i &  FO &  1.5 &  A &  $-$2.249 &  $-$3.428 &  0.042 &  0.090 &  0.172 &  0.980 \\
 0.004 &  0.25 &  z, i$-$z &  FO &  1.5 &  A &  $-$1.929 &  $-$3.314 &  0.046 &  0.097 &  0.186 &  0.975 \\
 0.004 &  0.25 &  y, g$-$y &  FO &  1.5 &  A &  $-$2.095 &  $-$3.338 &  0.044 &  0.095 &  0.181 &  0.976 \\
 0.004 &  0.25 &  g, u$-$g &  FO &  1.7 &  A &  $-$5.301 &  $-$3.539 &  0.044 &  0.117 &  0.145 &  0.983 \\
 0.004 &  0.25 &  r, g$-$r  &  FO &  1.7 &  A &  $-$2.416 &  $-$3.728 &  0.011 &  0.029 &  0.036 &  0.999 \\
 0.004 &  0.25 &  i, g$-$i &  FO &  1.7 &  A &  $-$2.188 &  $-$3.678 &  0.012 &  0.032 &  0.040 &  0.999 \\
 0.004 &  0.25 &  z, i$-$z &  FO &  1.7 &  A &  $-$1.872 &  $-$3.570 &  0.021 &  0.056 &  0.070 &  0.996 \\
 0.004 &  0.25 &  y, g$-$y &  FO &  1.7 &  A &  $-$2.037 &  $-$3.589 &  0.018 &  0.049 &  0.060 &  0.997 \\
 0.004 &  0.25 &  g, u$-$g &  FO &  1.9 &  A &  $-$5.289 &  $-$3.574 &  0.038 &  0.108 &  0.102 &  0.991 \\
 0.004 &  0.25 &  r, g$-$r  &  FO &  1.9 &  A &  $-$2.414 &  $-$3.758 &  0.007 &  0.020 &  0.018 &  1.000 \\
 0.004 &  0.25 &  i, g$-$i &  FO &  1.9 &  A &  $-$2.185 &  $-$3.712 &  0.009 &  0.026 &  0.024 &  1.000 \\
 0.004 &  0.25 &  z, i$-$z &  FO &  1.9 &  A &  $-$1.863 &  $-$3.611 &  0.019 &  0.054 &  0.051 &  0.998 \\
 0.004 &  0.25 &  y, g$-$y &  FO &  1.9 &  A &  $-$2.030 &  $-$3.628 &  0.016 &  0.045 &  0.042 &  0.998 \\
 0.004 &  0.25 &  g, u$-$g &  FO &  1.5 &  B &  $-$5.159 &  $-$3.351 &  0.095 &  0.204 &  0.197 &  0.941 \\
 0.004 &  0.25 &  r, g$-$r  &  FO &  1.5 &  B &  $-$2.274 &  $-$3.604 &  0.062 &  0.133 &  0.128 &  0.977 \\
 0.004 &  0.25 &  i, g$-$i &  FO &  1.5 &  B &  $-$2.051 &  $-$3.540 &  0.061 &  0.133 &  0.128 &  0.977 \\
 0.004 &  0.25 &  z, i$-$z &  FO &  1.5 &  B &  $-$1.748 &  $-$3.386 &  0.068 &  0.147 &  0.141 &  0.969 \\
 0.004 &  0.25 &  y, g$-$y &  FO &  1.5 &  B &  $-$1.906 &  $-$3.426 &  0.065 &  0.140 &  0.135 &  0.972 \\
 0.004 &  0.25 &  g, u$-$g &  FO &  1.7 &  B &  $-$5.222 &  $-$3.213 &  0.099 &  0.227 &  0.188 &  0.939 \\
 0.004 &  0.25 &  r, g$-$r  &  FO &  1.7 &  B &  $-$2.303 &  $-$3.537 &  0.073 &  0.168 &  0.139 &  0.972 \\
 0.004 &  0.25 &  i, g$-$i &  FO &  1.7 &  B &  $-$2.079 &  $-$3.477 &  0.072 &  0.166 &  0.137 &  0.971 \\
 0.004 &  0.25 &  z, i$-$z &  FO &  1.7 &  B &  $-$1.778 &  $-$3.321 &  0.075 &  0.173 &  0.143 &  0.966 \\
 0.004 &  0.25 &  y, g$-$y &  FO &  1.7 &  B &  $-$1.937 &  $-$3.358 &  0.074 &  0.169 &  0.140 &  0.968 \\
 0.004 &  0.25 &  g, u$-$g &  FO &  1.9 &  B &  $-$5.152 &  $-$3.574 &  0.070 &  0.216 &  0.097 &  0.986 \\
 0.004 &  0.25 &  r, g$-$r  &  FO &  1.9 &  B &  $-$2.287 &  $-$3.720 &  0.010 &  0.032 &  0.014 &  1.000 \\
 0.004 &  0.25 &  i, g$-$i &  FO &  1.9 &  B &  $-$2.057 &  $-$3.675 &  0.014 &  0.043 &  0.019 &  0.999 \\
 0.004 &  0.25 &  z, i$-$z &  FO &  1.9 &  B &  $-$1.737 &  $-$3.576 &  0.030 &  0.094 &  0.042 &  0.997 \\
 0.004 &  0.25 &  y, g$-$y &  FO &  1.9 &  B &  $-$1.902 &  $-$3.595 &  0.026 &  0.079 &  0.035 &  0.998 \\
 0.004 &  0.25 &  g, u$-$g &  FO &  1.5 &  C &  $-$5.122 &  $-$3.455 &  0.086 &  0.183 &  0.063 &  0.989 \\
 0.004 &  0.25 &  r, g$-$r  &  FO &  1.5 &  C &  $-$2.171 &  $-$3.641 &  0.013 &  0.028 &  0.010 &  1.000 \\
 0.004 &  0.25 &  i, g$-$i &  FO &  1.5 &  C &  $-$1.949 &  $-$3.592 &  0.018 &  0.040 &  0.014 &  1.000 \\
 0.004 &  0.25 &  z, i$-$z &  FO &  1.5 &  C &  $-$1.669 &  $-$3.467 &  0.044 &  0.095 &  0.033 &  0.997 \\
 0.004 &  0.25 &  y, g$-$y &  FO &  1.5 &  C &  $-$1.819 &  $-$3.495 &  0.035 &  0.076 &  0.026 &  0.998 \\
\hline\hline
\end{tabular}}
\end{table*}

\begin{table*}
\caption{\label{pw_lmc}The coefficients of the PW ($W$ = $a$ + $b$ ($\log P$)) relations for $Z$=$0.02$, $Y$= $0.28$. The last two columns represent the root-mean-square deviation ($\sigma$) and the R-squared ($R^2$) coefficients.}
\centering
\scalebox{0.75}{
\begin{tabular}{cccccccccccc}
\hline\hline
Z&Y&band&mode&$\alpha_{ml}$&ML&a&b&$\sigma_{a}$&$\sigma_{b}$&$\sigma$&$R^2$\\
\hline\hline
 0.008 &  0.25 &  g, u$-$g &   F &  1.5 &  A &  $-$4.170 &  $-$4.611 &  0.076 &  0.061 &  0.210 &  0.987 \\
 0.008 &  0.25 &  r, g$-$r &   F &  1.5 &  A &  $-$2.020 &  $-$3.659 &  0.015 &  0.012 &  0.042 &  0.999 \\
 0.008 &  0.25 &  i, g$-$i &   F &  1.5 &  A &  $-$1.838 &  $-$3.446 &  0.033 &  0.026 &  0.090 &  0.995 \\
 0.008 &  0.25 &  z, i$-$z &   F &  1.5 &  A &  $-$1.529 &  $-$3.160 &  0.064 &  0.052 &  0.177 &  0.980 \\
 0.008 &  0.25 &  y, g$-$y &   F &  1.5 &  A &  $-$1.675 &  $-$3.282 &  0.051 &  0.041 &  0.140 &  0.988 \\
 0.008 &  0.25 &  g, u$-$g &   F &  1.7 &  A &  $-$4.297 &  $-$4.475 &  0.061 &  0.051 &  0.137 &  0.993 \\
 0.008 &  0.25 &  r, g$-$r &   F &  1.7 &  A &  $-$2.020 &  $-$3.684 &  0.012 &  0.010 &  0.027 &  1.000 \\
 0.008 &  0.25 &  i, g$-$i &   F &  1.7 &  A &  $-$1.836 &  $-$3.483 &  0.027 &  0.022 &  0.060 &  0.998 \\
 0.008 &  0.25 &  z, i$-$z &   F &  1.7 &  A &  $-$1.538 &  $-$3.201 &  0.055 &  0.046 &  0.124 &  0.989 \\
 0.008 &  0.25 &  y, g$-$y &   F &  1.7 &  A &  $-$1.680 &  $-$3.321 &  0.043 &  0.036 &  0.096 &  0.994 \\
 0.008 &  0.25 &  g, u$-$g &   F &  1.9 &  A &  $-$4.495 &  $-$4.301 &  0.047 &  0.043 &  0.094 &  0.996 \\
 0.008 &  0.25 &  r, g$-$r &   F &  1.9 &  A &  $-$2.001 &  $-$3.729 &  0.011 &  0.010 &  0.022 &  1.000 \\
 0.008 &  0.25 &  i, g$-$i &   F &  1.9 &  A &  $-$1.804 &  $-$3.551 &  0.020 &  0.018 &  0.040 &  0.999 \\
 0.008 &  0.25 &  z, i$-$z &   F &  1.9 &  A &  $-$1.507 &  $-$3.282 &  0.042 &  0.038 &  0.084 &  0.995 \\
 0.008 &  0.25 &  y, g$-$y &   F &  1.9 &  A &  $-$1.651 &  $-$3.395 &  0.032 &  0.029 &  0.065 &  0.997 \\
 0.008 &  0.25 &  g, u$-$g &   F &  1.5 &  B &  $-$3.927 &  $-$4.716 &  0.097 &  0.068 &  0.301 &  0.980 \\
 0.008 &  0.25 &  r, g$-$r &   F &  1.5 &  B &  $-$1.905 &  $-$3.620 &  0.018 &  0.012 &  0.055 &  0.999 \\
 0.008 &  0.25 &  i, g$-$i &   F &  1.5 &  B &  $-$1.738 &  $-$3.395 &  0.035 &  0.024 &  0.108 &  0.995 \\
 0.008 &  0.25 &  z, i$-$z &   F &  1.5 &  B &  $-$1.457 &  $-$3.089 &  0.064 &  0.045 &  0.200 &  0.980 \\
 0.008 &  0.25 &  y, g$-$y &   F &  1.5 &  B &  $-$1.592 &  $-$3.219 &  0.052 &  0.036 &  0.160 &  0.988 \\
 0.008 &  0.25 &  g, u$-$g &   F &  1.7 &  B &  $-$4.046 &  $-$4.616 &  0.078 &  0.056 &  0.237 &  0.989 \\
 0.008 &  0.25 &  r, g$-$r &   F &  1.7 &  B &  $-$1.918 &  $-$3.621 &  0.016 &  0.011 &  0.048 &  0.999 \\
 0.008 &  0.25 &  i, g$-$i &   F &  1.7 &  B &  $-$1.750 &  $-$3.400 &  0.030 &  0.022 &  0.092 &  0.997 \\
 0.008 &  0.25 &  z, i$-$z &   F &  1.7 &  B &  $-$1.480 &  $-$3.091 &  0.054 &  0.039 &  0.165 &  0.988 \\
 0.008 &  0.25 &  y, g$-$y &   F &  1.7 &  B &  $-$1.612 &  $-$3.222 &  0.044 &  0.032 &  0.134 &  0.993 \\
 0.008 &  0.25 &  g, u$-$g &   F &  1.9 &  B &  $-$4.277 &  $-$4.413 &  0.057 &  0.043 &  0.150 &  0.995 \\
 0.008 &  0.25 &  r, g$-$r &   F &  1.9 &  B &  $-$1.900 &  $-$3.660 &  0.014 &  0.010 &  0.036 &  1.000 \\
 0.008 &  0.25 &  i, g$-$i &   F &  1.9 &  B &  $-$1.715 &  $-$3.461 &  0.024 &  0.019 &  0.064 &  0.999 \\
 0.008 &  0.25 &  z, i$-$z &   F &  1.9 &  B &  $-$1.436 &  $-$3.166 &  0.043 &  0.033 &  0.114 &  0.994 \\
 0.008 &  0.25 &  y, g$-$y &   F &  1.9 &  B &  $-$1.574 &  $-$3.290 &  0.035 &  0.027 &  0.092 &  0.997 \\
 0.008 &  0.25 &  g, u$-$g &   F &  1.5 &  C &  $-$3.965 &  $-$4.520 &  0.090 &  0.060 &  0.258 &  0.984 \\
 0.008 &  0.25 &  r, g$-$r &   F &  1.5 &  C &  $-$1.733 &  $-$3.632 &  0.016 &  0.011 &  0.046 &  0.999 \\
 0.008 &  0.25 &  i, g$-$i &   F &  1.5 &  C &  $-$1.536 &  $-$3.437 &  0.033 &  0.022 &  0.094 &  0.996 \\
 0.008 &  0.25 &  z, i$-$z &   F &  1.5 &  C &  $-$1.227 &  $-$3.163 &  0.063 &  0.043 &  0.181 &  0.984 \\
 0.008 &  0.25 &  y, g$-$y &   F &  1.5 &  C &  $-$1.374 &  $-$3.280 &  0.050 &  0.034 &  0.144 &  0.991 \\
 0.008 &  0.25 &  g, u$-$g &   F &  1.7 &  C &  $-$4.059 &  $-$4.447 &  0.075 &  0.051 &  0.207 &  0.990 \\
 0.008 &  0.25 &  r, g$-$r &   F &  1.7 &  C &  $-$1.767 &  $-$3.619 &  0.014 &  0.010 &  0.039 &  0.999 \\
 0.008 &  0.25 &  i, g$-$i &   F &  1.7 &  C &  $-$1.580 &  $-$3.422 &  0.030 &  0.020 &  0.084 &  0.997 \\
 0.008 &  0.25 &  z, i$-$z &   F &  1.7 &  C &  $-$1.299 &  $-$3.135 &  0.057 &  0.039 &  0.158 &  0.988 \\
 0.008 &  0.25 &  y, g$-$y &   F &  1.7 &  C &  $-$1.436 &  $-$3.257 &  0.046 &  0.031 &  0.126 &  0.993 \\
 0.008 &  0.25 &  g, u$-$g &   F &  1.9 &  C &  $-$4.098 &  $-$4.441 &  0.067 &  0.046 &  0.176 &  0.994 \\
 0.008 &  0.25 &  r, g$-$r &   F &  1.9 &  C &  $-$1.787 &  $-$3.612 &  0.015 &  0.010 &  0.039 &  1.000 \\
 0.008 &  0.25 &  i, g$-$i &   F &  1.9 &  C &  $-$1.613 &  $-$3.405 &  0.026 &  0.018 &  0.068 &  0.998 \\
 0.008 &  0.25 &  z, i$-$z &   F &  1.9 &  C &  $-$1.355 &  $-$3.097 &  0.045 &  0.031 &  0.118 &  0.994 \\
 0.008 &  0.25 &  y, g$-$y &   F &  1.9 &  C &  $-$1.483 &  $-$3.226 &  0.037 &  0.025 &  0.097 &  0.996 \\
 0.008 &  0.25 &  g, u$-$g &  FO &  1.5 &  A &  $-$5.629 &  $-$3.525 &  0.045 &  0.101 &  0.217 &  0.974 \\
 0.008 &  0.25 &  r, g$-$r &  FO &  1.5 &  A &  $-$2.637 &  $-$3.578 &  0.026 &  0.058 &  0.125 &  0.991 \\
 0.008 &  0.25 &  i, g$-$i &  FO &  1.5 &  A &  $-$2.368 &  $-$3.505 &  0.025 &  0.058 &  0.124 &  0.991 \\
 0.008 &  0.25 &  z, i$-$z &  FO &  1.5 &  A &  $-$1.995 &  $-$3.374 &  0.029 &  0.067 &  0.143 &  0.987 \\
 0.008 &  0.25 &  y, g$-$y &  FO &  1.5 &  A &  $-$2.186 &  $-$3.406 &  0.028 &  0.064 &  0.138 &  0.988 \\
 0.008 &  0.25 &  g, u$-$g &  FO &  1.7 &  A &  $-$5.552 &  $-$3.672 &  0.039 &  0.095 &  0.154 &  0.985 \\
 0.008 &  0.25 &  r, g$-$r &  FO &  1.7 &  A &  $-$2.569 &  $-$3.731 &  0.021 &  0.050 &  0.081 &  0.996 \\
 0.008 &  0.25 &  i, g$-$i &  FO &  1.7 &  A &  $-$2.303 &  $-$3.656 &  0.021 &  0.051 &  0.082 &  0.996 \\
 0.008 &  0.25 &  z, i$-$z &  FO &  1.7 &  A &  $-$1.933 &  $-$3.518 &  0.026 &  0.063 &  0.102 &  0.993 \\
 0.008 &  0.25 &  y, g$-$y &  FO &  1.7 &  A &  $-$2.122 &  $-$3.553 &  0.024 &  0.058 &  0.095 &  0.994 \\
 0.008 &  0.25 &  g, u$-$g &  FO &  1.9 &  A &  $-$5.557 &  $-$3.735 &  0.034 &  0.161 &  0.099 &  0.984 \\
 0.008 &  0.25 &  r, g$-$r &  FO &  1.9 &  A &  $-$2.556 &  $-$3.841 &  0.007 &  0.032 &  0.019 &  0.999 \\
 0.008 &  0.25 &  i, g$-$i &  FO &  1.9 &  A &  $-$2.290 &  $-$3.784 &  0.008 &  0.038 &  0.023 &  0.999 \\
 0.008 &  0.25 &  z, i$-$z &  FO &  1.9 &  A &  $-$1.924 &  $-$3.690 &  0.016 &  0.078 &  0.048 &  0.996 \\
 0.008 &  0.25 &  y, g$-$y &  FO &  1.9 &  A &  $-$2.113 &  $-$3.700 &  0.014 &  0.067 &  0.041 &  0.997 \\
 0.008 &  0.25 &  g, u$-$g &  FO &  1.5 &  B &  $-$5.424 &  $-$3.680 &  0.051 &  0.136 &  0.144 &  0.977 \\
 0.008 &  0.25 &  r, g$-$r &  FO &  1.5 &  B &  $-$2.408 &  $-$3.757 &  0.012 &  0.032 &  0.034 &  0.999 \\
 0.008 &  0.25 &  i, g$-$i &  FO &  1.5 &  B &  $-$2.148 &  $-$3.669 &  0.015 &  0.039 &  0.041 &  0.998 \\
 0.008 &  0.25 &  z, i$-$z &  FO &  1.5 &  B &  $-$1.799 &  $-$3.501 &  0.029 &  0.078 &  0.082 &  0.992 \\
 0.008 &  0.25 &  y, g$-$y &  FO &  1.5 &  B &  $-$1.980 &  $-$3.547 &  0.024 &  0.065 &  0.068 &  0.994 \\
 0.008 &  0.25 &  g, u$-$g &  FO &  1.7 &  B &  $-$5.406 &  $-$3.697 &  0.048 &  0.133 &  0.113 &  0.986 \\
 0.008 &  0.25 &  r, g$-$r &  FO &  1.7 &  B &  $-$2.404 &  $-$3.773 &  0.008 &  0.021 &  0.018 &  1.000 \\
 0.008 &  0.25 &  i, g$-$i &  FO &  1.7 &  B &  $-$2.143 &  $-$3.689 &  0.011 &  0.031 &  0.027 &  0.999 \\
 0.008 &  0.25 &  z, i$-$z &  FO &  1.7 &  B &  $-$1.789 &  $-$3.530 &  0.026 &  0.073 &  0.062 &  0.995 \\
 0.008 &  0.25 &  y, g$-$y &  FO &  1.7 &  B &  $-$1.971 &  $-$3.573 &  0.021 &  0.059 &  0.050 &  0.997 \\
 0.008 &  0.25 &  g, u$-$g &  FO &  1.9 &  B &  $-$5.430 &  $-$3.572 &  0.056 &  0.325 &  0.108 &  0.960 \\
 0.008 &  0.25 &  r, g$-$r &  FO &  1.9 &  B &  $-$2.420 &  $-$3.747 &  0.008 &  0.045 &  0.015 &  0.999 \\
 0.008 &  0.25 &  i, g$-$i &  FO &  1.9 &  B &  $-$2.156 &  $-$3.679 &  0.011 &  0.064 &  0.021 &  0.998 \\
 0.008 &  0.25 &  z, i$-$z &  FO &  1.9 &  B &  $-$1.795 &  $-$3.541 &  0.026 &  0.151 &  0.050 &  0.991 \\
 0.008 &  0.25 &  y, g$-$y &  FO &  1.9 &  B &  $-$1.982 &  $-$3.566 &  0.022 &  0.126 &  0.042 &  0.994 \\
 0.008 &  0.25 &  g, u$-$g &  FO &  1.5 &  C &  $-$5.402 &  $-$3.481 &  0.078 &  0.189 &  0.102 &  0.974 \\
 0.008 &  0.25 &  r, g$-$r &  FO &  1.5 &  C &  $-$2.312 &  $-$3.648 &  0.018 &  0.044 &  0.024 &  0.999 \\
 0.008 &  0.25 &  i, g$-$i &  FO &  1.5 &  C &  $-$2.058 &  $-$3.556 &  0.023 &  0.057 &  0.031 &  0.998 \\
 0.008 &  0.25 &  z, i$-$z &  FO &  1.5 &  C &  $-$1.739 &  $-$3.351 &  0.048 &  0.117 &  0.063 &  0.989 \\
 0.008 &  0.25 &  y, g$-$y &  FO &  1.5 &  C &  $-$1.909 &  $-$3.410 &  0.040 &  0.096 &  0.052 &  0.993 \\
\hline\hline
\end{tabular}}
\end{table*}

\begin{table*}
\caption{\label{pw_mw}The coefficients of the PW ($W$ = $a$ + $b$ ($\log P$)) relations for $Z$=$0.03$, $Y$= $0.28$. The last two columns represent the root-mean-square deviation ($\sigma$) and the R-squared ($R^2$) coefficients.}
\centering
\scalebox{0.75}{
\begin{tabular}{cccccccccccc}
\hline\hline
Z&Y&band&mode&$\alpha_{ml}$&ML&a&b&$\sigma_{a}$&$\sigma_{b}$&$\sigma$&$R^2$\\
\hline\hline
 0.02 &  0.28 &  g, u$-$g &   F &  1.5 &  A &  $-$4.511 &  $-$5.009 &  0.103 &  0.085 &  0.324 &  0.981 \\
 0.02 &  0.28 &  r, g$-$r &   F &  1.5 &  A &  $-$2.167 &  $-$3.708 &  0.013 &  0.011 &  0.042 &  0.999 \\
 0.02 &  0.28 &  i, g$-$i &   F &  1.5 &  A &  $-$1.894 &  $-$3.434 &  0.026 &  0.022 &  0.083 &  0.997 \\
 0.02 &  0.28 &  z, i$-$z &   F &  1.5 &  A &  $-$1.435 &  $-$3.127 &  0.044 &  0.036 &  0.138 &  0.991 \\
 0.02 &  0.28 &  y, g$-$y &   F &  1.5 &  A &  $-$1.654 &  $-$3.244 &  0.037 &  0.031 &  0.116 &  0.994 \\
 0.02 &  0.28 &  g, u$-$g &   F &  1.7 &  A &  $-$4.681 &  $-$4.868 &  0.087 &  0.078 &  0.217 &  0.992 \\
 0.02 &  0.28 &  r, g$-$r &   F &  1.7 &  A &  $-$2.146 &  $-$3.768 &  0.013 &  0.011 &  0.031 &  1.000 \\
 0.02 &  0.28 &  i, g$-$i &   F &  1.7 &  A &  $-$1.872 &  $-$3.501 &  0.022 &  0.020 &  0.056 &  0.999 \\
 0.02 &  0.28 &  z, i$-$z &   F &  1.7 &  A &  $-$1.433 &  $-$3.177 &  0.034 &  0.030 &  0.084 &  0.997 \\
 0.02 &  0.28 &  y, g$-$y &   F &  1.7 &  A &  $-$1.645 &  $-$3.300 &  0.029 &  0.026 &  0.073 &  0.998 \\
 0.02 &  0.28 &  g, u$-$g &   F &  1.9 &  A &  $-$5.110 &  $-$3.920 &  0.019 &  0.046 &  0.017 &  1.000 \\
 0.02 &  0.28 &  r, g$-$r &   F &  1.9 &  A &  $-$2.144 &  $-$3.763 &  0.004 &  0.011 &  0.004 &  1.000 \\
 0.02 &  0.28 &  i, g$-$i &   F &  1.9 &  A &  $-$1.803 &  $-$3.664 &  0.009 &  0.023 &  0.009 &  1.000 \\
 0.02 &  0.28 &  z, i$-$z &   F &  1.9 &  A &  $-$1.303 &  $-$3.515 &  0.021 &  0.053 &  0.020 &  0.999 \\
 0.02 &  0.28 &  y, g$-$y &   F &  1.9 &  A &  $-$1.543 &  $-$3.562 &  0.017 &  0.043 &  0.016 &  1.000 \\
 0.02 &  0.28 &  g, u$-$g &   F &  1.5 &  B &  $-$4.129 &  $-$5.241 &  0.121 &  0.086 &  0.396 &  0.978 \\
 0.02 &  0.28 &  r, g$-$r &   F &  1.5 &  B &  $-$2.067 &  $-$3.644 &  0.018 &  0.013 &  0.059 &  0.999 \\
 0.02 &  0.28 &  i, g$-$i &   F &  1.5 &  B &  $-$1.816 &  $-$3.352 &  0.031 &  0.022 &  0.102 &  0.996 \\
 0.02 &  0.28 &  z, i$-$z &   F &  1.5 &  B &  $-$1.381 &  $-$3.029 &  0.048 &  0.034 &  0.158 &  0.990 \\
 0.02 &  0.28 &  y, g$-$y &   F &  1.5 &  B &  $-$1.589 &  $-$3.154 &  0.041 &  0.029 &  0.135 &  0.993 \\
 0.02 &  0.28 &  g, u$-$g &   F &  1.7 &  B &  $-$4.237 &  $-$5.189 &  0.105 &  0.074 &  0.294 &  0.990 \\
 0.02 &  0.28 &  r, g$-$r &   F &  1.7 &  B &  $-$2.078 &  $-$3.652 &  0.025 &  0.017 &  0.069 &  0.999 \\
 0.02 &  0.28 &  i, g$-$i &   F &  1.7 &  B &  $-$1.824 &  $-$3.362 &  0.033 &  0.023 &  0.093 &  0.998 \\
 0.02 &  0.28 &  z, i$-$z &   F &  1.7 &  B &  $-$1.385 &  $-$3.041 &  0.041 &  0.029 &  0.115 &  0.995 \\
 0.02 &  0.28 &  y, g$-$y &   F &  1.7 &  B &  $-$1.597 &  $-$3.163 &  0.038 &  0.027 &  0.106 &  0.996 \\
 0.02 &  0.28 &  g, u$-$g &   F &  1.9 &  B &  $-$4.961 &  $-$3.930 &  0.035 &  0.086 &  0.024 &  0.999 \\
 0.02 &  0.28 &  r, g$-$r &   F &  1.9 &  B &  $-$1.994 &  $-$3.753 &  0.008 &  0.020 &  0.006 &  1.000 \\
 0.02 &  0.28 &  i, g$-$i &   F &  1.9 &  B &  $-$1.658 &  $-$3.647 &  0.019 &  0.047 &  0.013 &  1.000 \\
 0.02 &  0.28 &  z, i$-$z &   F &  1.9 &  B &  $-$1.165 &  $-$3.490 &  0.044 &  0.109 &  0.031 &  0.997 \\
 0.02 &  0.28 &  y, g$-$y &   F &  1.9 &  B &  $-$1.402 &  $-$3.539 &  0.036 &  0.088 &  0.025 &  0.998 \\
 0.02 &  0.28 &  g, u$-$g &   F &  1.5 &  C &  $-$4.090 &  $-$5.103 &  0.147 &  0.101 &  0.427 &  0.970 \\
 0.02 &  0.28 &  r, g$-$r &   F &  1.5 &  C &  $-$1.922 &  $-$3.640 &  0.020 &  0.014 &  0.057 &  0.999 \\
 0.02 &  0.28 &  i, g$-$i &   F &  1.5 &  C &  $-$1.661 &  $-$3.363 &  0.037 &  0.026 &  0.108 &  0.996 \\
 0.02 &  0.28 &  z, i$-$z &   F &  1.5 &  C &  $-$1.223 &  $-$3.051 &  0.060 &  0.041 &  0.172 &  0.986 \\
 0.02 &  0.28 &  y, g$-$y &   F &  1.5 &  C &  $-$1.431 &  $-$3.172 &  0.051 &  0.035 &  0.147 &  0.991 \\
 0.02 &  0.28 &  g, u$-$g &   F &  1.7 &  C &  $-$4.211 &  $-$5.008 &  0.103 &  0.074 &  0.263 &  0.989 \\
 0.02 &  0.28 &  r, g$-$r &   F &  1.7 &  C &  $-$1.918 &  $-$3.665 &  0.020 &  0.015 &  0.052 &  0.999 \\
 0.02 &  0.28 &  i, g$-$i &   F &  1.7 &  C &  $-$1.662 &  $-$3.386 &  0.029 &  0.021 &  0.074 &  0.998 \\
 0.02 &  0.28 &  z, i$-$z &   F &  1.7 &  C &  $-$1.233 &  $-$3.067 &  0.042 &  0.030 &  0.106 &  0.995 \\
 0.02 &  0.28 &  y, g$-$y &   F &  1.7 &  C &  $-$1.441 &  $-$3.188 &  0.037 &  0.026 &  0.093 &  0.997 \\
 0.02 &  0.28 &  g, u$-$g &   F &  1.9 &  C &  $-$4.702 &  $-$4.292 &  0.076 &  0.120 &  0.070 &  0.995 \\
 0.02 &  0.28 &  r, g$-$r &   F &  1.9 &  C &  $-$1.822 &  $-$3.821 &  0.013 &  0.020 &  0.012 &  1.000 \\
 0.02 &  0.28 &  i, g$-$i &   F &  1.9 &  C &  $-$1.534 &  $-$3.609 &  0.020 &  0.032 &  0.019 &  1.000 \\
 0.02 &  0.28 &  z, i$-$z &   F &  1.9 &  C &  $-$1.125 &  $-$3.281 &  0.042 &  0.067 &  0.039 &  0.998 \\
 0.02 &  0.28 &  y, g$-$y &   F &  1.9 &  C &  $-$1.326 &  $-$3.405 &  0.034 &  0.054 &  0.032 &  0.998 \\
 0.02 &  0.28 &  g, u$-$g &  FO &  1.5 &  A &  $-$5.819 &  $-$4.000 &  0.024 &  0.093 &  0.079 &  0.993 \\
 0.02 &  0.28 &  r, g$-$r &  FO &  1.5 &  A &  $-$2.705 &  $-$3.935 &  0.003 &  0.011 &  0.009 &  1.000 \\
 0.02 &  0.28 &  i, g$-$i &  FO &  1.5 &  A &  $-$2.367 &  $-$3.836 &  0.006 &  0.022 &  0.019 &  1.000 \\
 0.02 &  0.28 &  z, i$-$z &  FO &  1.5 &  A &  $-$1.897 &  $-$3.676 &  0.017 &  0.064 &  0.054 &  0.996 \\
 0.02 &  0.28 &  y, g$-$y &  FO &  1.5 &  A &  $-$2.128 &  $-$3.720 &  0.013 &  0.051 &  0.043 &  0.998 \\
 0.02 &  0.28 &  g, u$-$g &  FO &  1.7 &  A &  $-$5.832 &  $-$3.991 &  0.031 &  0.190 &  0.058 &  0.991 \\
 0.02 &  0.28 &  r, g$-$r &  FO &  1.7 &  A &  $-$2.728 &  $-$3.915 &  0.003 &  0.021 &  0.006 &  1.000 \\
 0.02 &  0.28 &  i, g$-$i &  FO &  1.7 &  A &  $-$2.390 &  $-$3.850 &  0.007 &  0.041 &  0.012 &  1.000 \\
 0.02 &  0.28 &  z, i$-$z &  FO &  1.7 &  A &  $-$1.920 &  $-$3.762 &  0.018 &  0.110 &  0.033 &  0.997 \\
 0.02 &  0.28 &  y, g$-$y &  FO &  1.7 &  A &  $-$2.152 &  $-$3.774 &  0.015 &  0.091 &  0.028 &  0.998 \\
 0.02 &  0.28 &  g, u$-$g &  FO &  1.5 &  B &  $-$5.702 &  $-$3.975 &  0.044 &  0.239 &  0.105 &  0.972 \\
 0.02 &  0.28 &  r, g$-$r &  FO &  1.5 &  B &  $-$2.567 &  $-$3.942 &  0.009 &  0.049 &  0.021 &  0.999 \\
 0.02 &  0.28 &  i, g$-$i &  FO &  1.5 &  B &  $-$2.233 &  $-$3.830 &  0.013 &  0.072 &  0.031 &  0.997 \\
 0.02 &  0.28 &  z, i$-$z &  FO &  1.5 &  B &  $-$1.778 &  $-$3.636 &  0.028 &  0.155 &  0.068 &  0.986 \\
 0.02 &  0.28 &  y, g$-$y &  FO &  1.5 &  B &  $-$2.004 &  $-$3.689 &  0.024 &  0.129 &  0.056 &  0.990 \\
 0.02 &  0.28 &  g, u$-$g &  FO &  1.7 &  B &  $-$5.597 &  $-$1.018 &  0.011 &  0.195 &  0.007 &  0.932 \\
 0.02 &  0.28 &  r, g$-$r &  FO &  1.7 &  B &  $-$2.571 &  $-$3.221 &  0.010 &  0.185 &  0.006 &  0.993 \\
 0.02 &  0.28 &  i, g$-$i &  FO &  1.7 &  B &  $-$2.226 &  $-$2.912 &  0.007 &  0.128 &  0.004 &  0.996 \\
 0.02 &  0.28 &  z, i$-$z &  FO &  1.7 &  B &  $-$1.728 &  $-$1.884 &  0.001 &  0.019 &  0.001 &  1.000 \\
 0.02 &  0.28 &  y, g$-$y &  FO &  1.7 &  B &  $-$1.966 &  $-$2.144 &  0.002 &  0.041 &  0.001 &  0.999 \\
 0.02 &  0.28 &  g, u$-$g &  FO &  1.5 &  C &  $-$5.680 &  $-$3.645 &  0.072 &  0.306 &  0.075 &  0.973 \\
 0.02 &  0.28 &  r, g$-$r &  FO &  1.5 &  C &  $-$2.480 &  $-$3.723 &  0.022 &  0.093 &  0.023 &  0.997 \\
 0.02 &  0.28 &  i, g$-$i &  FO &  1.5 &  C &  $-$2.146 &  $-$3.646 &  0.029 &  0.124 &  0.030 &  0.995 \\
 0.02 &  0.28 &  z, i$-$z &  FO &  1.5 &  C &  $-$1.706 &  $-$3.486 &  0.055 &  0.235 &  0.057 &  0.982 \\
 0.02 &  0.28 &  y, g$-$y &  FO &  1.5 &  C &  $-$1.927 &  $-$3.528 &  0.047 &  0.201 &  0.049 &  0.987 \\
\hline\hline
\end{tabular}}
\end{table*}

\begin{table*}
\caption{\label{pw_supersolar}The coefficients of the PW ($W$ = $a$ + $b$ ($\log P$)) relations for $Z$=$0.008$, $Y$= $0.25$. The last two columns represent the root-mean-square deviation ($\sigma$) and the R-squared ($R^2$) coefficients.}
\centering
\scalebox{0.75}{
\begin{tabular}{cccccccccccc}
\hline\hline
Z&Y&band&mode&$\alpha_{ml}$&ML&a&b&$\sigma_{a}$&$\sigma_{b}$&$\sigma$&$R^2$\\
\hline\hline
  0.03 &  0.28 &  g, u$-$g &   F &  1.5 &  A &  $-$4.693 &  $-$5.150 &  0.102 &  0.087 &  0.301 &  0.984 \\
 0.03 &  0.28 &  r, g$-$r &   F &  1.5 &  A &  $-$2.292 &  $-$3.714 &  0.015 &  0.012 &  0.043 &  0.999 \\
 0.03 &  0.28 &  i, g$-$i &   F &  1.5 &  A &  $-$1.984 &  $-$3.414 &  0.025 &  0.021 &  0.073 &  0.998 \\
 0.03 &  0.28 &  z, i$-$z &   F &  1.5 &  A &  $-$1.469 &  $-$3.092 &  0.038 &  0.032 &  0.111 &  0.994 \\
 0.03 &  0.28 &  y, g$-$y &   F &  1.5 &  A &  $-$1.711 &  $-$3.213 &  0.033 &  0.028 &  0.096 &  0.996 \\
 0.03 &  0.28 &  g, u$-$g &   F &  1.7 &  A &  $-$4.831 &  $-$5.090 &  0.070 &  0.068 &  0.152 &  0.996 \\
 0.03 &  0.28 &  r, g$-$r &   F &  1.7 &  A &  $-$2.279 &  $-$3.776 &  0.020 &  0.019 &  0.043 &  0.999 \\
 0.03 &  0.28 &  i, g$-$i &   F &  1.7 &  A &  $-$1.970 &  $-$3.475 &  0.025 &  0.024 &  0.054 &  0.999 \\
 0.03 &  0.28 &  z, i$-$z &   F &  1.7 &  A &  $-$1.469 &  $-$3.131 &  0.026 &  0.025 &  0.057 &  0.999 \\
 0.03 &  0.28 &  y, g$-$y &   F &  1.7 &  A &  $-$1.709 &  $-$3.258 &  0.026 &  0.025 &  0.056 &  0.999 \\
 0.03 &  0.28 &  g, u$-$g &   F &  1.5 &  B &  $-$4.083 &  $-$5.559 &  0.145 &  0.100 &  0.382 &  0.978 \\
 0.03 &  0.28 &  r, g$-$r &   F &  1.5 &  B &  $-$2.258 &  $-$3.601 &  0.023 &  0.016 &  0.059 &  0.999 \\
 0.03 &  0.28 &  i, g$-$i &   F &  1.5 &  B &  $-$1.968 &  $-$3.290 &  0.032 &  0.022 &  0.084 &  0.997 \\
 0.03 &  0.28 &  z, i$-$z &   F &  1.5 &  B &  $-$1.483 &  $-$2.945 &  0.050 &  0.035 &  0.133 &  0.991 \\
 0.03 &  0.28 &  y, g$-$y &   F &  1.5 &  B &  $-$1.708 &  $-$3.080 &  0.042 &  0.029 &  0.112 &  0.994 \\
 0.03 &  0.28 &  g, u$-$g &   F &  1.7 &  B &  $-$4.214 &  $-$5.488 &  0.119 &  0.080 &  0.221 &  0.993 \\
 0.03 &  0.28 &  r, g$-$r &   F &  1.7 &  B &  $-$2.263 &  $-$3.621 &  0.035 &  0.023 &  0.064 &  0.999 \\
 0.03 &  0.28 &  i, g$-$i &   F &  1.7 &  B &  $-$1.979 &  $-$3.304 &  0.039 &  0.026 &  0.073 &  0.998 \\
 0.03 &  0.28 &  z, i$-$z &   F &  1.7 &  B &  $-$1.464 &  $-$2.982 &  0.044 &  0.030 &  0.082 &  0.997 \\
 0.03 &  0.28 &  y, g$-$y &   F &  1.7 &  B &  $-$1.706 &  $-$3.103 &  0.041 &  0.028 &  0.077 &  0.997 \\
 0.03 &  0.28 &  g, u$-$g &   F &  1.5 &  C &  $-$3.941 &  $-$5.495 &  0.191 &  0.122 &  0.428 &  0.967 \\
 0.03 &  0.28 &  r, g$-$r &   F &  1.5 &  C &  $-$2.151 &  $-$3.569 &  0.023 &  0.015 &  0.052 &  0.999 \\
 0.03 &  0.28 &  i, g$-$i &   F &  1.5 &  C &  $-$1.836 &  $-$3.279 &  0.039 &  0.025 &  0.088 &  0.996 \\
 0.03 &  0.28 &  z, i$-$z &   F &  1.5 &  C &  $-$1.312 &  $-$2.968 &  0.067 &  0.043 &  0.151 &  0.986 \\
 0.03 &  0.28 &  y, g$-$y &   F &  1.5 &  C &  $-$1.553 &  $-$3.089 &  0.055 &  0.035 &  0.124 &  0.991 \\
 0.03 &  0.28 &  g, u$-$g &   F &  1.7 &  C &  $-$4.033 &  $-$5.457 &  0.154 &  0.094 &  0.263 &  0.988 \\
 0.03 &  0.28 &  r, g$-$r &   F &  1.7 &  C &  $-$2.203 &  $-$3.554 &  0.029 &  0.018 &  0.050 &  0.999 \\
 0.03 &  0.28 &  i, g$-$i &   F &  1.7 &  C &  $-$1.891 &  $-$3.259 &  0.039 &  0.024 &  0.066 &  0.998 \\
 0.03 &  0.28 &  z, i$-$z &   F &  1.7 &  C &  $-$1.321 &  $-$2.978 &  0.054 &  0.033 &  0.093 &  0.995 \\
 0.03 &  0.28 &  y, g$-$y &   F &  1.7 &  C &  $-$1.584 &  $-$3.084 &  0.047 &  0.029 &  0.080 &  0.996 \\
\hline\hline
\end{tabular}}
\end{table*}

\begin{table*}
\caption{\label{pwz_lsst}The coefficients of the PWZ ($W$ = $a$ + $b$ ($\log P$) + c (Fe/H)) relations for F and FO-mode CCs. The last two columns represent the root-mean-square deviation ($\sigma$) and the R-squared ($R^2$) coefficients.}
\centering
\scalebox{0.75}{
\begin{tabular}{cccccccccccc}
\hline\hline
band&mode&$\alpha_{ml}$&ML&a&b&c&$\sigma_{a}$&$\sigma_{b}$&$\sigma_{c}$&$\sigma$&$R^2$\\
\hline\hline
 r, g$-$r &   F &  1.5 &  A &  $-$2.235 &  $-$3.662 &  $-$0.599 &  0.009 &  0.008 &  0.010 &  0.055 &  0.999 \\
 i, g$-$i &   F &  1.5 &  A &  $-$1.920 &  $-$3.430 &  $-$0.226 &  0.015 &  0.012 &  0.016 &  0.089 &  0.996 \\
 z, i$-$z &   F &  1.5 &  A &  $-$1.438 &  $-$3.146 &   $-$0.181 &  0.028 &  0.022 &  0.029 &  0.161 &  0.986 \\
 y, g$-$y &   F &  1.5 &  A &  $-$1.658 &  $-$3.261 &   $-$0.021 &  0.022 &  0.018 &  0.023 &  0.130 &  0.992 \\
 r, g$-$r &   F &  1.7 &  A &  $-$2.244 &  $-$3.688 &  $-$0.618 &  0.011 &  0.009 &  0.012 &  0.052 &  0.999 \\
 i, g$-$i &   F &  1.7 &  A &  $-$1.914 &  $-$3.483 &  $-$0.259 &  0.014 &  0.012 &  0.016 &  0.069 &  0.998 \\
 z, i$-$z &   F &  1.7 &  A &  $-$1.425 &  $-$3.211 &   $-$0.160 &  0.027 &  0.023 &  0.031 &  0.128 &  0.992 \\
 y, g$-$y &   F &  1.7 &  A &  $-$1.650 &  $-$3.321 &  $-$0.011 &  0.021 &  0.018 &  0.024 &  0.102 &  0.995 \\
 r, g$-$r &   F &  1.9 &  A &  $-$2.259 &  $-$3.674 &  $-$0.570 &  0.016 &  0.012 &  0.026 &  0.046 &  0.999 \\
 i, g$-$i &   F &  1.9 &  A &  $-$1.918 &  $-$3.542 &  $-$0.320 &  0.020 &  0.015 &  0.032 &  0.057 &  0.998 \\
 z, i$-$z &   F &  1.9 &  A &  $-$1.433 &  $-$3.334 &  $-$0.104 &  0.037 &  0.027 &  0.059 &  0.105 &  0.994 \\
 y, g$-$y &   F &  1.9 &  A &  $-$1.660 &  $-$3.422 &  $-$0.014 &  0.029 &  0.021 &  0.046 &  0.083 &  0.997 \\
 r, g$-$r &   F &  1.5 &  B &  $-$2.138 &  $-$3.605 &  $-$0.591 &  0.011 &  0.008 &  0.010 &  0.065 &  0.999 \\
 i, g$-$i &   F &  1.5 &  B &  $-$1.827 &  $-$3.362 &  $-$0.181 &  0.018 &  0.012 &  0.017 &  0.105 &  0.995 \\
 z, i$-$z &   F &  1.5 &  B &  $-$1.364 &  $-$3.060 &   $-$0.253 &  0.031 &  0.021 &  0.029 &  0.183 &  0.984 \\
 y, g$-$y &   F &  1.5 &  B &  $-$1.574 &  $-$3.183 &   $-$0.085 &  0.026 &  0.017 &  0.024 &  0.149 &  0.990 \\
 r, g$-$r &   F &  1.7 &  B &  $-$2.156 &  $-$3.610 &  $-$0.607 &  0.012 &  0.008 &  0.012 &  0.061 &  0.999 \\
 i, g$-$i &   F &  1.7 &  B &  $-$1.826 &  $-$3.381 &  $-$0.188 &  0.018 &  0.012 &  0.018 &  0.092 &  0.997 \\
 z, i$-$z &   F &  1.7 &  B &  $-$1.350 &  $-$3.089 &   $-$0.259 &  0.030 &  0.020 &  0.030 &  0.153 &  0.990 \\
 y, g$-$y &   F &  1.7 &  B &  $-$1.566 &  $-$3.208 &   $-$0.084 &  0.025 &  0.016 &  0.025 &  0.127 &  0.994 \\
 r, g$-$r &   F &  1.9 &  B &  $-$2.162 &  $-$3.621 &  $-$0.601 &  0.017 &  0.010 &  0.028 &  0.055 &  0.999 \\
 i, g$-$i &   F &  1.9 &  B &  $-$1.817 &  $-$3.457 &  $-$0.292 &  0.023 &  0.014 &  0.037 &  0.074 &  0.998 \\
 z, i$-$z &   F &  1.9 &  B &  $-$1.341 &  $-$3.202 &   $-$0.192 &  0.042 &  0.024 &  0.066 &  0.131 &  0.993 \\
 y, g$-$y &   F &  1.9 &  B &  $-$1.563 &  $-$3.311 &  $-$0.078 &  0.033 &  0.019 &  0.053 &  0.104 &  0.996 \\
 r, g$-$r &   F &  1.5 &  C &  $-$1.988 &  $-$3.604 &  $-$0.588 &  0.012 &  0.008 &  0.010 &  0.061 &  0.999 \\
 i, g$-$i &   F &  1.5 &  C &  $-$1.647 &  $-$3.385 &  $-$0.164 &  0.020 &  0.013 &  0.017 &  0.105 &  0.995 \\
 z, i$-$z &   F &  1.5 &  C &  $-$1.154 &  $-$3.109 &   $-$0.285 &  0.035 &  0.023 &  0.030 &  0.184 &  0.983 \\
 y, g$-$y &   F &  1.5 &  C &  $-$1.377 &  $-$3.222 &   $-$0.011 &  0.029 &  0.019 &  0.025 &  0.150 &  0.989 \\
 r, g$-$r &   F &  1.7 &  C &  $-$2.020 &  $-$3.599 &  $-$0.617 &  0.012 &  0.007 &  0.010 &  0.055 &  0.999 \\
 i, g$-$i &   F &  1.7 &  C &  $-$1.672 &  $-$3.386 &  $-$0.184 &  0.019 &  0.012 &  0.017 &  0.088 &  0.997 \\
 z, i$-$z &   F &  1.7 &  C &  $-$1.184 &  $-$3.110 &   $-$0.273 &  0.032 &  0.020 &  0.029 &  0.151 &  0.989 \\
 y, g$-$y &   F &  1.7 &  C &  $-$1.404 &  $-$3.222 &   $-$0.097 &  0.026 &  0.017 &  0.024 &  0.124 &  0.993 \\
 r, g$-$r &   F &  1.9 &  C &  $-$2.025 &  $-$3.597 &  $-$0.605 &  0.016 &  0.009 &  0.024 &  0.052 &  0.999 \\
 i, g$-$i &   F &  1.9 &  C &  $-$1.687 &  $-$3.421 &  $-$0.278 &  0.022 &  0.013 &  0.034 &  0.074 &  0.998 \\
 z, i$-$z &   F &  1.9 &  C &  $-$1.227 &  $-$3.148 &   $-$0.123 &  0.040 &  0.023 &  0.061 &  0.133 &  0.993 \\
 y, g$-$y &   F &  1.9 &  C &  $-$1.441 &  $-$3.265 &  $-$0.052 &  0.032 &  0.019 &  0.048 &  0.106 &  0.996 \\
 r, g$-$r &  FO &  1.5 &  A &  $-$2.778 &  $-$3.570 &  $-$0.438 &  0.028 &  0.046 &  0.058 &  0.141 &  0.986 \\
 i, g$-$i &  FO &  1.5 &  A &  $-$2.438 &  $-$3.504 &  $-$0.270 &  0.028 &  0.046 &  0.057 &  0.140 &  0.986 \\
 z, i$-$z &  FO &  1.5 &  A &  $-$1.968 &  $-$3.377 &  $-$0.143 &  0.031 &  0.051 &  0.064 &  0.156 &  0.982 \\
 y, g$-$y &  FO &  1.5 &  A &  $-$2.200 &  $-$3.406 &  $-$0.142 &  0.030 &  0.049 &  0.061 &  0.151 &  0.984 \\
 r, g$-$r &  FO &  1.7 &  A &  $-$2.735 &  $-$3.743 &  $-$0.449 &  0.022 &  0.030 &  0.044 &  0.064 &  0.997 \\
 i, g$-$i &  FO &  1.7 &  A &  $-$2.401 &  $-$3.679 &  $-$0.290 &  0.022 &  0.030 &  0.045 &  0.065 &  0.997 \\
 z, i$-$z &  FO &  1.7 &  A &  $-$1.935 &  $-$3.557 &  $-$0.168 &  0.030 &  0.041 &  0.060 &  0.088 &  0.994 \\
 y, g$-$y &  FO &  1.7 &  A &  $-$2.166 &  $-$3.584 &  $-$0.165 &  0.027 &  0.037 &  0.055 &  0.080 &  0.995 \\
 r, g$-$r &  FO &  1.9 &  A &  $-$2.756 &  $-$3.783 &  $-$0.496 &  0.018 &  0.019 &  0.033 &  0.021 &  1.000 \\
 i, g$-$i &  FO &  1.9 &  A &  $-$2.440 &  $-$3.734 &  $-$0.371 &  0.022 &  0.022 &  0.039 &  0.025 &  0.999 \\
 z, i$-$z &  FO &  1.9 &  A &  $-$2.017 &  $-$3.635 &  $-$0.227 &  0.043 &  0.044 &  0.078 &  0.050 &  0.997 \\
 y, g$-$y &  FO &  1.9 &  A &  $-$2.233 &  $-$3.650 &  $-$0.296 &  0.036 &  0.037 &  0.065 &  0.042 &  0.998 \\
 r, g$-$r &  FO &  1.5 &  B &  $-$2.606 &  $-$3.712 &  $-$0.520 &  0.026 &  0.058 &  0.057 &  0.089 &  0.990 \\
 i, g$-$i &  FO &  1.5 &  B &  $-$2.273 &  $-$3.632 &  $-$0.350 &  0.027 &  0.059 &  0.058 &  0.090 &  0.990 \\
 z, i$-$z &  FO &  1.5 &  B &  $-$1.824 &  $-$3.467 &  $-$0.126 &  0.033 &  0.073 &  0.072 &  0.113 &  0.984 \\
 y, g$-$y &  FO &  1.5 &  B &  $-$2.048 &  $-$3.511 &  $-$0.226 &  0.031 &  0.067 &  0.066 &  0.104 &  0.986 \\
 r, g$-$r &  FO &  1.7 &  B &  $-$2.616 &  $-$3.668 &  $-$0.505 &  0.044 &  0.081 &  0.092 &  0.101 &  0.988 \\
 i, g$-$i &  FO &  1.7 &  B &  $-$2.286 &  $-$3.596 &  $-$0.346 &  0.043 &  0.081 &  0.092 &  0.100 &  0.988 \\
 z, i$-$z &  FO &  1.7 &  B &  $-$1.839 &  $-$3.441 &  $-$0.133 &  0.048 &  0.090 &  0.102 &  0.111 &  0.985 \\
 y, g$-$y &  FO &  1.7 &  B &  $-$2.064 &  $-$3.480 &  $-$0.230 &  0.046 &  0.086 &  0.098 &  0.106 &  0.986 \\
 r, g$-$r &  FO &  1.9 &  B &  $-$2.602 &  $-$3.729 &  $-$0.454 &  0.018 &  0.025 &  0.034 &  0.015 &  1.000 \\
 i, g$-$i &  FO &  1.9 &  B &  $-$2.286 &  $-$3.677 &  $-$0.328 &  0.024 &  0.034 &  0.046 &  0.020 &  0.999 \\
 z, i$-$z &  FO &  1.9 &  B &  $-$1.864 &  $-$3.564 &  $-$0.177 &  0.055 &  0.079 &  0.106 &  0.047 &  0.996 \\
 y, g$-$y &  FO &  1.9 &  B &  $-$2.080 &  $-$3.585 &  $-$0.250 &  0.046 &  0.066 &  0.088 &  0.039 &  0.997 \\
 r, g$-$r &  FO &  1.5 &  C &  $-$2.494 &  $-$3.658 &  $-$0.471 &  0.011 &  0.029 &  0.021 &  0.021 &  0.999 \\
 i, g$-$i &  FO &  1.5 &  C &  $-$2.161 &  $-$3.581 &  $-$0.292 &  0.014 &  0.038 &  0.028 &  0.028 &  0.998 \\
 z, i$-$z &  FO &  1.5 &  C &  $-$1.726 &  $-$3.407 &  $-$0.136 &  0.028 &  0.076 &  0.056 &  0.056 &  0.992 \\
 y, g$-$y &  FO &  1.5 &  C &  $-$1.946 &  $-$3.455 &  $-$0.149 &  0.023 &  0.063 &  0.046 &  0.046 &  0.995 \\
 r, g$-$r &  FO &  1.7 &  C &  $-$2.504 &  $-$3.652 &  $-$0.475 &  0.017 &  0.034 &  0.032 &  0.019 &  0.999 \\
 i, g$-$i &  FO &  1.7 &  C &  $-$2.175 &  $-$3.579 &  $-$0.310 &  0.021 &  0.042 &  0.039 &  0.023 &  0.999 \\
 z, i$-$z &  FO &  1.7 &  C &  $-$1.750 &  $-$3.400 &  $-$0.179 &  0.041 &  0.081 &  0.074 &  0.043 &  0.995 \\
 y, g$-$y &  FO &  1.7 &  C &  $-$1.967 &  $-$3.448 &  $-$0.183 &  0.034 &  0.068 &  0.062 &  0.036 &  0.996 \\
 r, g$-$r &  FO &  1.9 &  C &  $-$2.550 &  $-$2.834 &  $-$0.212 &  0.025 &  0.266 &  0.079 &  0.004 &  0.997 \\
 i, g$-$i &  FO &  1.9 &  C &  $-$2.238 &  $-$2.634 &  $-$0.131 &  0.021 &  0.217 &  0.064 &  0.003 &  0.999 \\
 z, i$-$z &  FO &  1.9 &  C &  $-$1.872 &  $-$1.675 &   $-$0.384 &  0.007 &  0.077 &  0.023 &  0.001 &  1.000 \\
 y, g$-$y &  FO &  1.9 &  C &  $-$2.071 &  $-$1.941 &   $-$0.231 &  0.012 &  0.126 &  0.037 &  0.002 &  1.000 \\
\hline\hline
\end{tabular}}
\end{table*}

\section{The PAC and PACZ relations in the Rubin-LSST filters} 
The PAC coefficients in the Rubin-LSST filters for the various investigated chemical compositions are listed in Table \ref{pac_f_fo_allz}. The PACZ coefficients are reported in table \ref{pacz_f_fo_allz}.

\begin{table*}
\caption{\label{pac_f_fo_allz} The coefficients of PAC relations for both F and FO-modes and canonical and noncanonical ML assumptions, expressed as $ \log t$ = $a$ + $b$ $\log P$ + c $\text{CI}$, are provided for $Z$ = $0.004$, $Z$ = $0.008$, $Z$ = $0.02$ and $Z$ = $0.03$, in the selected Rubin-LSST filter combinations (see column 3 of the table). The last two columns represent the root-mean-square deviation ($\sigma$) and the R-squared ($R^2$) coefficients.}
\centering
\scalebox{0.75}{
\begin{tabular}{cccccccccccc}
\hline\hline
Z&Y&color&mode&ML&a&b&c&$\sigma_{a}$&$\sigma_{b}$&$\sigma_{c}$&$\sigma$\\
\hline
0.004&0.25&u$-$r&F&A&8.622&$-$0.562&$-$0.174&0.022&0.029&0.019&0.119\\
0.004&0.25&u$-$g&F&A&8.637&$-$0.561&$-$0.250&0.024&0.029&0.028&0.119\\
0.004&0.25&g$-$r&F&A&8.583&$-$0.575&$-$0.550&0.018&0.026&0.057&0.118\\
0.004&0.25&g$-$i&F&A&8.569&$-$0.576&$-$0.396&0.017&0.026&0.041&0.118\\
0.004&0.25&i$-$z&F&A&8.434&$-$0.569&$-$1.484&0.012&0.025&0.238&0.117\\
0.004&0.25&u$-$r&F&B&8.098&$-$0.397&0.013&0.008&0.015&0.007&0.107\\
0.004&0.25&u$-$g&F&B&8.097&$-$0.425&0.035&0.008&0.016&0.010&0.107\\
0.004&0.25&g$-$r&F&B&8.110&$-$0.339&$-$0.066&0.008&0.014&0.025&0.107\\
0.004&0.25&g$-$i&F&B&8.104&$-$0.356&$-$0.023&0.008&0.014&0.018&0.107\\
0.004&0.25&i$-$z&F&B&8.114&$-$0.407&0.272&0.009&0.013&0.096&0.107\\
0.008&0.25&u$-$r&F&A&8.470&$-$0.288&$-$0.218&0.016&0.033&0.014&0.123\\
0.008&0.25&u$-$g&F&A&8.472&$-$0.289&$-$0.299&0.016&0.034&0.019&0.123\\
0.008&0.25&g$-$r&F&A&8.465&$-$0.303&$-$0.776&0.015&0.030&0.045&0.121\\
0.008&0.25&g$-$i&F&A&8.440&$-$0.319&$-$0.546&0.015&0.030&0.033&0.122\\
0.008&0.25&i$-$z&F&A&8.215&$-$0.342&$-$1.844&0.019&0.030&0.181&0.123\\
0.008&0.25&u$-$r&F&B&8.646&$-$0.761&$-$0.021&0.015&0.034&0.017&0.118\\
0.008&0.25&u$-$g&F&B&8.638&$-$0.787&$-$0.010&0.015&0.034&0.024&0.118\\
0.008&0.25&g$-$r&F&B&8.657&$-$0.702&$-$0.188&0.013&0.031&0.058&0.117\\
0.008&0.25&g$-$i&F&B&8.650&$-$0.713&$-$0.122&0.012&0.031&0.042&0.117\\
0.008&0.25&i$-$z&F&B&8.614&$-$0.748&$-$0.422&0.016&0.030&0.224&0.118\\
0.02&0.28&u$-$r&F&A&8.363&$-$0.744&0.023&0.023&0.026&0.016&0.081\\
0.02&0.28&u$-$g&F&A&8.362&$-$0.743&0.031&0.024&0.026&0.022&0.081\\
0.02&0.28&g$-$r&F&A&8.365&$-$0.743&0.084&0.021&0.025&0.057&0.081\\
0.02&0.28&g$-$i&F&A&8.369&$-$0.742&0.061&0.019&0.024&0.041&0.081\\
0.02&0.28&i$-$z&F&A&8.398&$-$0.734&0.266&0.008&0.022&0.208&0.081\\
0.02&0.28&u$-$r&F&B&8.485&$-$0.697&0.022&0.015&0.019&0.009&0.089\\
0.02&0.28&u$-$g&F&B&8.488&$-$0.693&0.026&0.014&0.019&0.013&0.090\\
0.02&0.28&g$-$r&F&B&8.476&$-$0.706&0.106&0.015&0.017&0.034&0.089\\
0.02&0.28&g$-$i&F&B&8.485&$-$0.700&0.068&0.013&0.017&0.024&0.089\\
0.02&0.28&i$-$z&F&B&8.516&$-$0.682&0.201&0.011&0.016&0.117&0.090\\
0.03&0.28&u$-$r&F&A&8.279&$-$0.751&0.045&0.019&0.026&0.014&0.076\\
0.03&0.28&u$-$g&F&A&8.279&$-$0.748&0.059&0.020&0.026&0.019&0.076\\
0.03&0.28&g$-$r&F&A&8.281&$-$0.753&0.172&0.018&0.025&0.048&0.076\\
0.03&0.28&g$-$i&F&A&8.290&$-$0.751&0.125&0.016&0.024&0.035&0.076\\
0.03&0.28&i$-$z&F&A&8.353&$-$0.735&0.561&0.010&0.022&0.178&0.076\\
0.03&0.28&u$-$r&F&B&8.287&$-$0.770&0.048&0.017&0.023&0.012&0.076\\
0.03&0.28&u$-$g&F&B&8.287&$-$0.768&0.065&0.018&0.024&0.017&0.076\\
0.03&0.28&g$-$r&F&B&8.290&$-$0.772&0.185&0.016&0.022&0.042&0.076\\
0.03&0.28&g$-$i&F&B&8.299&$-$0.769&0.133&0.014&0.022&0.031&0.076\\
0.03&0.28&i$-$z&F&B&8.367&$-$0.752&0.592&0.010&0.020&0.156&0.076\\
0.004&0.25&u$-$r&FO&A&7.765&$-$1.178&0.540&0.169&0.103&0.157&0.086\\
0.004&0.25&u$-$g&FO&A&6.795&$-$1.349&1.684&0.410&0.135&0.445&0.086\\
0.004&0.25&g$-$r&FO&A&8.225&$-$1.092&0.773&0.043&0.087&0.240&0.087\\
0.004&0.25&g$-$i&FO&A&8.263&$-$1.090&0.517&0.034&0.087&0.161&0.087\\
0.004&0.25&i$-$z&FO&A&8.449&$-$1.144&1.068&0.039&0.096&1.193&0.086\\
0.004&0.25&u$-$r&FO&B&8.034&$-$0.347&0.095&0.014&0.018&0.010&0.023\\
0.004&0.25&u$-$g&FO&B&8.002&$-$0.355&0.158&0.016&0.018&0.017&0.023\\
0.004&0.25&g$-$r&FO&B&8.082&$-$0.334&0.235&0.012&0.017&0.027&0.023\\
0.004&0.25&g$-$i&FO&B&8.090&$-$0.333&0.166&0.011&0.017&0.019&0.023\\
0.004&0.25&i$-$z&FO&B&8.144&$-$0.336&1.117&0.012&0.017&0.129&0.023\\
0.008&0.25&u$-$r&FO&A&7.998&$-$0.426&0.052&0.106&0.059&0.082&0.068\\
0.008&0.25&u$-$g&FO&A&7.959&$-$0.433&0.103&0.159&0.065&0.156&0.068\\
0.008&0.25&g$-$r&FO&A&8.038&$-$0.419&0.102&0.049&0.053&0.173&0.068\\
0.008&0.25&g$-$i&FO&A&8.043&$-$0.419&0.071&0.042&0.053&0.121&0.068\\
0.008&0.25&i$-$z&FO&A&8.072&$-$0.429&0.653&0.028&0.058&0.913&0.068\\
0.008&0.25&u$-$r&FO&B&7.925&$-$0.590&0.316&0.136&0.108&0.114&0.063\\
0.008&0.25&u$-$g&FO&B&7.849&$-$0.549&0.463&0.214&0.118&0.219&0.064\\
0.008&0.25&g$-$r&FO&B&8.132&$-$0.605&0.797&0.053&0.094&0.231&0.062\\
0.008&0.25&g$-$i&FO&B&8.167&$-$0.610&0.566&0.042&0.094&0.160&0.062\\
0.008&0.25&i$-$z&FO&B&8.384&$-$0.636&1.033&0.031&0.100&1.135&0.062\\
0.02&0.28&u$-$r&FO&A&8.102&$-$0.443&0.016&0.151&0.108&0.107&0.056\\
0.02&0.28&u$-$g&FO&A&8.109&$-$0.437&0.014&0.216&0.118&0.190&0.056\\
0.02&0.28&g$-$r&FO&A&8.108&$-$0.448&0.061&0.070&0.097&0.242&0.056\\
0.02&0.28&g$-$i&FO&A&8.111&$-$0.449&0.046&0.057&0.096&0.170&0.056\\
0.02&0.28&i$-$z&FO&A&8.137&$-$0.461&0.536&0.037&0.101&1.288&0.056\\
0.02&0.28&u$-$r&FO&B&7.761&$-$0.716&0.308&0.456&0.190&0.309&0.041\\
0.02&0.28&u$-$g&FO&B&7.350&$-$0.823&0.739&0.632&0.217&0.540&0.040\\
0.02&0.28&g$-$r&FO&B&8.102&$-$0.608&0.370&0.214&0.151&0.703&0.041\\
0.02&0.28&g$-$i&FO&B&8.134&$-$0.599&0.232&0.172&0.148&0.493&0.041\\
0.02&0.28&i$-$z&FO&B&8.233&$-$0.579&1.069&0.070&0.172&3.769&0.041\\
\hline\hline
\end{tabular}}
\end{table*}

\begin{table*}
\caption{\label{pacz_f_fo_allz}The coefficients of the F
and FO PACZ relations $\log t$ = $a$ + $b \log P$ + $c$ CI + d [Fe/H] ), derived by adopting both {\sl cases A} and {\sl B} ML relations.}
\centering
\begin{tabular}{ccccccccccc}
\hline\hline
color&ML&a&b&c&d&$\sigma_{a}$&$\sigma_{b}$&$\sigma_{c}$&$\sigma_{d}$&$\sigma$\\
\hline
Fundamental mode\\
\hline
u$-$r&A&8.545&$-$0.583&$-$0.109&0.097&0.011&0.015&0.008&0.009&0.115\\
u$-$g&A&8.543&$-$0.595&$-$0.141&0.100&0.011&0.015&0.011&0.009&0.116\\
g$-$r&A&8.540&$-$0.562&$-$0.434&0.082&0.009&0.014&0.026&0.008&0.114\\
g$-$i&A&8.519&$-$0.566&$-$0.312&0.062&0.008&0.014&0.019&0.007&0.114\\
i$-$z&A&8.367&$-$0.586&$-$1.578&0.002&0.007&0.013&0.100&0.007&0.114\\
u-r&B&8.459&-0.735&0.017&-0.202&0.012&0.018&0.009&0.015&0.122\\
u-g&B&8.454&-0.743&0.029&-0.207&0.012&0.018&0.012&0.015&0.122\\
g-r&B&8.471&-0.715&0.023&-0.190&0.011&0.017&0.032&0.014&0.122\\
g-i&B&8.474&-0.710&0.009&-0.187&0.010&0.017&0.023&0.013&0.122\\
i-z&B&8.478&-0.707&0.022&-0.185&0.008&0.016&0.116&0.011&0.122\\
\hline
First overtone mode\\
\hline
u$-$r&A&7.897&$-$0.762&0.199&$-$0.202&0.105&0.044&0.069&0.045&0.081\\
u$-$g&A&7.618&$-$0.796&0.480&$-$0.281&0.157&0.045&0.130&0.056&0.080\\
g$-$r&A&8.120&$-$0.719&0.253&$-$0.130&0.043&0.041&0.137&0.031&0.081\\
g$-$i&A&8.142&$-$0.714&0.159&$-$0.118&0.034&0.041&0.094&0.027&0.081\\
i$-$z&A&8.214&$-$0.714&0.930&$-$0.085&0.020&0.048&0.702&0.016&0.082\\
u-r&B&8.099&-0.693&0.160&-0.185&0.042&0.035&0.029&0.030&0.069\\
u-g&B&8.058&-0.686&0.236&-0.199&0.055&0.036&0.049&0.033&0.069\\
g-r&B&8.196&-0.695&0.454&-0.152&0.022&0.032&0.072&0.026&0.068\\
g-i&B&8.219&-0.698&0.332&-0.137&0.018&0.032&0.051&0.026&0.068\\
i-z&B&8.375&-0.714&2.475&-0.065&0.013&0.031&0.339&0.025&0.067\\
\hline\hline
\end{tabular}
\end{table*}

\section*{Acknowledgements}
We are grateful to the anonymous Referee for their insightful suggestions, which have greatly improved the overall quality and readability of the manuscript.
G.D.S. and M.M. thank the ISSI International Team project {\it SHoT: The Stellar Path to the $H_{0}$ Tension in the Gaia, TESS, LSST and JWST Era} (PI: G. Clementini).
G.D.S. thanks Istituto Nazionale di Fisica Nucleare (INFN), Naples section, specific initiatives QGSKY and Moonlight2 and
Gaia DPAC funds from Istituto Nazionale di Astrofisica (INAF) and the Agenzia Spaziale Italiana (ASI) (PI: M.G. Lattanzi). 
SC acknowledges support from PRIN MIUR2022 Progetto "CHRONOS" (PI: S. Cassisi) finanziato dall'Unione Europea - Next Generation EU.
This paper was based upon work from COST Action CA21136-Addressing observational tensions in cosmology with systematics and fundamental physics (CosmoVerse), supported by COST (European Cooperation in Science and Technology).
This project has received funding from the European Union’s Horizon 2020 research and innovation program under the Marie Sklodowska-Curie grant agreement No. 886298.

\section*{Data Availability}
Data availability statement: The data underlying this article are available in the article and its online supplementary material.



\bibliographystyle{mnras}
\bibliography{desomma_main} 


\bsp	
\label{lastpage}
\end{document}